\documentclass[aps,prb,reprint,twocolumn,showpacs,superscriptaddress,nofootinbib]{revtex4-2}
\usepackage{physics}
\usepackage{amsmath}
\usepackage{graphicx}
\usepackage{lmodern}
\usepackage{amsmath}
\usepackage{color}
\usepackage{amssymb}
\usepackage{bm}
\usepackage[caption=false]{subfig}
\usepackage{braket}
\usepackage{multirow}
\usepackage{tikz}
\usetikzlibrary{quantikz}
\usepackage{mathtools}
\usepackage{afterpage}
\usepackage{dsfont}
\usepackage{multibib}
\newcites{SM}{References for Supplement}

\usepackage{tikz}

\newcommand{\circledsmall}[2][1]{%
  \tikz[baseline=(char.base)]{
    \node[draw,circle,inner sep=#1pt] (char) {\scriptsize #2};
  }%
}

\usepackage[compact]{titlesec}         
\titlespacing{\section}{5pt}{5pt}{5pt} 
\AtBeginDocument{
  \setlength\abovedisplayskip{5pt}
  \setlength\belowdisplayskip{5pt}}

\newcommand{\be}{\begin{equation}}
\newcommand{\ee}{\end{equation}}

\usepackage[dvipsnames]{xcolor}
\usepackage[colorlinks=true]{hyperref}
\hypersetup{
    colorlinks=true,
    linkcolor=blue,
    citecolor=blue,
    urlcolor=blue} 

\makeatletter
\def\maketitle{
\@author@finish
\title@column\titleblock@produce
\suppressfloats[t]}
\makeatother

\begin{document}
\title{Observation of disorder-induced superfluidity}
\author{Google Quantum AI and collaborators\,$^\dagger$}

\begin{abstract}
The emergence of states with long-range correlations in a disordered landscape is rare, as disorder typically suppresses the particle mobility required for long-range coherence. But when more than two energy levels are available per site, disorder can induce resonances that locally enhance mobility. Here we explore phases arising from the interplay between disorder, kinetic energy, and interactions on a superconducting processor with qutrit readout and control. Compressibility measurements distinguish an incompressible Mott insulator from surrounding compressible phases and reveal signatures of glassiness, reflected in non-ergodic behavior. Spatially-resolved two-point correlator measurements identify regions of the phase diagram with a non-vanishing condensate fraction. We also visualize the spectrum by measuring the dynamical structure factor. A linearly-dispersing phonon mode materializes in the superfluid, appearing even when disorder is introduced to the clean Mott insulator. Our results provide strong experimental evidence for disorder-induced superfluidity.
\end{abstract}

\maketitle

Condensate formation, as in superfluids or superconductors, necessitates the onset of long-range phase coherence. This coherence manifests as the macroscopic occupation of a single quantum state\,\cite{Leggett_1999_Superfluidity}. Such states are generally fragile to spatial inhomogeneities that relax momentum and reduce the condensate fraction; impurities or boundary effects will scramble the phase of a propagating particle, resulting in predominantly destructive interference between paths and driving the system into an insulating state. The localization of the many-body wavefunction is generally believed to be incompatible with extended phases of matter such as superfluidity. Nonetheless, in a multi-level bosonic system, disorder could lead to local condensate formation by enhancing resonant tunneling between nearby sites. These processes can give rise to rare-region effects, where atypical pockets of nearly resonant sites form local coherent puddles within a globally disordered system. Here, we show that these local mechanisms can give rise to a macroscopic superfluid characterized by long-range phase coherence and linearly-dispersing phonon modes. 

In this work, we study the low-energy phase diagram of strongly interacting bosons on a two-dimensional lattice. Independent tuning of on-site disorder $W$ and nearest-neighbor hopping $J$ (Fig.~1) allows us to traverse the insulator–superfluid transition and probe distinct insulating phases. The interplay between these parameters has been extensively explored through analytical \cite{Fisher_PR_1989,Pollet_PRL_2009}, numerical \cite{Monien_PRB_1999,CapogrossoPRA2008,Prokofev_PRB2009,dang2009}, and experimental studies in both solid-state \cite{Crowell1997,amo2009superfluidity,saxberg2022disorder} and ultracold-atom platforms \cite{Greiner2002,zwierlein2006direct,deissler2010delocalization,d2014observation,Yu2024,koehn2025,russ2025}. 

At low disorder and when the kinetic energy is much smaller than the interaction energy, $J/U \ll 1$, the system is in a strongly-correlated Mott insulator\,(MI) phase, characterized by a finite gap to single-particle excitations. As hopping between adjacent sites increases, the system undergoes a quantum phase transition to a superfluid\,(SF) phase, where bosons can move freely. At low hopping $J$, increasing disorder drives a transition from the MI to the Bose glass\,(BG) phase—a gapless insulator whose properties and boundaries remain under debate \cite{Fallani2007,Prokofev_PRB2009,Pollet_PRL_2009,Prokofev_PRL_2011,yu2012bose,Meldgin_naturePhysics_2016}. Superfluidity---flow without viscosity---is inherently a dynamical phenomenon. Thus, while long-range phase coherence is a prerequisite for superfluidity, frequency-resolved probes are necessary for providing unambiguous evidence for the phase. Direct experimental detection of disorder-induced superfluidity has remained elusive, largely due to the difficulty of resolving dynamical quantities in quantum simulators or synthesizing materials with tunable disorder. 

\begin{figure}[th!]
\centering
\includegraphics[width=0.3\textwidth]{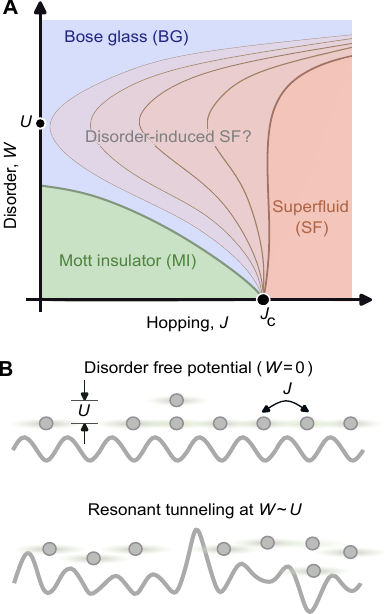}
\caption{\small{\textbf{Phases of interacting bosons in a disordered landscape.} (\textbf{A})\,Heuristic phase diagram as a function of hopping ($J$) and disorder strength ($W$), with $U$ the on-site interaction. Theory and numerics \cite{Prokofev_PRB2009, Prokofev_PRL_2011} point to a `re-entrant' disorder-mediated superfluid phase above the Mott insulator; however, no evidence for this has been observed experimentally. The precise shape of the phase diagram is therefore unknown. (\textbf{B})\,For $W=0$, once $J$ exceeds a critical value $J_\text{c}$, bosons hop freely between sites and form a coherent superfluid. Although disorder localizes single-particle states, particles may undergo short-range resonant tunneling---which could, in principle, promote superfluidity.}
\label{fig:main_phase_diagram}}
\end{figure}

\begin{figure*}[th!]
\centering
\includegraphics[width=0.68\textwidth]{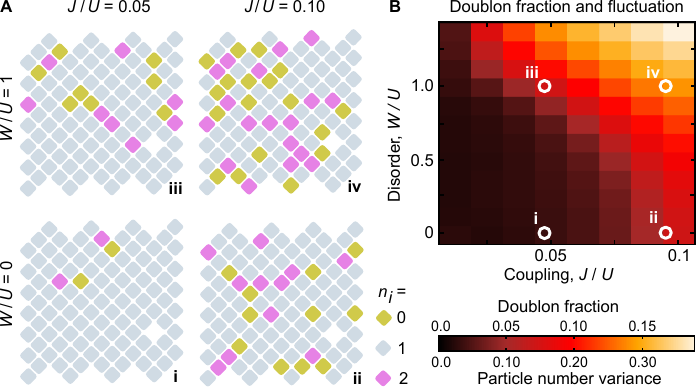}
\caption{\small{\textbf{Formation of particle-hole excitations.} (\textbf{A})\,Typical population measurement instances, showing dependence of doublon ($n=2$) generation 
on different excitation hopping $J$ and disorder $W$ values on a 2D grid of qutrits with $U/(2\pi)=190$~MHz. As $J/U$ is increased, more particle-hole pairs are formed. Data is post-selected to enforce number conservation, yielding matched holon and doublon counts.(\textbf{B})\,Measured doublon fraction in the $J$-$W$ parameter plane, averaged over 10,000 measurement instances and 10 disorder realizations over a $4 \times 9$ grid of qutrits. Since $\langle n \rangle = 1$, the spatially averaged excitation number variance, $\langle n^2 \rangle - \langle n \rangle^2$, can be directly obtained by multiplying the measured doublon fraction by 2. }
\label{fig:main_doublon_frac}}
\end{figure*}

The transmon-based quantum processor used here is assembled from a two-dimensional array of coupled nonlinear superconducting resonators. The elementary excitations---microwave photons in each resonator---obey Bose-Einstein statistics. Any integer number of microwave photons can, in principle, occupy the same resonator. The physics of these strongly interacting bosons can be effectively captured by the Bose-Hubbard model,

\begin{equation}
    H= -J\sum_{\langle i,j\rangle}\left(a^\dagger_i a_j + a^\dagger_j a_i\right) + \frac{U}{2}\sum_i n_i (n_i-1) - \sum_i \mu_i n_i,
\label{eq:BoseHubb}
 \end{equation}
where $a^\dagger_i = |1\rangle \langle 0| + \sqrt{2}\, |2\rangle \langle 1| + \dots$ describes a creation operator for a boson at site $i$. Hopping between nearest neighboring sites is mediated by tunable coupling elements\,\cite{gmon_PRL_2014} with coupling constant $J$; $n_i\equiv a^\dagger_i a_i$ is the number operator at site $i$; $U$ is the on-site Hubbard repulsion; and the on-site potential, $\mu_i$, is drawn from a uniform distribution centered at some fixed chemical potential $\mu$ with width $W$: $\mu_i \sim \mu + \text{Unif}[-W/2,W/2]$. By engineering the initial states and adiabatic protocols, we restrict the dynamics to a regime where at most two excitations reside on any given site (Fig.\,2), effectively treating the resonators as `qutrits'. This occupancy constraint is enforced by the strength of the transmon nonlinearity. 

\vspace{1mm}
The ground state phase diagram of this model was first characterized analytically by Fisher \textit{et al.} \cite{Fisher_PR_1989} and later quantified numerically \cite{Monien_PRB_1999,CapogrossoPRA2008}. In the clean two-dimensional ($2D$) system, a phase transition between the Mott insulator (MI) and superfluid (SF) is expected at $J_\text{c}/ U\simeq 0.06$. The existence of a re-entrant superfluid `finger' above small $J$, most prominent for values of $W\sim U$, has also been proposed on the basis of Monte-Carlo data\,\cite{Prokofev_PRB2009, Prokofev_PRL_2011,morone2024reentrantphasetransitionsinduced}. Early studies of the disordered superfluid-to-insulator transition were performed on porous helium-4 films \cite{Crowell1997} and in granular superconductors. Both the clean \cite{Greiner2002} and disordered \cite{Yu2024,koehn2025,russ2025} Bose-Hubbard models have also been realized in optical lattices, and the essential characteristics, such as phase coherence in the superfluid and non-ergodicity in the Bose glass, have been reported \cite{Esslinger_PRL_2004,Campbell2006_science, gemelke2009situ,Bakr2010_science,Sherson2010_nature,weitenberg2011single,Endres2011observation,Wei_PRX_2023}. The existence of a re-entrant superfluid phase mediated by disorder has been disputed, with most works \cite{Sanchez-Palencia2010} finding that increased disorder leads to increased dissipation and localization incompatible with superfluidity in one spatial dimension $(1D)$  \cite{Fallani2007,d2014observation}, $2D$ \cite{Krinner2013,Allard2012}, and $3D$ \cite{White2009,Pasienski2010}. There are some indications, however, that repulsive interactions can promote delocalization and phase coherence\,\cite{deissler2010delocalization}. To our knowledge, there exists no definitive experimental evidence of re-entrant disorder-induced superfluidity in the Bose-Hubbard model. 

\begin{figure*}[th!]
\centering
\includegraphics[width=0.99\textwidth]{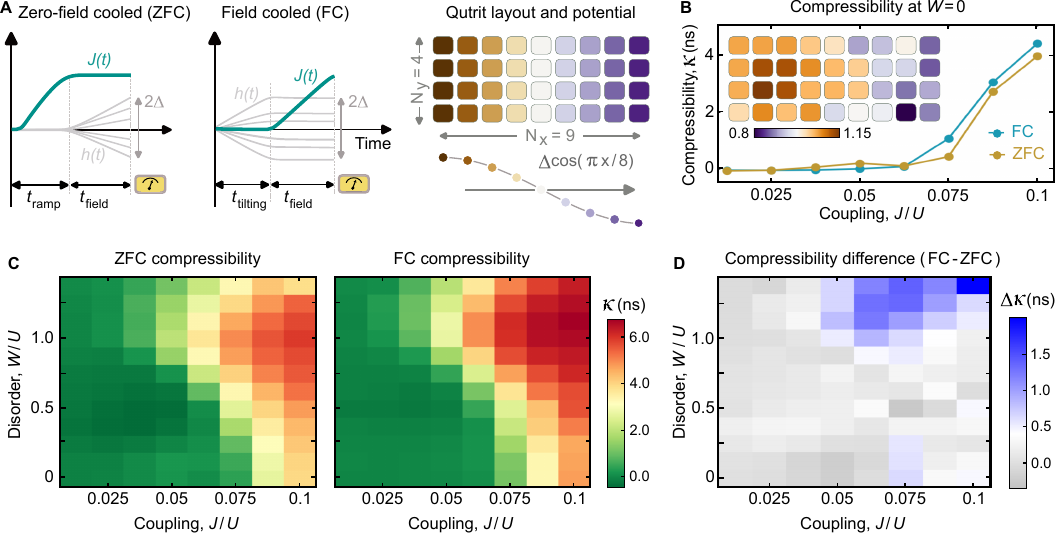}
\caption{\small{\textbf{Mapping the compressibility in the $J-W$ plane} (\textbf{A})\,Annealing ramp protocols, analogous to `field-cooling' (FC) and `zero-field-cooling' (ZFC) experiments in glasses. In the ZFC case, the potential gradient is applied after the state has been adiabatically prepared; for the FC ramps, the gradient is applied during state preparation, with $t_\text{ramp}=100\text{ ns}$ and $t_\text{field}=250 \text{ ns}$. In a $N_x=9$ by $N_y=4$ grid, potential qutrit energies are ramped to $\Delta \cos( \pi x/8)$, with the strength of the perturbation $\Delta/(2\pi) = 30 \text{ MHz}$. (\textbf{B})\,Extracted compressibility, using Eq. (2), at zero disorder vs $J$ from $ \langle n \rangle $ measurements with both FC and ZFC protocols. The inset shows $ \langle n_i \rangle $ at $J/U=0.1$ and $W=0$ for a state prepared with the FC protocol. (\textbf{C})\,Measured compressibility in the  $J-W$ plane, displaying transition to a compressible state as $W$ and $J$ are increased. 150 disorder realizations were taken. (\textbf{D})\, Difference in compressibility, $\Delta \kappa$, between FC and ZFC protocols, showing emergence of ergodicity breaking ($\Delta \kappa\neq 0$).}
\label{fig:main_compressibility}}
\end{figure*}

\vspace{1mm}
To visualize particle-hole formation, we initialize a $2D$ array of qutrits at commensurate unit filling, $n=1$, by placing a single excitation on each site: $|111...\rangle$. To remain in the low-energy subspace, we adiabatically ramp to target Hamiltonian parameters\,(see Supp.). In Fig.\,2 we show typical boson populations as a function of $J$ and $W$. We observe an increased density of doublons (the $|2\rangle$ qutrit state), matched by holon formation (the $|0\rangle $ state) on other sites, with the total density of particle-hole excitations growing with the disorder strength \(W\) or coupling \(J\) (brighter colors in panel~B). The increased doublon–holon incidence as a function of $W$ can be understood as arising when detunings between nearby sites approach $U$ and hence permit tunneling, making particle–hole pairs energetically favorable.

The particle-number fluctuation $\langle n^2 \rangle - \langle n \rangle^2$, exactly equal to twice the doublon fraction if restricting ourselves to the qutrit subspace, can also be readily extracted from our measurements.  This quantity, considered as a function of $J$ and $W$, reflects the basic properties of the MI and SF phases: the MI phase has a finite gap to single-particle excitations, which remain well localized on individual lattice sites, whereas the SF phase is gapless and supports freely propagating excitations with well-defined momenta. This reasoning is commonly used to identify phases in cold-atom systems\,\cite{Greiner2002}.

\begin{figure*}[th!]
\centering
\includegraphics[width=0.97\textwidth]{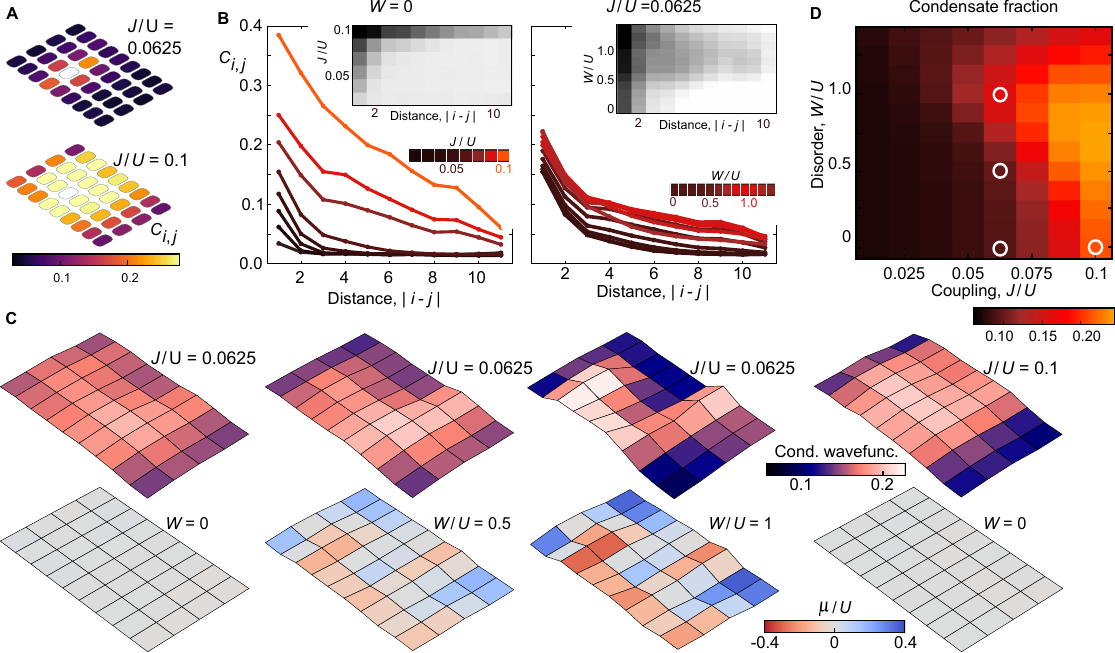}
\caption{\small{\textbf{Condensate wavefunction and fraction} (\textbf{A}) Two instances of $C_{ij}$ measurement for a fixed $i$ (white square) for two $J/U$ values of 0.0625 and 0.1, setting $W=0$, over a 4 by 9 grid of qutrits. (\textbf{B}) $C_{ij}$ measurement averaged over all $i$ and $j$ pairs vs. Manhattan distances for (left) $W=0$ and several $J$ values, and (right) $J/U=0.0625$ and several $W$ values, averaged over 10 disorder realizations. The insets show the same data in grayscale. (\textbf{C}) Measurement of $C_{ij}$ for all pairs allows construction of the SPDM\,(not shown). The largest eigenstates of the SPDM for 4 different values of $J$ and $W$ are shown. (\textbf{D}) The largest eigenvalue of constructed SPDMs, based on $C_{ij}$ measurement for all pairs, vs. $J$ and $W$. When $W \neq 0$, 10 disorder realizations are averaged. }
\label{fig:main_cond_frac}}
\end{figure*}

\begin{figure*}[th!]
\centering
\includegraphics[width=0.98\textwidth]{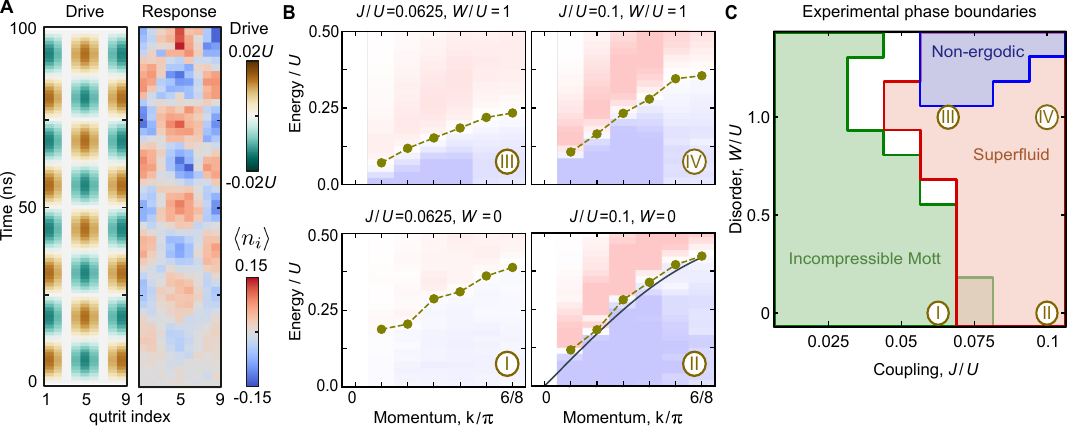}
\caption{\small{\textbf{Direct measurement of dynamical structure factor.} (\textbf{A}) On a 4-by-9 grid of qutrits, after ramping to the final $J$ and $W$ values, a spatiotemporal potential modulation pattern was applied, and $\langle n \rangle$ was measured at each site\,(see Supp.). A typical drive and response for $k=2\pi/8$ momentum and $\omega/(2\pi)=40$~MHz frequency is shown. The drive is spatially uniform over the shorter dimension and sinusoidal in the other dimension. The response shown here has subtracted from it the response with no drive and is averaged over the smaller spatial dimension. (\textbf{B}) Reactive part of the space- and time-resolved $\langle n \rangle$, after averaging over the shorter dimension ($L=4$). 10 disorder realizations are taken for experiments where $W\neq 0$. The locations of the zeros (where the system is maximally absorptive, corresponding to its fundamental modes) are indicated by brown dots. Theory prediction using the Gutzwiller ansatz (see Supp.) is plotted in black for the disorder-free experiments. \,(\textbf{C}) Phase boundaries were delineated by binarization of the compressibility, the difference between ZFC and FC\,(Fig.\,3), and the condensation fraction measurements\,(Fig. 4). The 4 parameter points used in Fig.\,5\,B are shown with encircled numbers. }
\label{fig:main_dyn_response}}
\end{figure*}

On-site particle-number fluctuations $\langle n_i^2 \rangle - \langle n_i \rangle^2$, as shown in Fig. 2B, may reflect either particle mobility or the underlying disordered potential landscape. As such, this metric is not an appropriate diagnostic for the compressibility in the presence of intermediate-to-strong disorder (see Supp. for further discussion). We instead quantify the compressibility $\kappa$ by measuring the system's response, $\langle \delta n_i\rangle$, to an applied potential gradient, $\delta\mu$. A non-zero compressibility, $\kappa\equiv \left(\partial n/\partial\mu\right) > 0$, indicates a finite density of excitations at zero energy, thus allowing us to distinguish the gapped MI from surrounding compressible phases. Crucially, in the presence of disorder, the observables might depend on the path taken through phase space during preparation of the Hamiltonian. If two different state preparation protocols \,(Fig. 3A) yield different values of the compressibility, $\kappa_\text{path 1}\neq \kappa_\text{path 2}$, then the system is non-ergodic over experimentally-accessible timescales. This observation allows us to distinguish between non-ergodic glass-like states and surrounding ergodic states such as MI and SF. Accordingly, we implement two protocols: (1) prepare the state by first bringing qutrits to resonance and ramping up $J(t)$ before applying the gradient, or (2) apply the cosine perturbation first and then prepare the state---in the presence of the applied field---by ramping up $J(t)$. We refer to these as “zero-field cooled’’ (ZFC) and “field-cooled’’ (FC), respectively, in analogy to classical experiments on glassy systems \cite{fischer_hertz_text}.

To measure $\kappa$, a static perturbation $\mu_i \to\mu_i + \Delta \cos(\pi\,\bm{x}_i/L)$ is applied to each of the on-site potentials, and the response is measured:
\begin{equation}
\kappa = \frac{2}{N\Delta}\sum_i\langle n_i\rangle\cos(\pi\,\bm{x}_i/L).
\end{equation}
In Fig.\,3B, we plot the compressibility obtained using both the FC and ZFC protocols as a function of $J$ when $W=0$. In panel (C), we extend the measurements to the full $J-W$ plane, and observe that the onset of compressibility coincides with the emergence of particle-hole excitations as seen in Fig.\,2. The incompressible phase (green area) is gapped; a finite amount of energy, of order $U$, is required to bridge the gap above the ground state. This gapped insulator can be identified as a Mott, i.e. `correlated', insulator.  While both protocols indicate the existence of a compressible state as a function of either increasing $J$ or $W$, the reported value of the compressibility differs at large disorder (blue region in Fig.\,3D). This difference is a hallmark of glassy behavior and possibly signals a breakdown of ergodicity\,\cite{MezardParisiVirasoro}, at least on the timescales accessible to this experiment. Note that $\kappa$ is the static, long-wavelength limit of the dynamical structure factor, which we define and measure in Fig.\,5. 

\vspace{1mm}
A distinguishing property of superfluidity is the macroscopic occupation of a single quantum state, reflected in the emergence of a non-zero condensate fraction $n_0\gg1/N$~($N$ being the total number of particles). To obtain this quantity, we construct the spatially-resolved two-point correlator, defined

\vspace{-3mm}
\begin{equation}
    C_{ij} \equiv \langle a^\dagger_i a_j\rangle, 
\end{equation}
\noindent where $C_{ij}$ is the single-particle density matrix (SPDM) whose largest eigenvalue, $\lambda_0$, gives the condensate fraction $n_0 \equiv \lambda_0/N$ \cite{Yang_RMP_1962}. We interpret $n_0$ as the fraction of particles in the macroscopically occupied $k\to0$ mode, serving as a proxy for phase coherence. In Fig.\,4A we show two typical instances of $C_{ij}$ in a disorder-free landscape, fixing $i$ to a given lattice site (marked in white) and showing its correlation with all other sites $j$. We plot results for $J/U = 0.0625$\,(top), in the MI, and $J /U=0.1 $\,(bottom), which is deep in the SF phase. The two-point correlator shows rapid spatial decay in the MI, whereas in the SF the correlations are stronger and decay more slowly. In Fig.\,4B we measure $C_{ij}$ for all $i,j$ pairs and plot the correlator as a function of Manhattan distance for various values of $J$ at $W=0$ (left) and for a ranges of $W$ values at fixed $J= 0.0625U$\,(right). In the MI phase, the correlations decay rapidly to a constant background (left plot, dark lines), whereas $C_{ij}$ becomes significantly longer-ranged as $J$ approaches $0.1U$ (left plot, brighter colors). The cut at $J/U=0.0625$ shows a non-monotonic dependence on disorder strength. As seen more clearly in the inset plots, $C_{ij}$ developed longer-ranged behavior around $W \sim U$. This suggests that when the disorder is on the order of $U$, the generation of particle-hole pairs is promoted by resonances between sites with mutual detuning $U$; this, in turn, can lead to the emergence of long-range phase coherence.

By measuring $C_{ij}$ for all $i,j$ pairs in a $4 \times 9$ lattice of qutrits, we construct the full SPDM. Its largest normalized eigenvalue and corresponding eigenfunction yield the condensate fraction and condensate wavefunction, respectively. In Fig.\,4C we show four example condensate wavefunctions, matched to the disorder landscapes that produced them in the bottom row. A clear correlation between the on-site disorder and the spatial variation of the wavefunction is observed. The condensation pattern at $J/U=0.1$ in the disorder-free case (far right) resembles a standing wave, as expected. When $W\neq 0$, we find that the condensate wavefunction tracks the underlying disordered landscape. In (D) we summarize these data by plotting the condensation fraction, obtained from SPDM measurements, across the $J$–$W$ plane. This plot confirms the signatures hinted at in other panels: the emergence of a condensate with increasing disorder, peaked for disorder strength close to the Hubbard $U$.

Superfluidity is distinguished by its dynamical and topological properties, including frictionless flow and the emergence of collective excitations. This contrasts with Bose–Einstein condensation, a ground-state phenomenon that can arise even in the absence of interactions. Among the most notable of these dynamical features is the emergence of sound—gapless, linearly-propagating Bogoliubov modes. While a direct measurement of the excitation spectrum is difficult, we can nevertheless probe the dynamical structure factor,
\begin{equation}
    \chi^R(x_i,t;x_j,t') = -i\Theta(t-t')\langle[ n(x_i,t),n(x_j,t')]\rangle,
\end{equation}
where $\Theta$ is the Heaviside step function, by driving the system with a standing wave in the horizontal direction at fixed wave-vector $k = 2\pi m/L$ and frequency $\omega$\,(Fig.\,5\,A), $\mu_j \rightarrow A_d\sin(\omega t)\cos(k x_j)$ with $L=8$. In the condensate, the poles of the structure factor coincide with those of the phonon mode to all orders in perturbation theory. This surprising coincidence in the spectra derives from the condensate-induced intermixing of single-particle and density excitations \cite{Griffin1993}. The dynamical structure factor thus reveals a branch of the quasiparticle spectrum, offering complementary evidence for superfluidity beyond that captured by the condensate fraction. 

In Fig.\,5\,B we plot the real part of the Fourier-transformed dynamical structure factor for different values of $J$ and $W$. By locating the zero-crossing of the reactive response, we can extract the points in $(k,\omega)$ space that are most absorptive; in other words, the modes of the system. We observe very little response in the Mott insulator\,($\circledsmall{I}$). By contrast, the superfluid exhibits a linearly dispersing mode, as shown in $\circledsmall{II}$. At weak hopping and near-critical disorder, i.e. $J/U=0.065$ and $W/U \approx 1$, a linearly-dispersing mode appears once more $\circledsmall{III}$. We also take data deep in the disordered superfluid ($J/U=0.1$, $W/U = 1$), showing that the superfluid is robust to the disorder strengths considered here (\,$\circledsmall{IV}$). Note that the slope of the disordered superfluid, corresponding to the speed of sound $c_s$ of the phonon mode, is depressed relative to that of the clean superfluid. This is because $c_s$ is related to the superfluid density $n_s$ and the compressibility $\kappa$ through $c^2_s \sim n_s/\kappa$. Disorder is expected to reduce the superfluid density while maintaining or increasing the compressibility, thereby resulting in a suppressed speed of sound. 

We conclude by constructing a phase diagram based on all the experiments presented here (Fig.\,5C). While precise delineation of the phase boundaries remains challenging, we identify three distinct regimes, each associated with a corresponding phase in the thermodynamic limit. Note that the transition from the gapped Mott insulator (MI) to the gapless Bose glass (BG) is of the Griffiths type---dominated by rare-region effects---and lies beyond our resolution due to the finite size of the system \cite{sengupta2007}. 

Collectively, our results provide a clear picture of disorder-induced superfluidity. Starting from a MI, one can seed resonances with on-site disorder. The formation of particle-hole excitations is most favorable as $W$ approaches $U$, as evinced both directly and indirectly through the compressibility and condensate fraction experiments. These mobile particle-hole excitations destroy the MI and temporarily keep Anderson localization at bay. For weak disorder, the resonances are confined to small `puddles' or `grains' of locally mobile bosons. Near the critical disorder strength, $W_c\sim U$, the puddles grow large enough to establish global phase coherence, reminiscent of granular superconductor physics. 

The physics described here is relevant to a broad class of systems governed by the dirty superfluid–insulator transition, including thin films, porous media, Josephson junction arrays, and granular superconductors. This transition may also play a key role in understanding more exotic phases such as anomalous metals and the elusive Bose metal\,\cite{Kapitulnik2019,YangScience2019}. We conclude by noting a suggestive parallel with the high-$T_c$ cuprates: the superconducting (superfluid) dome in the cuprate (disordered Bose–Hubbard) phase diagram emerges from doping holes (introducing disorder) into a MI. The superconducting phase near the critical doping point has been proposed to be granular, and doping away from it is analogous to adding disorder \cite{ramshaw_kivelson2025}. 

\vspace{20mm}

\begin{acknowledgments}
We would like to thank M.P.A. Fisher, N. Prokof'ev, E. Berg, S. Raghu, C. Xu, V. Khemani, A. Stern, A. Vishwanath, P. Nosov, A. Pandey, A. Sarma, and N. Myerson-Jain for insightful conversations. TGK acknowledges support from the Gordon and Betty Moore Foundation under GBMF7392 and from the National Science Foundation under grant PHY-2309135 to the Kavli Institute for Theoretical Physics (KITP).  EJM acknowledges support from
the National Science Foundation under Grant No. PHY-2409403.
\end{acknowledgments}

\vspace{35mm}
\onecolumngrid

\begin{flushleft}

{\hypertarget{authorlist}{${}^\dagger$}  \small Google Quantum AI and Collaborators}

\bigskip

    \renewcommand{\author}[2]{#1\textsuperscript{\textrm{\scriptsize #2}}}
    \renewcommand{\affiliation}[2]{\textsuperscript{\textrm{\scriptsize #1} #2} \\}
    \newcommand{\corrauthora}[2]{#1$^{\textrm{\scriptsize #2}, \hyperlink{corra}{\ddagger}}$}
    \newcommand{\corrauthorb}[2]{#1$^{\textrm{\scriptsize #2}, \hyperlink{corrb}{\mathsection}}$}

\begin{footnotesize}

\newcommand{\xGoogle}{\affiliation{1}{Google Research, Mountain View, CA, USA}}
\newcommand{\xStanford}{\affiliation{2}{Department of Applied Physics, Stanford University, Stanford, California 94305, USA}}

\newcommand{\xUMD}{\affiliation{3}{Department of Physics, University of Maryland, College Park, MD}}

\newcommand{\xKings}{\affiliation{4}{Department of Physics, King’s College London, Strand, London, WC2R 2LS, UK}}

\newcommand{\xStanfordPhys}{\affiliation{5}{Department of Physics, Stanford University, Stanford California 94305, USA}}

\newcommand{\xCornell}{\affiliation{6}{School of Applied and Engineering Physics, Cornell University, Ithaca, New York 14853, USA}}

\newcommand{\xPrinceton}{\affiliation{7}{Department of Physics, Princeton University, Princeton, NJ}}

\newcommand{\xUConnStorrs}{\affiliation{8}{Department of Physics, University of Connecticut, Storrs, CT}}

\newcommand{\xUMass}{\affiliation{9}{Department of Electrical and Computer Engineering, University of Massachusetts, Amherst, MA}}

\newcommand{\xUCSB}{\affiliation{10}{Department of Physics, University of California, Santa Barbara, CA}}

\newcommand{\xAuburnECE}{\affiliation{11}{Department of Electrical and Computer Engineering, Auburn University, Auburn, AL}}

\newcommand{\xKITP}{\affiliation{12}{Kavli Institute for Theoretical Physics, University of California, Santa Barbara, CA}}


\newcommand{\Google}{1}
\newcommand{\Stanford}{2}
\newcommand{\UMD}{3}
\newcommand{\Kings}{4}
\newcommand{\StanfordPhys}{5}
\newcommand{\Cornell}{6}
\newcommand{\Princeton}{7}
\newcommand{\UConnStorrs}{8}
\newcommand{\UMass}{9}
\newcommand{\UCSB}{10}
\newcommand{\AuburnECE}{11}
\newcommand{\KITP}{12}

\corrauthora{N. S.~Ticea}{\Google, \!\Stanford},
\corrauthora{E. Portolés}{\Google},
\corrauthora{E. Rosenberg}{\Google},
\corrauthora{A. Schuckert}{\UMD},
\corrauthora{A. Szasz}{\Google},
\author{P.~Praneel}{\Google, \!\Cornell},
\author{B. Kobrin}{\Google},
\author{N. Pomata}{\UMD},
\author{C.~Miao}{\Google, \!\StanfordPhys},
\author{S.~Kumar}{\Google, \!\Princeton},
\author{E.~Crane}{\Kings},
\author{I. Drozdov}{\Google,\! \UConnStorrs},
\author{Y. Lensky}{\Google},
\author{T. G. Kiely}{\KITP},
\author{S.~Gonz\'alez-Garc\'ia}{\Google,\! \UCSB},
% ...
\corrauthorb{A. Petukhov}{\Google},
\corrauthorb{M. Hafezi}{\UMD},
\corrauthorb{P. Roushan}{\Google}

\bigskip

\xGoogle
\xStanford
\xUMD
\xKings
\xStanfordPhys
\xCornell
\xUConnStorrs
\xUMass
\xUCSB
\xAuburnECE
\xKITP

\vspace{2mm}

{\hypertarget{corra}{${}^\ddagger$} These authors contributed equally to this work.}\\

{\hypertarget{corrb}{${}^\mathsection$} Corresponding authors: hafezi@umd.edu  and pedramr@google.com}\\

\end{footnotesize}
\end{flushleft}
\twocolumngrid

\bibliographystyle{apsrev4-2}

\let\oldaddcontentsline\addcontentsline
\renewcommand{\addcontentsline}[3]{}
\bibliography{References}

\begin{thebibliography}{37}%
\makeatletter
\providecommand \@ifxundefined [1]{%
 \@ifx{#1\undefined}
}%
\providecommand \@ifnum [1]{%
 \ifnum #1\expandafter \@firstoftwo
 \else \expandafter \@secondoftwo
 \fi
}%
\providecommand \@ifx [1]{%
 \ifx #1\expandafter \@firstoftwo
 \else \expandafter \@secondoftwo
 \fi
}%
\providecommand \natexlab [1]{#1}%
\providecommand \enquote  [1]{``#1''}%
\providecommand \bibnamefont  [1]{#1}%
\providecommand \bibfnamefont [1]{#1}%
\providecommand \citenamefont [1]{#1}%
\providecommand \href@noop [0]{\@secondoftwo}%
\providecommand \href [0]{\begingroup \@sanitize@url \@href}%
\providecommand \@href[1]{\@@startlink{#1}\@@href}%
\providecommand \@@href[1]{\endgroup#1\@@endlink}%
\providecommand \@sanitize@url [0]{\catcode `\\12\catcode `\$12\catcode `\&12\catcode `\#12\catcode `\^12\catcode `\_12\catcode `\%12\relax}%
\providecommand \@@startlink[1]{}%
\providecommand \@@endlink[0]{}%
\providecommand \url  [0]{\begingroup\@sanitize@url \@url }%
\providecommand \@url [1]{\endgroup\@href {#1}{\urlprefix }}%
\providecommand \urlprefix  [0]{URL }%
\providecommand \Eprint [0]{\href }%
\providecommand \doibase [0]{https://doi.org/}%
\providecommand \selectlanguage [0]{\@gobble}%
\providecommand \bibinfo  [0]{\@secondoftwo}%
\providecommand \bibfield  [0]{\@secondoftwo}%
\providecommand \translation [1]{[#1]}%
\providecommand \BibitemOpen [0]{}%
\providecommand \bibitemStop [0]{}%
\providecommand \bibitemNoStop [0]{.\EOS\space}%
\providecommand \EOS [0]{\spacefactor3000\relax}%
\providecommand \BibitemShut  [1]{\csname bibitem#1\endcsname}%
\let\auto@bib@innerbib\@empty
\bibitem [{\citenamefont {Kjaergaard}\ \emph {et~al.}(2020)\citenamefont {Kjaergaard}, \citenamefont {Schwartz}, \citenamefont {Braum\"{u}ller}, \citenamefont {Krantz}, \citenamefont {Wang}, \citenamefont {Gustavsson},\ and\ \citenamefont {Oliver}}]{Kjaergaard2020}%
  \BibitemOpen
  \bibfield  {author} {\bibinfo {author} {\bibfnamefont {M.}~\bibnamefont {Kjaergaard}}, \bibinfo {author} {\bibfnamefont {M.~E.}\ \bibnamefont {Schwartz}}, \bibinfo {author} {\bibfnamefont {J.}~\bibnamefont {Braum\"{u}ller}}, \bibinfo {author} {\bibfnamefont {P.}~\bibnamefont {Krantz}}, \bibinfo {author} {\bibfnamefont {J.~I.-J.}\ \bibnamefont {Wang}}, \bibinfo {author} {\bibfnamefont {S.}~\bibnamefont {Gustavsson}},\ and\ \bibinfo {author} {\bibfnamefont {W.~D.}\ \bibnamefont {Oliver}},\ }\bibfield  {title} {\bibinfo {title} {Superconducting {Q}ubits: {C}urrent {S}tate of {P}lay},\ }\href {https://doi.org/10.1146/annurev-conmatphys-031119-050605} {\bibfield  {journal} {\bibinfo  {journal} {Annual Review of Condensed Matter Physics}\ }\textbf {\bibinfo {volume} {11}},\ \bibinfo {pages} {369–395} (\bibinfo {year} {2020})}\BibitemShut {NoStop}%
\bibitem [{\citenamefont {Yanay}\ \emph {et~al.}(2020)\citenamefont {Yanay}, \citenamefont {Braum\"{u}ller}, \citenamefont {Gustavsson}, \citenamefont {Oliver},\ and\ \citenamefont {Tahan}}]{Yanay2020}%
  \BibitemOpen
  \bibfield  {author} {\bibinfo {author} {\bibfnamefont {Y.}~\bibnamefont {Yanay}}, \bibinfo {author} {\bibfnamefont {J.}~\bibnamefont {Braum\"{u}ller}}, \bibinfo {author} {\bibfnamefont {S.}~\bibnamefont {Gustavsson}}, \bibinfo {author} {\bibfnamefont {W.~D.}\ \bibnamefont {Oliver}},\ and\ \bibinfo {author} {\bibfnamefont {C.}~\bibnamefont {Tahan}},\ }\bibfield  {title} {\bibinfo {title} {Two-dimensional hard-core {B}ose–{H}ubbard model with superconducting qubits},\ }\href {https://doi.org/10.1038/s41534-020-0269-1} {\bibfield  {journal} {\bibinfo  {journal} {npj Quantum Information}\ }\textbf {\bibinfo {volume} {6}},\ \bibinfo {pages} {58} (\bibinfo {year} {2020})}\BibitemShut {NoStop}%
\bibitem [{\citenamefont {Koch}\ \emph {et~al.}(2007)\citenamefont {Koch}, \citenamefont {Yu}, \citenamefont {Gambetta}, \citenamefont {Houck}, \citenamefont {Schuster}, \citenamefont {Majer}, \citenamefont {Blais}, \citenamefont {Devoret}, \citenamefont {Girvin},\ and\ \citenamefont {Schoelkopf}}]{Koch2007}%
  \BibitemOpen
  \bibfield  {author} {\bibinfo {author} {\bibfnamefont {J.}~\bibnamefont {Koch}}, \bibinfo {author} {\bibfnamefont {T.~M.}\ \bibnamefont {Yu}}, \bibinfo {author} {\bibfnamefont {J.}~\bibnamefont {Gambetta}}, \bibinfo {author} {\bibfnamefont {A.~A.}\ \bibnamefont {Houck}}, \bibinfo {author} {\bibfnamefont {D.~I.}\ \bibnamefont {Schuster}}, \bibinfo {author} {\bibfnamefont {J.}~\bibnamefont {Majer}}, \bibinfo {author} {\bibfnamefont {A.}~\bibnamefont {Blais}}, \bibinfo {author} {\bibfnamefont {M.~H.}\ \bibnamefont {Devoret}}, \bibinfo {author} {\bibfnamefont {S.~M.}\ \bibnamefont {Girvin}},\ and\ \bibinfo {author} {\bibfnamefont {R.~J.}\ \bibnamefont {Schoelkopf}},\ }\bibfield  {title} {\bibinfo {title} {Charge-insensitive qubit design derived from the {C}ooper pair box},\ }\href {https://doi.org/10.1103/PhysRevA.76.042319} {\bibfield  {journal} {\bibinfo  {journal} {Physical Review A}\ }\textbf {\bibinfo {volume} {76}},\ \bibinfo {pages} {042319} (\bibinfo {year} {2007})}\BibitemShut {NoStop}%
\bibitem [{\citenamefont {Yan}\ \emph {et~al.}(2018)\citenamefont {Yan}, \citenamefont {Krantz}, \citenamefont {Sung}, \citenamefont {Kjaergaard}, \citenamefont {Campbell}, \citenamefont {Orlando}, \citenamefont {Gustavsson},\ and\ \citenamefont {Oliver}}]{Yan2018}%
  \BibitemOpen
  \bibfield  {author} {\bibinfo {author} {\bibfnamefont {F.}~\bibnamefont {Yan}}, \bibinfo {author} {\bibfnamefont {P.}~\bibnamefont {Krantz}}, \bibinfo {author} {\bibfnamefont {Y.}~\bibnamefont {Sung}}, \bibinfo {author} {\bibfnamefont {M.}~\bibnamefont {Kjaergaard}}, \bibinfo {author} {\bibfnamefont {D.~L.}\ \bibnamefont {Campbell}}, \bibinfo {author} {\bibfnamefont {T.~P.}\ \bibnamefont {Orlando}}, \bibinfo {author} {\bibfnamefont {S.}~\bibnamefont {Gustavsson}},\ and\ \bibinfo {author} {\bibfnamefont {W.~D.}\ \bibnamefont {Oliver}},\ }\bibfield  {title} {\bibinfo {title} {Tunable coupling scheme for implementing high-fidelity two-qubit gates},\ }\href {https://doi.org/10.1103/PhysRevApplied.10.054062} {\bibfield  {journal} {\bibinfo  {journal} {Physical Review Applied}\ }\textbf {\bibinfo {volume} {10}},\ \bibinfo {pages} {054062} (\bibinfo {year} {2018})}\BibitemShut {NoStop}%
\bibitem [{\citenamefont {Arute}\ \emph {et~al.}(2019)\citenamefont {Arute} \emph {et~al.}}]{Arute2019supremacy}%
  \BibitemOpen
  \bibfield  {author} {\bibinfo {author} {\bibfnamefont {F.}~\bibnamefont {Arute}} \emph {et~al.} (\bibinfo {collaboration} {Google Quantum AI}),\ }\bibfield  {title} {\bibinfo {title} {Quantum supremacy using a programmable superconducting processor},\ }\href {https://doi.org/10.1038/s41586-019-1666-5} {\bibfield  {journal} {\bibinfo  {journal} {Nature}\ }\textbf {\bibinfo {volume} {574}},\ \bibinfo {pages} {510} (\bibinfo {year} {2019})}\BibitemShut {NoStop}%
\bibitem [{\citenamefont {Bengtsson}\ \emph {et~al.}(2024)\citenamefont {Bengtsson}, \citenamefont {Opremcak}, \citenamefont {Khezri}, \citenamefont {Sank}, \citenamefont {Bourassa}, \citenamefont {Satzinger}, \citenamefont {Hong}, \citenamefont {Erickson}, \citenamefont {Lester}, \citenamefont {Miao}, \citenamefont {Korotkov}, \citenamefont {Kelly}, \citenamefont {Chen},\ and\ \citenamefont {Klimov}}]{Bengtsson_2024}%
  \BibitemOpen
  \bibfield  {author} {\bibinfo {author} {\bibfnamefont {A.}~\bibnamefont {Bengtsson}}, \bibinfo {author} {\bibfnamefont {A.}~\bibnamefont {Opremcak}}, \bibinfo {author} {\bibfnamefont {M.}~\bibnamefont {Khezri}}, \bibinfo {author} {\bibfnamefont {D.}~\bibnamefont {Sank}}, \bibinfo {author} {\bibfnamefont {A.}~\bibnamefont {Bourassa}}, \bibinfo {author} {\bibfnamefont {K.~J.}\ \bibnamefont {Satzinger}}, \bibinfo {author} {\bibfnamefont {S.}~\bibnamefont {Hong}}, \bibinfo {author} {\bibfnamefont {C.}~\bibnamefont {Erickson}}, \bibinfo {author} {\bibfnamefont {B.~J.}\ \bibnamefont {Lester}}, \bibinfo {author} {\bibfnamefont {K.~C.}\ \bibnamefont {Miao}}, \bibinfo {author} {\bibfnamefont {A.~N.}\ \bibnamefont {Korotkov}}, \bibinfo {author} {\bibfnamefont {J.}~\bibnamefont {Kelly}}, \bibinfo {author} {\bibfnamefont {Z.}~\bibnamefont {Chen}},\ and\ \bibinfo {author} {\bibfnamefont {P.~V.}\ \bibnamefont {Klimov}},\ }\bibfield  {title} {\bibinfo {title} {Model-based optimization of superconducting qubit readout},\
  }\href {https://doi.org/10.1103/PhysRevLetters132.100603} {\bibfield  {journal} {\bibinfo  {journal} {Physical Review Letters}\ }\textbf {\bibinfo {volume} {132}},\ \bibinfo {pages} {100603} (\bibinfo {year} {2024})}\BibitemShut {NoStop}%
\bibitem [{\citenamefont {Walter}\ \emph {et~al.}(2017)\citenamefont {Walter}, \citenamefont {Kurpiers}, \citenamefont {Gasparinetti}, \citenamefont {Magnard}, \citenamefont {Poto\ifmmode~\check{c}\else \v{c}\fi{}nik}, \citenamefont {Salath\'e}, \citenamefont {Pechal}, \citenamefont {Mondal}, \citenamefont {Oppliger}, \citenamefont {Eichler},\ and\ \citenamefont {Wallraff}}]{PhysRevApplied.7.054020}%
  \BibitemOpen
  \bibfield  {author} {\bibinfo {author} {\bibfnamefont {T.}~\bibnamefont {Walter}}, \bibinfo {author} {\bibfnamefont {P.}~\bibnamefont {Kurpiers}}, \bibinfo {author} {\bibfnamefont {S.}~\bibnamefont {Gasparinetti}}, \bibinfo {author} {\bibfnamefont {P.}~\bibnamefont {Magnard}}, \bibinfo {author} {\bibfnamefont {A.}~\bibnamefont {Poto\ifmmode~\check{c}\else \v{c}\fi{}nik}}, \bibinfo {author} {\bibfnamefont {Y.}~\bibnamefont {Salath\'e}}, \bibinfo {author} {\bibfnamefont {M.}~\bibnamefont {Pechal}}, \bibinfo {author} {\bibfnamefont {M.}~\bibnamefont {Mondal}}, \bibinfo {author} {\bibfnamefont {M.}~\bibnamefont {Oppliger}}, \bibinfo {author} {\bibfnamefont {C.}~\bibnamefont {Eichler}},\ and\ \bibinfo {author} {\bibfnamefont {A.}~\bibnamefont {Wallraff}},\ }\bibfield  {title} {\bibinfo {title} {Rapid high-fidelity single-shot dispersive readout of superconducting qubits},\ }\href {https://doi.org/10.1103/PhysRevApplied.7.054020} {\bibfield  {journal} {\bibinfo  {journal} {Physical Review Appl.}\ }\textbf
  {\bibinfo {volume} {7}},\ \bibinfo {pages} {054020} (\bibinfo {year} {2017})}\BibitemShut {NoStop}%
\bibitem [{\citenamefont {Blais}\ \emph {et~al.}(2021)\citenamefont {Blais}, \citenamefont {Grimsmo}, \citenamefont {Girvin},\ and\ \citenamefont {Wallraff}}]{Blais2021QED}%
  \BibitemOpen
  \bibfield  {author} {\bibinfo {author} {\bibfnamefont {A.}~\bibnamefont {Blais}}, \bibinfo {author} {\bibfnamefont {A.~L.}\ \bibnamefont {Grimsmo}}, \bibinfo {author} {\bibfnamefont {S.~M.}\ \bibnamefont {Girvin}},\ and\ \bibinfo {author} {\bibfnamefont {A.}~\bibnamefont {Wallraff}},\ }\bibfield  {title} {\bibinfo {title} {Circuit quantum electrodynamics},\ }\href {https://doi.org/10.1103/RevModPhys.93.025005} {\bibfield  {journal} {\bibinfo  {journal} {Reviews of Modern Physics}\ }\textbf {\bibinfo {volume} {93}},\ \bibinfo {pages} {025005} (\bibinfo {year} {2021})}\BibitemShut {NoStop}%
\bibitem [{\citenamefont {Andersen}(2025)}]{Andersen2025}%
  \BibitemOpen
  \bibfield  {author} {\bibinfo {author} {\bibfnamefont {T.~I.}\ \bibnamefont {Andersen}},\ }\bibfield  {title} {\bibinfo {title} {Thermalization and criticality on an analogue--digital quantum simulator},\ }\href {https://doi.org/10.1038/s41586-024-08460-3} {\bibfield  {journal} {\bibinfo  {journal} {Nature}\ }\textbf {\bibinfo {volume} {638}},\ \bibinfo {pages} {79} (\bibinfo {year} {2025})}\BibitemShut {NoStop}%
\bibitem [{\citenamefont {Pe\~na Ardila}\ \emph {et~al.}(2018)\citenamefont {Pe\~na Ardila}, \citenamefont {Heyl},\ and\ \citenamefont {Eckardt}}]{Ardila2018}%
  \BibitemOpen
  \bibfield  {author} {\bibinfo {author} {\bibfnamefont {L.~A.}\ \bibnamefont {Pe\~na Ardila}}, \bibinfo {author} {\bibfnamefont {M.}~\bibnamefont {Heyl}},\ and\ \bibinfo {author} {\bibfnamefont {A.}~\bibnamefont {Eckardt}},\ }\bibfield  {title} {\bibinfo {title} {Measuring the single-particle density matrix for fermions and hard-core bosons in an optical lattice},\ }\href {https://doi.org/10.1103/PhysRevLetters121.260401} {\bibfield  {journal} {\bibinfo  {journal} {Physical Review Letters}\ }\textbf {\bibinfo {volume} {121}},\ \bibinfo {pages} {260401} (\bibinfo {year} {2018})}\BibitemShut {NoStop}%
\bibitem [{\citenamefont {Hubbard}(1964)}]{Hubbard1964}%
  \BibitemOpen
  \bibfield  {author} {\bibinfo {author} {\bibfnamefont {J.}~\bibnamefont {Hubbard}},\ }\bibfield  {title} {\bibinfo {title} {Electron correlations in narrow energy bands. {II}. the degenerate band case},\ }\href {http://www.jstor.org/stable/2414862} {\bibfield  {journal} {\bibinfo  {journal} {Proceedings of the Royal Society of London. Series A, Mathematical and Physical Sciences}\ }\textbf {\bibinfo {volume} {277}},\ \bibinfo {pages} {237} (\bibinfo {year} {1964})}\BibitemShut {NoStop}%
\bibitem [{\citenamefont {Foxen}\ \emph {et~al.}(2020)\citenamefont {Foxen} \emph {et~al.}}]{BrooksFSim}%
  \BibitemOpen
  \bibfield  {author} {\bibinfo {author} {\bibfnamefont {B.}~\bibnamefont {Foxen}} \emph {et~al.} (\bibinfo {collaboration} {Google AI Quantum}),\ }\bibfield  {title} {\bibinfo {title} {Demonstrating a continuous set of two-qubit gates for near-term quantum algorithms},\ }\href {https://doi.org/10.1103/PhysRevLetters125.120504} {\bibfield  {journal} {\bibinfo  {journal} {Physical Review Letters}\ }\textbf {\bibinfo {volume} {125}},\ \bibinfo {pages} {120504} (\bibinfo {year} {2020})}\BibitemShut {NoStop}%
\bibitem [{\citenamefont {White}(1992)}]{white_1992}%
  \BibitemOpen
  \bibfield  {author} {\bibinfo {author} {\bibfnamefont {S.~R.}\ \bibnamefont {White}},\ }\bibfield  {title} {\bibinfo {title} {Density matrix formulation for quantum renormalization groups},\ }\href {https://doi.org/10.1103/PhysRevLetters69.2863} {\bibfield  {journal} {\bibinfo  {journal} {Physical Review Letters}\ }\textbf {\bibinfo {volume} {69}},\ \bibinfo {pages} {2863} (\bibinfo {year} {1992})}\BibitemShut {NoStop}%
\bibitem [{\citenamefont {White}(1993)}]{white_1993}%
  \BibitemOpen
  \bibfield  {author} {\bibinfo {author} {\bibfnamefont {S.~R.}\ \bibnamefont {White}},\ }\bibfield  {title} {\bibinfo {title} {Density-matrix algorithms for quantum renormalization groups},\ }\href {https://doi.org/10.1103/PhysRevB.48.10345} {\bibfield  {journal} {\bibinfo  {journal} {Physical Review B}\ }\textbf {\bibinfo {volume} {48}},\ \bibinfo {pages} {10345} (\bibinfo {year} {1993})}\BibitemShut {NoStop}%
\bibitem [{\citenamefont {Schollw\"ock}(2011)}]{Schollwock_2011}%
  \BibitemOpen
  \bibfield  {author} {\bibinfo {author} {\bibfnamefont {U.}~\bibnamefont {Schollw\"ock}},\ }\bibfield  {title} {\bibinfo {title} {The density-matrix renormalization group in the age of matrix product states},\ }\href {https://doi.org/10.1016/j.aop.2010.09.012} {\bibfield  {journal} {\bibinfo  {journal} {Annals of Physics}\ }\textbf {\bibinfo {volume} {326}},\ \bibinfo {pages} {96} (\bibinfo {year} {2011})}\BibitemShut {NoStop}%
\bibitem [{\citenamefont {Deutsch}(1991)}]{Deutsch1991}%
  \BibitemOpen
  \bibfield  {author} {\bibinfo {author} {\bibfnamefont {J.~M.}\ \bibnamefont {Deutsch}},\ }\bibfield  {title} {\bibinfo {title} {Quantum statistical mechanics in a closed system},\ }\href {https://doi.org/10.1103/PhysRevA.43.2046} {\bibfield  {journal} {\bibinfo  {journal} {Physical Review A}\ }\textbf {\bibinfo {volume} {43}},\ \bibinfo {pages} {2046} (\bibinfo {year} {1991})}\BibitemShut {NoStop}%
\bibitem [{\citenamefont {Nozi{\`e}res}\ and\ \citenamefont {Pines}(2000)}]{PinesNozieres1999}%
  \BibitemOpen
  \bibfield  {author} {\bibinfo {author} {\bibfnamefont {P.}~\bibnamefont {Nozi{\`e}res}}\ and\ \bibinfo {author} {\bibfnamefont {D.}~\bibnamefont {Pines}},\ }\href {https://doi.org/10.1201/9780429495717} {\emph {\bibinfo {title} {Theory of Quantum Liquids}}}\ (\bibinfo  {publisher} {CRC Press},\ \bibinfo {year} {2000})\ \bibinfo {note} {reprint of 1966/1968 edition}\BibitemShut {NoStop}%
\bibitem [{\citenamefont {Negele}\ and\ \citenamefont {Orland}(1998)}]{NegeleOrland1998}%
  \BibitemOpen
  \bibfield  {author} {\bibinfo {author} {\bibfnamefont {J.~W.}\ \bibnamefont {Negele}}\ and\ \bibinfo {author} {\bibfnamefont {H.}~\bibnamefont {Orland}},\ }\href {http://www.worldcat.org/isbn/0738200522} {\emph {\bibinfo {title} {Quantum Many-particle Systems}}}\ (\bibinfo  {publisher} {Westview Press},\ \bibinfo {year} {1998})\BibitemShut {NoStop}%
\bibitem [{\citenamefont {Feynman}(1954)}]{Feynman1954}%
  \BibitemOpen
  \bibfield  {author} {\bibinfo {author} {\bibfnamefont {R.~P.}\ \bibnamefont {Feynman}},\ }\bibfield  {title} {\bibinfo {title} {Atomic theory of the two-fluid model of liquid helium},\ }\href {https://doi.org/10.1103/PhysRev.94.262} {\bibfield  {journal} {\bibinfo  {journal} {Physical Review}\ }\textbf {\bibinfo {volume} {94}},\ \bibinfo {pages} {262} (\bibinfo {year} {1954})}\BibitemShut {NoStop}%
\bibitem [{\citenamefont {\"Ostlund}\ and\ \citenamefont {Rommer}(1995)}]{Ostlund_1995}%
  \BibitemOpen
  \bibfield  {author} {\bibinfo {author} {\bibfnamefont {S.}~\bibnamefont {\"Ostlund}}\ and\ \bibinfo {author} {\bibfnamefont {S.}~\bibnamefont {Rommer}},\ }\bibfield  {title} {\bibinfo {title} {Thermodynamic limit of density matrix renormalization},\ }\href {https://doi.org/10.1103/PhysRevLetters75.3537} {\bibfield  {journal} {\bibinfo  {journal} {Physical Review Letters}\ }\textbf {\bibinfo {volume} {75}},\ \bibinfo {pages} {3537} (\bibinfo {year} {1995})}\BibitemShut {NoStop}%
\bibitem [{\citenamefont {Hauschild}\ and\ \citenamefont {Pollmann}(2018)}]{tenpy}%
  \BibitemOpen
  \bibfield  {author} {\bibinfo {author} {\bibfnamefont {J.}~\bibnamefont {Hauschild}}\ and\ \bibinfo {author} {\bibfnamefont {F.}~\bibnamefont {Pollmann}},\ }\bibfield  {title} {\bibinfo {title} {{Efficient numerical simulations with Tensor Networks: Tensor Network Python (TeNPy)}},\ }\href {https://doi.org/10.21468/SciPostPhysLectNotes.5} {\bibfield  {journal} {\bibinfo  {journal} {SciPost Phys. Lect. Notes}\ ,\ \bibinfo {pages} {5}} (\bibinfo {year} {2018})}\BibitemShut {NoStop}%
\bibitem [{\citenamefont {Haegeman}\ \emph {et~al.}(2016)\citenamefont {Haegeman}, \citenamefont {Lubich}, \citenamefont {Oseledets}, \citenamefont {Vandereycken},\ and\ \citenamefont {Verstraete}}]{Haegeman2016}%
  \BibitemOpen
  \bibfield  {author} {\bibinfo {author} {\bibfnamefont {J.}~\bibnamefont {Haegeman}}, \bibinfo {author} {\bibfnamefont {C.}~\bibnamefont {Lubich}}, \bibinfo {author} {\bibfnamefont {I.}~\bibnamefont {Oseledets}}, \bibinfo {author} {\bibfnamefont {B.}~\bibnamefont {Vandereycken}},\ and\ \bibinfo {author} {\bibfnamefont {F.}~\bibnamefont {Verstraete}},\ }\bibfield  {title} {\bibinfo {title} {Unifying time evolution and optimization with matrix product states},\ }\href {https://doi.org/10.1103/PhysRevB.94.165116} {\bibfield  {journal} {\bibinfo  {journal} {Physical Review B}\ }\textbf {\bibinfo {volume} {94}},\ \bibinfo {pages} {165116} (\bibinfo {year} {2016})}\BibitemShut {NoStop}%
\bibitem [{\citenamefont {Paeckel}\ \emph {et~al.}(2019)\citenamefont {Paeckel}, \citenamefont {Köhler}, \citenamefont {Swoboda}, \citenamefont {Manmana}, \citenamefont {Schollwöck},\ and\ \citenamefont {Hubig}}]{PAECKEL2019}%
  \BibitemOpen
  \bibfield  {author} {\bibinfo {author} {\bibfnamefont {S.}~\bibnamefont {Paeckel}}, \bibinfo {author} {\bibfnamefont {T.}~\bibnamefont {Köhler}}, \bibinfo {author} {\bibfnamefont {A.}~\bibnamefont {Swoboda}}, \bibinfo {author} {\bibfnamefont {S.~R.}\ \bibnamefont {Manmana}}, \bibinfo {author} {\bibfnamefont {U.}~\bibnamefont {Schollwöck}},\ and\ \bibinfo {author} {\bibfnamefont {C.}~\bibnamefont {Hubig}},\ }\bibfield  {title} {\bibinfo {title} {Time-evolution methods for matrix-product states},\ }\href {https://doi.org/https://doi.org/10.1016/j.aop.2019.167998} {\bibfield  {journal} {\bibinfo  {journal} {Annals of Physics}\ }\textbf {\bibinfo {volume} {411}},\ \bibinfo {pages} {167998} (\bibinfo {year} {2019})}\BibitemShut {NoStop}%
\bibitem [{\citenamefont {Hubig}\ \emph {et~al.}(2017)\citenamefont {Hubig}, \citenamefont {McCulloch},\ and\ \citenamefont {Schollw\"ock}}]{Hubig2017}%
  \BibitemOpen
  \bibfield  {author} {\bibinfo {author} {\bibfnamefont {C.}~\bibnamefont {Hubig}}, \bibinfo {author} {\bibfnamefont {I.~P.}\ \bibnamefont {McCulloch}},\ and\ \bibinfo {author} {\bibfnamefont {U.}~\bibnamefont {Schollw\"ock}},\ }\bibfield  {title} {\bibinfo {title} {Generic construction of efficient matrix product operators},\ }\href {https://doi.org/10.1103/PhysRevB.95.035129} {\bibfield  {journal} {\bibinfo  {journal} {Physical Review B}\ }\textbf {\bibinfo {volume} {95}},\ \bibinfo {pages} {035129} (\bibinfo {year} {2017})}\BibitemShut {NoStop}%
\bibitem [{\citenamefont {Sheshadri}\ \emph {et~al.}(1993)\citenamefont {Sheshadri}, \citenamefont {Krishnamurthy}, \citenamefont {Pandit},\ and\ \citenamefont {Ramakrishnan}}]{Sheshadri1993}%
  \BibitemOpen
  \bibfield  {author} {\bibinfo {author} {\bibfnamefont {K.}~\bibnamefont {Sheshadri}}, \bibinfo {author} {\bibfnamefont {H.~R.}\ \bibnamefont {Krishnamurthy}}, \bibinfo {author} {\bibfnamefont {R.}~\bibnamefont {Pandit}},\ and\ \bibinfo {author} {\bibfnamefont {T.~V.}\ \bibnamefont {Ramakrishnan}},\ }\bibfield  {title} {\bibinfo {title} {Superfluid and insulating phases in an interacting-boson model: Mean-field theory and the {RPA}},\ }\href {https://doi.org/10.1209/0295-5075/22/4/004} {\bibfield  {journal} {\bibinfo  {journal} {Europhysics Letters}\ }\textbf {\bibinfo {volume} {22}},\ \bibinfo {pages} {257} (\bibinfo {year} {1993})}\BibitemShut {NoStop}%
\bibitem [{\citenamefont {van Oosten}\ \emph {et~al.}(2005)\citenamefont {van Oosten}, \citenamefont {Dickerscheid}, \citenamefont {Farid}, \citenamefont {van~der Straten},\ and\ \citenamefont {Stoof}}]{vanOosten2005}%
  \BibitemOpen
  \bibfield  {author} {\bibinfo {author} {\bibfnamefont {D.}~\bibnamefont {van Oosten}}, \bibinfo {author} {\bibfnamefont {D.~B.~M.}\ \bibnamefont {Dickerscheid}}, \bibinfo {author} {\bibfnamefont {B.}~\bibnamefont {Farid}}, \bibinfo {author} {\bibfnamefont {P.}~\bibnamefont {van~der Straten}},\ and\ \bibinfo {author} {\bibfnamefont {H.~T.~C.}\ \bibnamefont {Stoof}},\ }\bibfield  {title} {\bibinfo {title} {Inelastic light scattering from a {M}ott insulator},\ }\href {https://doi.org/10.1103/PhysRevA.71.021601} {\bibfield  {journal} {\bibinfo  {journal} {Physical Review A}\ }\textbf {\bibinfo {volume} {71}},\ \bibinfo {pages} {021601} (\bibinfo {year} {2005})}\BibitemShut {NoStop}%
\bibitem [{\citenamefont {Trefzger}\ \emph {et~al.}(2008)\citenamefont {Trefzger}, \citenamefont {Menotti},\ and\ \citenamefont {Lewenstein}}]{Trefzger2008}%
  \BibitemOpen
  \bibfield  {author} {\bibinfo {author} {\bibfnamefont {C.}~\bibnamefont {Trefzger}}, \bibinfo {author} {\bibfnamefont {C.}~\bibnamefont {Menotti}},\ and\ \bibinfo {author} {\bibfnamefont {M.}~\bibnamefont {Lewenstein}},\ }\bibfield  {title} {\bibinfo {title} {Ultracold dipolar gas in an optical lattice: The fate of metastable states},\ }\href {https://doi.org/10.1103/PhysRevA.78.043604} {\bibfield  {journal} {\bibinfo  {journal} {Physical Review A}\ }\textbf {\bibinfo {volume} {78}},\ \bibinfo {pages} {043604} (\bibinfo {year} {2008})}\BibitemShut {NoStop}%
\bibitem [{\citenamefont {Krutitsky}\ and\ \citenamefont {Navez}(2011)}]{Krutitsky2011}%
  \BibitemOpen
  \bibfield  {author} {\bibinfo {author} {\bibfnamefont {K.~V.}\ \bibnamefont {Krutitsky}}\ and\ \bibinfo {author} {\bibfnamefont {P.}~\bibnamefont {Navez}},\ }\bibfield  {title} {\bibinfo {title} {Excitation dynamics in a lattice bose gas within the time-dependent {G}utzwiller mean-field approach},\ }\href {https://doi.org/10.1103/PhysRevA.84.033602} {\bibfield  {journal} {\bibinfo  {journal} {Physical Review A}\ }\textbf {\bibinfo {volume} {84}},\ \bibinfo {pages} {033602} (\bibinfo {year} {2011})}\BibitemShut {NoStop}%
\bibitem [{\citenamefont {Altman}\ and\ \citenamefont {Auerbach}(2002)}]{Altman2002}%
  \BibitemOpen
  \bibfield  {author} {\bibinfo {author} {\bibfnamefont {E.}~\bibnamefont {Altman}}\ and\ \bibinfo {author} {\bibfnamefont {A.}~\bibnamefont {Auerbach}},\ }\bibfield  {title} {\bibinfo {title} {Oscillating superfluidity of bosons in optical lattices},\ }\href {https://doi.org/10.1103/PhysRevLetters89.250404} {\bibfield  {journal} {\bibinfo  {journal} {Physical Review Letters}\ }\textbf {\bibinfo {volume} {89}},\ \bibinfo {pages} {250404} (\bibinfo {year} {2002})}\BibitemShut {NoStop}%
\bibitem [{\citenamefont {Huber}\ \emph {et~al.}(2007)\citenamefont {Huber}, \citenamefont {Altman}, \citenamefont {B\"uchler},\ and\ \citenamefont {Blatter}}]{Huber2007}%
  \BibitemOpen
  \bibfield  {author} {\bibinfo {author} {\bibfnamefont {S.~D.}\ \bibnamefont {Huber}}, \bibinfo {author} {\bibfnamefont {E.}~\bibnamefont {Altman}}, \bibinfo {author} {\bibfnamefont {H.~P.}\ \bibnamefont {B\"uchler}},\ and\ \bibinfo {author} {\bibfnamefont {G.}~\bibnamefont {Blatter}},\ }\bibfield  {title} {\bibinfo {title} {Dynamical properties of ultracold bosons in an optical lattice},\ }\href {https://doi.org/10.1103/PhysRevB.75.085106} {\bibfield  {journal} {\bibinfo  {journal} {Physical Review B}\ }\textbf {\bibinfo {volume} {75}},\ \bibinfo {pages} {085106} (\bibinfo {year} {2007})}\BibitemShut {NoStop}%
\bibitem [{\citenamefont {Baker}\ \emph {et~al.}(2024)\citenamefont {Baker}, \citenamefont {Foley},\ and\ \citenamefont {S{\'e}n{\'e}chal}}]{Baker:2021fvd}%
  \BibitemOpen
  \bibfield  {author} {\bibinfo {author} {\bibfnamefont {T.~E.}\ \bibnamefont {Baker}}, \bibinfo {author} {\bibfnamefont {A.}~\bibnamefont {Foley}},\ and\ \bibinfo {author} {\bibfnamefont {D.}~\bibnamefont {S{\'e}n{\'e}chal}},\ }\bibfield  {title} {\bibinfo {title} {{Direct solution of multiple excitations in a matrix product state with block {L}anczos}},\ }\href {https://doi.org/10.1140/epjb/s10051-024-00702-7} {\bibfield  {journal} {\bibinfo  {journal} {Eur. Phys. J. B}\ }\textbf {\bibinfo {volume} {97}},\ \bibinfo {pages} {72} (\bibinfo {year} {2024})}\BibitemShut {NoStop}%
\bibitem [{\citenamefont {Gonz\'{a}lez-Garc\'{i}a}\ \emph {et~al.}(2025)\citenamefont {Gonz\'{a}lez-Garc\'{i}a}, \citenamefont {Szasz}, \citenamefont {Pagano}, \citenamefont {Kafri}, \citenamefont {Vidal},\ and\ \citenamefont {Paolo}}]{Gonzalez2025MTDMRGX}%
  \BibitemOpen
  \bibfield  {author} {\bibinfo {author} {\bibfnamefont {S.}~\bibnamefont {Gonz\'{a}lez-Garc\'{i}a}}, \bibinfo {author} {\bibfnamefont {A.}~\bibnamefont {Szasz}}, \bibinfo {author} {\bibfnamefont {A.}~\bibnamefont {Pagano}}, \bibinfo {author} {\bibfnamefont {D.}~\bibnamefont {Kafri}}, \bibinfo {author} {\bibfnamefont {G.}~\bibnamefont {Vidal}},\ and\ \bibinfo {author} {\bibfnamefont {A.~D.}\ \bibnamefont {Paolo}},\ }\href@noop {} {\bibinfo {title} {Multi-{T}arget {D}ensity {M}atrix {R}enormalization {G}roup {X} algorithm and its application to circuit quantum electrodynamics}} (\bibinfo {year} {2025}),\ \Eprint {https://arxiv.org/abs/2506.24109} {arXiv:2506.24109 [quant-ph]} \BibitemShut {NoStop}%
\bibitem [{\citenamefont {Fishman}\ \emph {et~al.}(2022)\citenamefont {Fishman}, \citenamefont {White},\ and\ \citenamefont {Stoudenmire}}]{itensor}%
  \BibitemOpen
  \bibfield  {author} {\bibinfo {author} {\bibfnamefont {M.}~\bibnamefont {Fishman}}, \bibinfo {author} {\bibfnamefont {S.~R.}\ \bibnamefont {White}},\ and\ \bibinfo {author} {\bibfnamefont {E.~M.}\ \bibnamefont {Stoudenmire}},\ }\bibfield  {title} {\bibinfo {title} {{The ITensor Software Library for Tensor Network Calculations}},\ }\href {https://doi.org/10.21468/SciPostPhysCodeb.4} {\bibfield  {journal} {\bibinfo  {journal} {SciPost Phys. Codebases}\ ,\ \bibinfo {pages} {4}} (\bibinfo {year} {2022})}\BibitemShut {NoStop}%
\bibitem [{\citenamefont {Stoudenmire}\ and\ \citenamefont {White}(2012)}]{Stoudenmire2012}%
  \BibitemOpen
  \bibfield  {author} {\bibinfo {author} {\bibfnamefont {E.}~\bibnamefont {Stoudenmire}}\ and\ \bibinfo {author} {\bibfnamefont {S.~R.}\ \bibnamefont {White}},\ }\bibfield  {title} {\bibinfo {title} {Studying two-dimensional systems with the density matrix renormalization group},\ }\href {https://doi.org/10.1146/annurev-conmatphys-020911-125018} {\bibfield  {journal} {\bibinfo  {journal} {Annual Review of Condensed Matter Physics}\ }\textbf {\bibinfo {volume} {3}},\ \bibinfo {pages} {111–128} (\bibinfo {year} {2012})}\BibitemShut {NoStop}%
\bibitem [{\citenamefont {Bravyi}\ \emph {et~al.}(2011)\citenamefont {Bravyi}, \citenamefont {DiVincenzo},\ and\ \citenamefont {Loss}}]{bravyi2011schrieffer}%
  \BibitemOpen
  \bibfield  {author} {\bibinfo {author} {\bibfnamefont {S.}~\bibnamefont {Bravyi}}, \bibinfo {author} {\bibfnamefont {D.~P.}\ \bibnamefont {DiVincenzo}},\ and\ \bibinfo {author} {\bibfnamefont {D.}~\bibnamefont {Loss}},\ }\bibfield  {title} {\bibinfo {title} {Schrieffer–{W}olff transformation for quantum many-body systems},\ }\href {https://doi.org/https://doi.org/10.1016/j.aop.2011.06.004} {\bibfield  {journal} {\bibinfo  {journal} {Annals of Physics}\ }\textbf {\bibinfo {volume} {326}},\ \bibinfo {pages} {2793} (\bibinfo {year} {2011})}\BibitemShut {NoStop}%
\bibitem [{\citenamefont {Gelfand}\ and\ \citenamefont {Singh}(2000)}]{gelfand2000high}%
  \BibitemOpen
  \bibfield  {author} {\bibinfo {author} {\bibfnamefont {M.~P.}\ \bibnamefont {Gelfand}}\ and\ \bibinfo {author} {\bibfnamefont {R.~R.~P.}\ \bibnamefont {Singh}},\ }\bibfield  {title} {\bibinfo {title} {High-order convergent expansions for quantum many particle systems},\ }\href {https://doi.org/10.1080/000187300243390} {\bibfield  {journal} {\bibinfo  {journal} {Advances in Physics}\ }\textbf {\bibinfo {volume} {49}},\ \bibinfo {pages} {93} (\bibinfo {year} {2000})}\BibitemShut {NoStop}%
\bibitem [{\citenamefont {Hörmann}\ and\ \citenamefont {Schmidt}(2023)}]{hormann2023projective}%
  \BibitemOpen
  \bibfield  {author} {\bibinfo {author} {\bibfnamefont {M.}~\bibnamefont {Hörmann}}\ and\ \bibinfo {author} {\bibfnamefont {K.~P.}\ \bibnamefont {Schmidt}},\ }\bibfield  {title} {\bibinfo {title} {{Projective cluster-additive transformation for quantum lattice models}},\ }\href {https://doi.org/10.21468/SciPostPhys.15.3.097} {\bibfield  {journal} {\bibinfo  {journal} {SciPost Phys.}\ }\textbf {\bibinfo {volume} {15}},\ \bibinfo {pages} {097} (\bibinfo {year} {2023})}\BibitemShut {NoStop}%
\end{thebibliography}
\let\addcontentsline\oldaddcontentsline

\clearpage

\newcommand{\beginsupplement}{%
    \pagebreak
    \onecolumngrid
    \renewcommand{\thefigure}{S\arabic{figure}}
    \renewcommand{\theequation}{S\arabic{equation}}
    \titlespacing*{\section} {0pt}{3.5ex plus 1ex minus .2ex}{2.3ex plus .2ex}
    \titlespacing*{\subsection} {0pt}{3.25ex plus 1ex minus .2ex}{1.5ex plus .2ex}
    \titlespacing*{\subsubsection}{0pt}{3.25ex plus 1ex minus .2ex}{1.5ex plus .2ex}
    \titleformat*{\section}{\filcenter\bfseries\MakeUppercase}
}

\title{Supplementary Materials for ``Observation of disorder-induced superfluidity''}
\date{\today}

\maketitle

\newpage

\beginsupplement
\tableofcontents


\part{\LARGE Experimental Protocols\label{part:experiment}}

\section{Transmon-based realization of the Bose-Hubbard model}\label{sec:BHM}

\begin{figure}[h]
    \centering
    \includegraphics[width=0.5\linewidth]{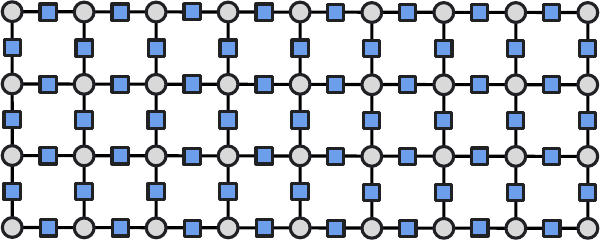}
    \caption{Geometric layout of transmons on our superconducting transmon processor.  Both gray squares and blue circles are transmons, but only the gray transmons forming a 4x9 square lattice are treated as active qudit degrees of freedom in our model. The blue transmons are ``couplers'' used to control the coupling strength between the connected qudits.}
    \label{fig:chip_qubit_coupler_layout}
\end{figure}

\begin{figure}[h]
\includegraphics[width=0.5\columnwidth]{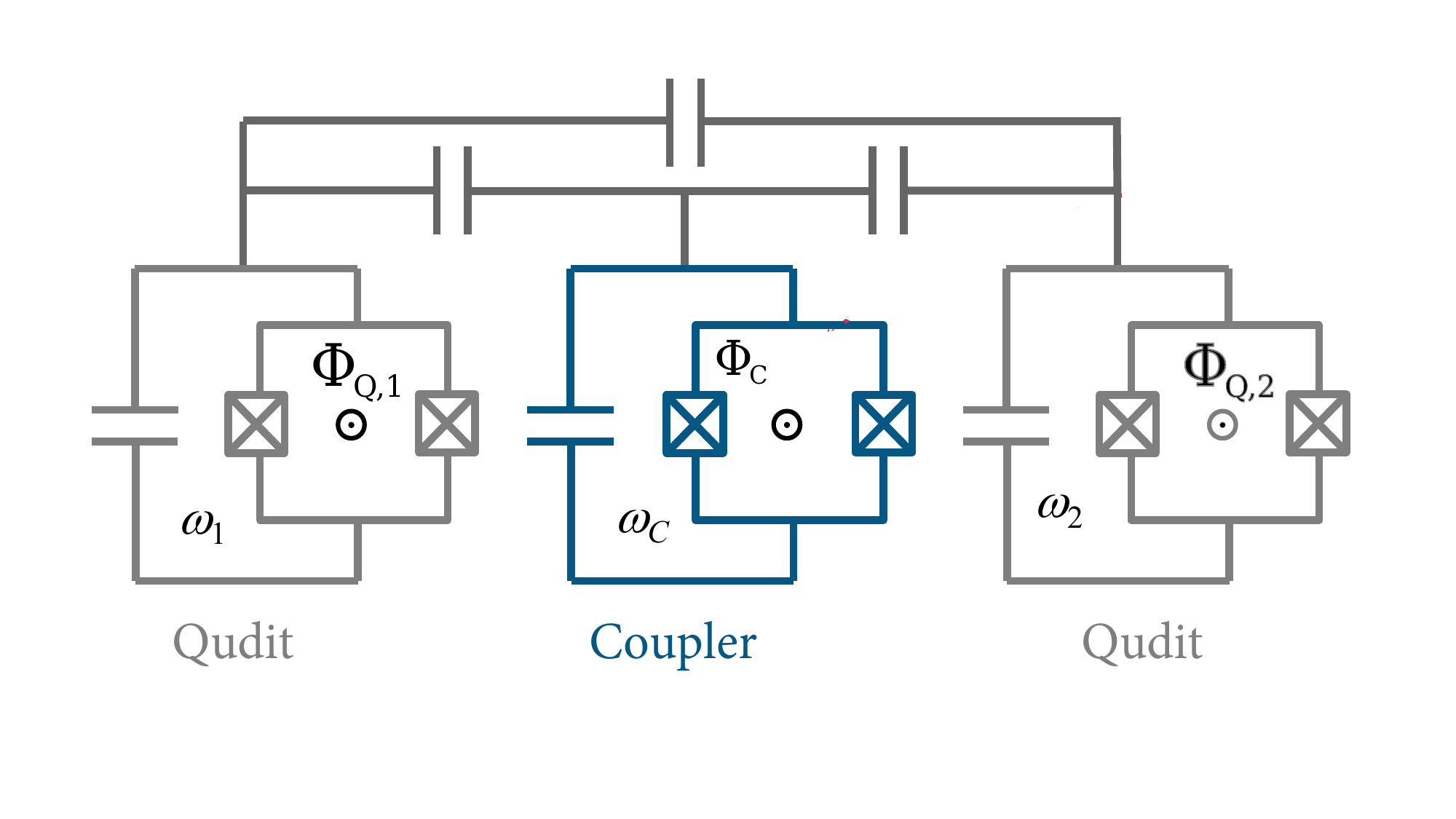}
\caption{Schematic circuit diagram of a transmon dimer with a tunable coupler.} \label{Fig:circuit_coupler}
\end{figure}

As reviewed in \citeSM{Kjaergaard2020,Yanay2020}, there are a large number of superconducting qubit architectures, all of which are suitable for implementing the Bose-Hubbard model.  Our processor is an array of tunable transmons, with a lattice layout partially depicted in Fig.~\ref{fig:chip_qubit_coupler_layout}.  In its simplest incarnation, a transmon is comprised of two Josephson junctions shunted with a capacitor, see Fig.~\ref{Fig:circuit_coupler}. The Hamiltonian of a single transmon can be approximated by that of the anharmonic Duffing oscillator~\citeSM{Koch2007}
with eigenstates labeled by an integer $n$, corresponding to an energy $E_n=\hbar \omega(n+1/2)+(\eta/2)n(n-1)+\cdots$, where the neglected terms are higher order in $n$. This, in turn, can be viewed as a Bose-Hubbard Hamiltonian in the atomic limit with Hubbard energy equal to the transmon anharmonicity $\eta$.  Thus, a transmon is a multi-level system (qudit) with Bose operators $\hat a$ and $\hat a^\dagger$ connecting the states in its Hilbert space: $\hat a |n\rangle=\sqrt{n} |n-1\rangle$, $\hat a^\dagger |n\rangle=\sqrt{(n+1)}|n\rangle.$ 
We interpret $n$ as the number of bosons (microwave photons) on a site, and introduce the number operator $\hat n=a^\dagger a$, satisfying $\hat n |n\rangle=n|n\rangle.$ 

If every transmon were treated as a qudit in the effective Bose-Hubbard model, the hopping terms $g(a_i^\dagger + a_i)(a_j^\dagger + a_j)$ would be generated by the capacitive coupling between the transmons.  Instead, we treat all the blue transmons in Fig.~\ref{fig:chip_qubit_coupler_layout} as ``couplers'' that have high frequency and therefore remain unpopulated\footnote{At idle the couplers are essentially unpopulated.  During the analog evolution, the effective qudit degrees of freedom do partially hybridize over the physical coupler transmons.  However, there are localized degrees of freedom adiabatically connected to the coupler modes at idle, which remain unpopulated.} and serve only to mediate the coupling between the qudits.  Adjacent qudits (gray transmons) are capacitively coupled to each other and also to the coupler between them.  The resulting coupling $g$ between the qudits is given by \citeSM{Yan2018}
\begin{equation}
\label{Eq:gmain}
    g \simeq \left( k_d - k_1k_2 \frac{\omega_q^2}{\omega_c^2 - \omega_q^2}\right)\frac{\sqrt{\omega_1\omega_2}}{2} \: ,
\end{equation}
where $\omega_q=(\omega_1+\omega_2)/2$ and dimensionless parameters $k_d$ and $k_i$ are coupling efficiencies that are functions of the circuit capacitances (Fig.~\ref{Fig:circuit_coupler}).  Here $k_d\ll k_1,k_2\ll 1$, and away from resonance $|g|\ll \omega_q$.

Both the qudit frequencies $\omega_i$ and the coupler frequencies $\omega_c$ are controllable via external fluxes, $\Phi_i$ and $\Phi_c$~\citeSM{Koch2007}.  Using an external flux bias to control $\Phi_c$ and therefore $\omega_c$, we can tune the hopping strength, Eq.~\eqref{Eq:gmain}, over a broad range.
In particular, the coupling can be turned off, $g=0$, at the coupler idle frequency, 
\begin{equation} \label{eq:omega_idle}
\omega_c^{\text{off}}=\omega_q\sqrt{1+k_1 k_2/k_d}.
\end{equation}
Note that the direct coupling $k_d$ is necessary for $g=0$ to be achievable.  When $\omega_c>\omega_c^{\text{off}}$ the effective coupling is negative, $g<0$, which is the regime we consider in the present experiment.  

 This procedure yields an effective Hamiltonian on the 4 x 9 qudit grid (gray transmons in Fig.~\ref{fig:chip_qubit_coupler_layout}) 
\begin{equation}\label{eq:transmon_bhm_v1}
    H_{\rm BH}^\prime=-g\sum_{\langle i,j\rangle}(a_i^\dagger - a_i)(a_j^\dagger - a_j)+\frac{\eta}{2}\sum_{i
    }n_i(n_i-1)+\sum_{i }\omega_i n_i.
\end{equation}
This Hamiltonian includes terms that do not conserve particle number ($a_i^\dagger a_j^\dagger$, etc.), and has a wide range of energy scales, with $\omega_i$ on the order of 6 GHz, $\eta$ on the order of 200 MHz, and $g$ on the order of 20 MHz or less.   However, the qudit frequencies $\omega_i=\omega_q+\delta_i$ are all relatively close to each other, $\delta_i\ll \omega_i$, where $\omega_q$ is the average qubit frequency. 
 As such, we can  transform into  a rotating frame with frequency $\omega_q$, \emph{i.e.} we work in the interaction picture with $H_0 = \omega_q \sum_i n_i$.  Then since $\omega_q\gg |g|$, we can make a rotating-wave approximation and drop the particle number non-conserving terms which take the system far off-resonance, arriving at the effective Hamiltonian for our experiments:
\begin{equation}
    H_{\rm BH}=g\sum_{\langle i,j\rangle}(a^\dagger_ia_j+{\rm H.c.})+\frac{\eta}{2}\sum_{i
    }n_i(n_i-1)+\sum_{i }\delta_in_i.
    \label{eq:transmon_bhm}
\end{equation}
Here $i$ indexes the transmons.  Operators $a_i^\dagger$, $a_i$, and $n_i=a_i^\dagger a_i$ are raising, lowering, and number operators for the $i$th transmon.  The first term encodes coupling between connected nearest-neighbors.  The non-linearity, $\eta<0$, is  fixed.   For state preparation and measurements one can also send microwave signals into the circuit which drive transitions, $H_{\rm ext}=\sum_i(\Omega_i(t) a_i+\Omega_i^*(t) a_i^\dagger)$.

Comparing with Eq.~(1) of the main text, we see that the coupled transmons reproduce the Bose-Hubbard model if we identify $J=-g$, $\mu_i=-\delta_i$, $U=\eta$.
Note that under this mapping, the Hubbard interaction $U$ is  negative but much of the interesting Bose-Hubbard physics requires $U>0$. 
The thermodynamic properties of the Bose-Hubbard model with an attractive interaction  are dramatically different from that of the repulsive model. In particular, the ground state of lattice bosons with an attractive interaction will exhibit a collapse: all bosons will seek to localize on a single lattice site.  
In the absence of a heat-bath, however, the sign of $U$ is largely irrelevant. Due to time-reversal symmetry one can map time evolution with a Hamiltonian $H$ onto evolution under $-H$.  The transformation $H\to-H$ flips the sign of $U$.  It also flips the signs of $g$ and $\delta_i$.  On a bipartite lattice the sign of $g$ can be restored via a gauge transformation (flipping the sign of the operators for one sublattice), and $\delta_i$ is under experimental control.  Thus one can use experiments on Hamiltonian~(Eq.~\ref{eq:transmon_bhm}) with $\eta<0$ to learn about the physics of Eq.~(1) in the main text with $U>0$. This connection between experiments involving $H$ and $-H$ is perhaps even more clear in the spectral domain.  The eigenstates of $H$ and $-H$ are the same after taking $E\to-E$.  Thus the ground state of $H$ maps onto the highest excited state of $-H$ and one can use the adiabatic theorem to produce this highest energy state starting from the product state $|11\dots 11\rangle$ when $g=0$, and then adiabatically modulating $g$ and possibly $\delta_i$ as needed.

\section{Qutrit calibration}\label{calibration}

In previous experiments using our quantum chip, we restricted ourselves (apart from leakage errors) to the $\{|0\rangle,\,\,|1\rangle\}$ qubit subspace.  Here we intentionally populate the $|2\rangle$ state and therefore develop additional calibration protocols to accurately control and readout the $|2\rangle$ state.

In order to control and readout the $\mid$2$\rangle$ state of our transmons, we have adapted the existing software infrastructure for the ($\mid$0$\rangle$, $\mid$1$\rangle$) subspace~\citeSM{Arute2019supremacy}. To begin with, we perform the same set of calibrations for control pulses as for the $\mid$0$\rangle\leftrightarrow\,\, \mid$1$\rangle$ transition, but for the $\mid$1$\rangle\leftrightarrow \,\,\mid$2$\rangle$ transition. This implies calibrating the frequency and amplitude of the pulses for the qubit idling at a given frequency in order to perform single qubit rotations accurately. 

We then calibrate the readout protocol to distinguish between the $\mid$0$\rangle$, $\mid$1$\rangle$, and $\mid$2$\rangle$ states. To this end, we adapted the optimization routine described in Ref. \citeSM{Bengtsson_2024} by factoring in the readout fidelity of the $\mid$2$\rangle$ state in the cost function of the routine. The primary challenge lies in choosing the duration of the readout pulse. When optimizing for readout fidelities, one wants to make the readout tone as long as possible to maximize the signal-to-noise ratio (SNR), but short enough to not be affected by $T_1$ errors where the transmon decays to another state~\citeSM{PhysRevApplied.7.054020}; the trade-off between these two effects leads to an optimal readout time.  However, the $T_1$ of the $\mid$2$\rangle$ state is $\sqrt{2}$ times shorter than that of the $\mid$1$\rangle$ state~\citeSM{Blais2021QED}.  Thus, a long pulse would maximize the separation between the $\mid$0$\rangle$ and $\mid$1$\rangle$ states, but the fidelity when distinguishing the $\mid$1$\rangle$ and $\mid$2$\rangle$ states would be limited by the shorter decay time of the $\mid$2$\rangle$ state. On the other hand, a short readout pulse would not suffer as much from the decay of the $\mid$2$\rangle$ state, but the separation between the $\mid$0$\rangle$ and $\mid$1$\rangle$ state readout clouds would be sub-optimal.  We choose a readout time that balances the fidelity in the $\mid$1$\rangle$ state and in the $\mid$2$\rangle$ state.  The resulting state preparation and measurement (SPAM) errors are shown in Table~\ref{table:SPAM}.  These readout fidelities could be optimized further, but this level of accuracy was sufficient for this project.

\begin{table}[h]
\centering
\begin{tabular}{cc|ccc}
& \multicolumn{1}{c}{} & \multicolumn{2}{c}{\textbf{$\,\,\,\,\,\,$ Measured (\%)}} \\
\multirow{4}{*}{\rotatebox[origin=c]{90}{\textbf{Prepared$\,\,\,\,$}}} & & $|0\rangle$ & $|1\rangle$ & $|2\rangle$ \\ \cline{2-5}
& $|0\rangle$ & 99.34 & 0.63 & 0.03 \\
& $|1\rangle$ & 1.36 & 98.49 & 0.15 \\
& $|2\rangle$ & 0.40 & 2.92 & 96.68 
\end{tabular}
\caption{State preparation and measurement errors including the $\mid$2$\rangle$ state.  We first prepare qutrits in either the $\mid$0$\rangle$, $\mid$1$\rangle$, or $\mid$2$\rangle$ states, then immediately measure them.  Here we report the \% measured in each of the three states.  In the absence of error, the result would be 100\% for each diagonal entry and 0 elsewhere.  For comparison, when we calibrate readout for the $\mid$0$\rangle$ and $\mid$1$\rangle$ states alone, the median readout fidelity is around 99.5\%.
\label{table:SPAM}}
\end{table}

Finally, the analog evolution part of the experimental procedure requires a precise calibration, detailed in Ref. \citeSM{Andersen2025}. No difference in this calibration procedure is required for operating the $\mid$2$\rangle$ state.

\section{Measuring two-point correlators}\label{sec:correlators}
\subsection{Measurement protocol}
An important quantity to measure in our experiments is the single-particle density matrix (SPDM)
\begin{align}
C_{ij}\equiv \langle a^\dagger_i a_j\rangle, 
\end{align}
where, as already introduced, $a_i$ is the annihilation operator for site $i$.  The matrix $C_{ij}$ encodes all single particle properties.  For example, its diagonal elements give the density on each site, and one can extract the kinetic energy as a sum of the nearest-neighbor off-diagonal elements.  The eigenvectors of $C_{ij}$ are the normal modes of the system, and the eigenvalues give the fraction of particles occupying each of these modes.  A Bose-Einstein condensate is defined by having one macroscopic eigenvalue; this eigenvalue, divided by the total number of particles, is the condensate fraction.  Despite its centrality, the single particle density matrix is hard to measure in most settings~\citeSM{Ardila2018}.

We have devised a protocol to measure $C_{ij}$ in our transmon implementation of the Bose Hubbard Model.  It requires turning off the coupling between the qutrits, isolating the sites, then applying a set of local microwave rotations.  One then measures the occupation numbers of each site.  
As described below, if the microwave gates are chosen appropriately, one can  then extract $C_{ij}\equiv \langle a^\dagger_i a_j\rangle$  from the number correlations.  

The key ingredients in our protocol are the Givens rotations between the number states $|m\rangle_i$ and $|n\rangle_i$ on site $i$:
\begin{equation}
    U^{mn}_{i,\alpha}(\varphi) = \exp\left(-i\frac{\alpha}{2}\left(e^{-i\varphi}X_i^{mn}+e^{i\varphi}X_i^{nm} \right) \right). 
\end{equation}
Here $X_i^{mn} = |m\rangle_i\, {}_i\!\langle n |$ are the bosonic Hubbard operators~\citeSM{Hubbard1964}, 
$m,n\in\{0,1,2\}$  are the number of particles on each site and $m\neq n$.  Viewing $m$ and $n$ as a two level system, and using the standard representation of the Bloch sphere, this corresponds to a rotation by $\alpha$ about a  vector in the $x-y$ plane, which makes an angle $\varphi$ with the $x$ axis.  These rotations can be implemented using the same pulse sequences which are used to perform traditional single qubit gates. 

We then note that within the space spanned by  the states $0,1,2$,  the operators $(n_i-1)$ and  $i(e^{-i\varphi} a_i-e^{-i\varphi} a_i^{\dagger})/\sqrt{3}$  have the same eigenvalues: -1, 0 and 1.  Hence there exists a unitary operator $V_i(\varphi)$ for which
\begin{equation}
    V_i^\dagger(\varphi)(n_i-1)V_i(\varphi) = \frac{i}{\sqrt{3}}\left(e^{-i\varphi}a_i - e^{i\varphi}a^\dagger_i \right)
    \label{Eq:Vi}
\end{equation}
This operator is unique up to a gauge transformation, corresponding to multiplying the rows by arbitrary phases.  After some non-trivial arithmetic one can show that this operator can be implemented as a three-part rotation in the $01$, $12$, and $01$ qutrit subspaces:
\begin{equation}
\label{Eq:Viangle}
    V_i(\varphi_i) = U_{i,\alpha}^{01}(\varphi_i)U_{i,\beta}^{12}(\varphi_i)U_{i,\gamma}^{01}(\varphi_i)
\end{equation}
with the choice of angles:  
\begin{equation}
    \begin{aligned}
        \alpha &= \pi/2&
        \quad\beta &= 2\arccos\left(\frac{1}{\sqrt{3}}\right)&
        \quad\gamma &= \pi/3.
    \end{aligned}
\end{equation}

Since the qutrits are isolated we can simultaneously apply microwave gates $V_i(\varphi_i)$ to a given pair of qutrits or to any subset including the entire system, i.e. perform the  
rotation of the system's state   $|\Psi\rangle$:
\begin{equation}
    |\Psi\rangle \to |\Psi_{R(\varphi)}\rangle = \prod_i V_i(\varphi_i)|\Psi\rangle,
\end{equation}
where index $i$ runs over all rotated sites. According to Eq.~\eqref{Eq:Vi},  measuring the density correlations then gives: 
\begin{equation}
\label{Eq:Correlator2}
    \langle \Psi_{R(\varphi)}|(n_i-1)(n_j-1)|\Psi_{R(\varphi)}\rangle = \frac{1}{3}\langle \Psi|\left(e^{i\varphi_i}a_i^\dagger -e^{-i\varphi_i} a_i \right)\left(e^{-i\varphi_j}a_j - e^{i\varphi_j}a^\dagger_j \right)|\Psi \rangle
\end{equation}
If $|\Psi\rangle$ belongs to a Fock space with fixed particle number, then 
\begin{equation}
    \langle \Psi |a_i a_j|\Psi\rangle = \langle \Psi |a_i^\dagger a_j^\dagger|\Psi\rangle =0.
\end{equation}
Finally, if the expectation value is parameterized as $\langle\Psi| a^\dagger_i a_j|\Psi\rangle = |C_{ij}|e^{i\phi_{ij}}$, we obtain the following relationship between the population measurement in the rotated state and the two-point correlator:
\begin{equation}
    \langle \Psi_{R(\varphi)}|(n_i-1)(n_j-1)|\Psi_{R(\varphi)}\rangle = \frac{2}{3} |C_{ij}|\cos(\varphi-\phi_{ij}). \label{eq:correlator}
\end{equation}
In principle, we could  measure the two-point correlator 
by sweeping the phase difference $\varphi=\varphi_i-\varphi_j$ between the microwave pulses applied to qutrits $i$ and $j$, 
measuring the density correlator, and fitting our data to Eq.~\eqref{eq:correlator}.
This procedure yields both
the amplitude $|C_{ij}|$  and phase $\phi_{ij}$.

In practice we measure a slightly more general operator.  We define
\begin{align}
W_i(\theta_i,\chi_i) &= 
U_{i,\alpha}^{01}(\varphi_i)U_{i,\beta}^{12}(\chi_i)U_{i,\gamma}^{01}(\varphi_i),
\label{Eq:Wi}
\end{align}
which differs from $V$ in that 
we have changed the phase angle of the second rotation.  This lets us correct for phases which accumulate  after the ramp-down of the qutrit couplings.  
For example, the isolated qutrits evolve under a Hamiltonian,
\begin{align}
{\cal H }_a &=\sum_i h_i(t)=\sum_i\left(\delta\omega_i(t) n_i+\frac{\eta_i}{2}n_i(n_i-1)\right),
\label{Eq:Ha}
\end{align}
where $\delta \omega_i(t)$ are the detunings and $\eta_i$ are the nonlinearities. The evolution operator $\exp(-i\int_0^\tau h_i(t)dt)$
does not commute with $V_i$ and induces phase shifts:
\begin{equation}
    V_i(\varphi_i)\rightarrow U_{i,\alpha}^{01}(\delta\varphi_i)U_{i,\beta}^{12}(\delta\varphi_i-\eta_i\tau)U_{i,\gamma}^{01}(\delta\varphi_i),
    \label{Eq:backshifts}
\end{equation}
where $\delta\varphi_i=\varphi-\int_0^\tau\delta\omega_i(t)dt$. The above operator in Eq.~\eqref{Eq:backshifts} has the same functional form as Eq.~\eqref{Eq:Wi}, which explains why the second phase $\chi_i$ is necessary if the application of the microwave gates is preceded by a free evolution under the ``atomic'' Hamiltonian~\eqref{Eq:Ha}.

In our measurement protocol, we  turn off the qutrit couplings, wait a time $\tau$, apply the $\hat W$ operators, then measure the occupations.   We can then extract the correlation function
\begin{equation}
Z_{ij}=
\langle \Psi (\tau)| 
\hat W_i^\dagger(\varphi_i,\chi_i)
\hat W_j^\dagger(\varphi_j,\chi_j)
(\hat n_i-1)(\hat n_j-1) 
\hat W_j(\varphi_j,\chi_j)
\hat W_i(\varphi_i,\chi_i)
 |{\Psi(\tau)}\rangle,
 \end{equation}
where $|\Psi(\tau)\rangle=\exp\left(-i\int_0^\tau {\cal H}_a(t)dt\right)|\Psi(0)\rangle$. Then the generalized Eq.~\eqref{eq:correlator} reads
\begin{equation}
\frac{3}{2}Z_{ij} = \rho^{01,10}_{ij} \cos(\varphi_i-\varphi_j-a_\tau)
 +\sqrt{2} \rho^{01,21}_{ij} \cos(\varphi_i-\chi_j-b_\tau) 
 +\sqrt{2} \rho^{21,01}_{ij} \cos(\varphi_j-\chi_i-c_\tau)
 + 2 \rho^{12,21}_{ij}\cos(\chi_i-\chi_j -d_\tau)
\end{equation}
where $\rho^{xy,zw}_{ij}=|\langle X_i^{xy} X_j^{zw}\rangle|$ are coherences between different occupation numbers.  
 The dynamical phases $a_\tau, b_\tau,c_\tau,$ and $d_\tau$ depend linearly on the wait time, and the qutrit parameters. 
 We vary the phases  $\varphi_i$ and $\chi_i$ to extract the coherences.  The modulus of the single particle density matrix is then
 \begin{align}
 |C_{ij}|=\rho _{ij}^{10,01}+\sqrt{2}\rho _{ij}^{12,10}+\sqrt{2}\rho _{ij}^{10,12}+2\rho _{ij}^{12,21}.
 \end{align}

\subsection{Benchmarking}
In order to validate our protocol for measuring correlators, $\langle a_i^\dagger a_j\rangle$, 
we use the two-qutrit circuit: 
\medskip
\begin{center}
\begin{quantikz}
\label{cirquit}
\lstick{$q0: |0\rangle$}&\qw\gategroup[2,steps=5,style={dashed,rounded
corners,fill=blue!20, inner
xsep=2pt},background,label style={label
position=below,anchor=north,yshift=-0.2cm}]{{\text
State preparation}}&\gate{U^{01}_\pi(0)}&\gate[2]{\rotatebox{90}{{\text fSim}}}&\gate{U_\pi^{12}(0)}&\gate{U_\pi^{01}(0)}&\gate{V(0)}&\meter{} \\
\lstick{$q1: |0\rangle$}&\qw&\qw&&\qw&\qw&\gate{V(\varphi)}&\meter{},
\end{quantikz}
\end{center} 
\medskip
That is, we are using a four-gate  sequence to prepare the target state, then applying our tomography pulses $V_i(\varphi_i)$ (see Eq.~\eqref{Eq:Viangle})  and measuring the populations. The  fSim gate~\citeSM{BrooksFSim}
is a fractional iSWAP  gate described by the unitary in the two-qutrit subspace spanned by the states $|10\rangle$ and $|01\rangle$:
\begin{equation}
\text{fSim}(\theta)=\left(\begin{array}{cc}\cos(\theta) & -i\sin(\theta)\\-i\sin(\theta) & \cos(\theta) \end{array}\right)
\end{equation}
The gate is implemented using a coupler pulse of duration $T$ when the qutrits are on resonance. This creates a time-dependent
coupling $g(t)$ (see Eq.~\eqref{Eq:gmain}) between the qutrits and enables a population swap. The swap angle $\theta$ depends on both the  shape and the strength of the pulse:
\begin{equation}
\label{Eq:theta1}
\theta=\int_0^Tg(t) dt \,.
\end{equation}
For instance, we may consider  pulses of the form
\begin{equation}
\label{g(t)}
g(t)=-J f(t,t_h,t_r)
\end{equation}
where $J$ is the maximum coupling strength and $f(t,t_h,t_r)$ is a flat-top (e.g. trapezoidal)
waveform, which is 1 over a hold interval of length $t_h$, and has equal ramp-up and ramp-down times $t_r$. Therefore, $T=t_h+2t_r$ and $\theta=\theta(J)=J(t_h+t_r)$ is just the area of the trapezoid. This relation is valid only if the qutrits are strictly on resonance. In reality it may not be the case due to some unwanted detunings that may inhibit a complete swap $\theta=\pi/2$. For this reason we need to characterize the fSim gate, i.e. independently measure $\theta(J)$ prior to performing our benchmarking protocol.
\begin{figure}[h!]
\centering
\includegraphics[width=0.5\textwidth]{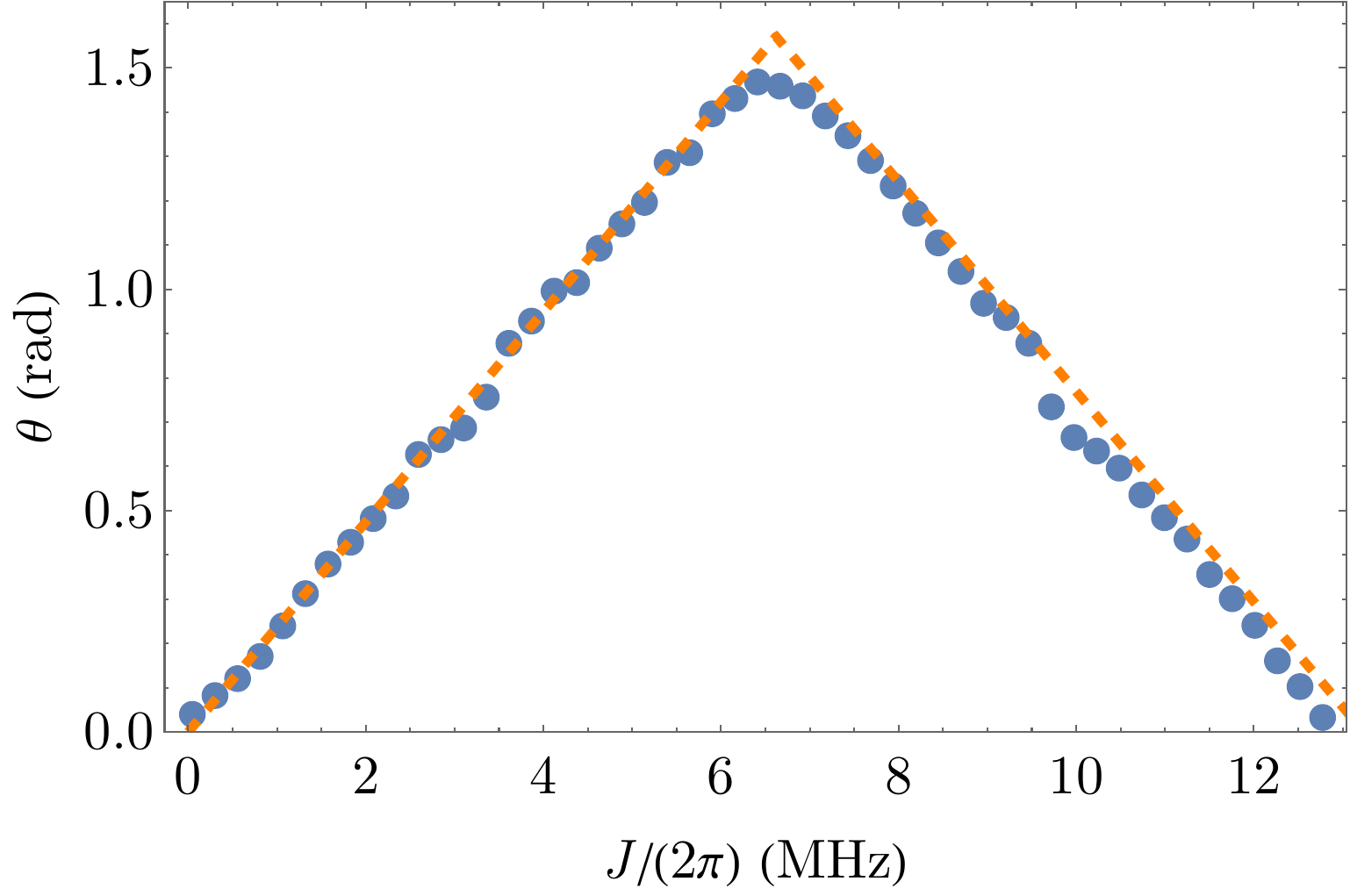}
\caption{Measured dependence $\theta(J)$ for the fSim gate used for benchmarking of our
qutrit tomography protocol (dark blue circles). The same dependence for an ideal gate is shown
for comparison (orange dashed line). By convention the swap angle is defined in the first 
quadrant, e.g. $\theta(J)=\pi-J(t_h+t_r)$ if $J(t_h+t_r)>\pi/2$ for the trapezoidal pulse.}
\label{Fig:Theta_J}
\end{figure}

As shown in the above circuit diagram, we first apply a $\pi$-pulse to qutrit $q0$  and initialize the system in the state $|10\rangle$. As a next step, we apply the entangling gate
$\text{fSim}(\theta)$  to create a superposition between $|10\rangle$ and $|01\rangle$ and then apply two consecutive $\pi$-pulses in 1-2  and 0-1 subspaces of the same qutrit. The resulting target state takes the form:
\begin{equation}
\label{eq:fsim_state}
|\psi\rangle= -i\sin\left(\theta(J)\right)|11\rangle+\cos\left(\theta(J)\right) |20\rangle
\end{equation}
Using this equation it is straightforward to find  the correlator in question:
\begin{equation}
\label{eq:fsim_correlator}
C_{01}(J) = \langle\psi|a_0^\dagger a_1|\psi\rangle=\frac{i}{\sqrt 2} \sin\!\left(2\,\theta(J)\right)
\end{equation}

\begin{figure*}[th!]
\centering
\includegraphics[width=0.5\textwidth]{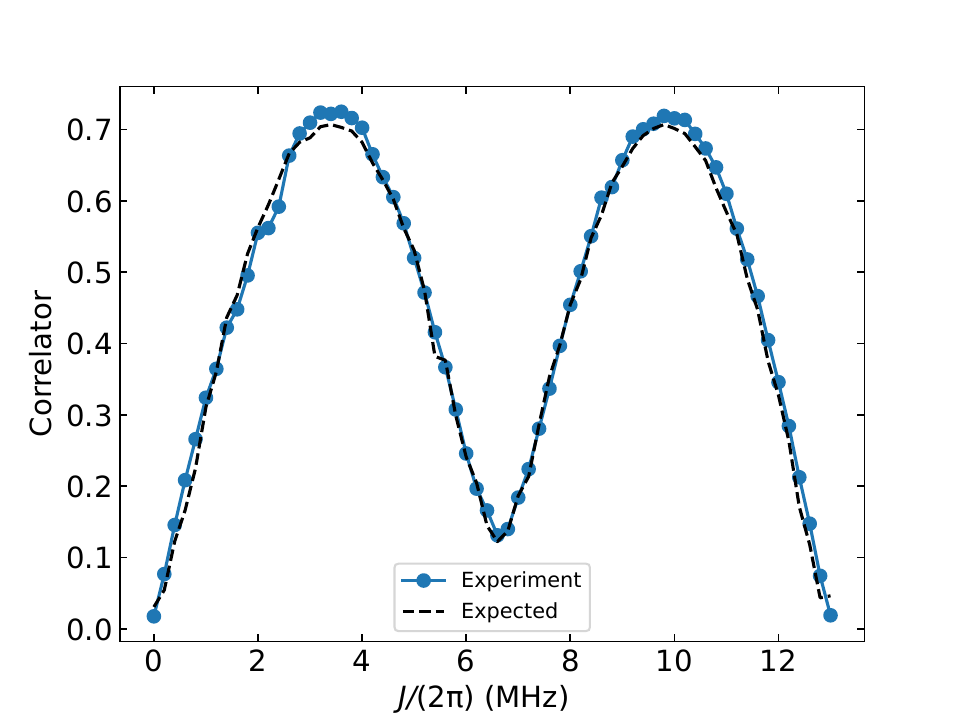}
\caption{Benchmarking using the fSim gate. Along the horizontal axis we show the  coupling strength $J$, while along the vertical axis we plot the absolute value of the correlator. We compare data (blue) to the predicted theoretical value (dashed black line) based on Eq.~\eqref{eq:fsim_correlator} and measured swap angles $\theta(J)$ (Fig.~\ref{Fig:Theta_J}).}  
\label{fig:benchmarking_fsim}
\end{figure*}

We implement and characterize the fSim gate for varying $\theta(J)$ values by turning on the the coupling between the two neighboring qutrits for a fixed duration and sweeping the coupling strength $J$ with qutrits nominally on resonance. Using our standard unitary tomography in qubit subspace, we measure the angle $\theta$ for each coupling strength.
The results of these measurements are shown in Fig.~\ref{Fig:Theta_J}. Using the characterized gates, we sweep $J$, prepare the state $|\psi \rangle$ (Eq.~\eqref{eq:fsim_state}), and measure the correlators using our protocol. We compare the measured correlators to those predicted from Eq.~\eqref{eq:fsim_correlator}
with measured $\theta(J)$ (Fig.~\ref{Fig:Theta_J}). Fig.~\ref{fig:benchmarking_fsim}
reveals a very good agreement between the measured and the predicted values of the correlator.

\section{Experimental protocol: adiabatic ground state preparation\label{sec:adiabaticity}}

In this section we describe our protocol for preparing low-energy states.  We also include some comparisons of measured energy with results from the numerical simulations described in Part~\ref{part:simulation} below, showing that we successfully prepare a state close to the ground state for a range of interaction strengths $J$ and disorder strengths $W$.

\subsection{State preparation protocol}

We follow a typical adiabatic state preparation procedure. To prepare the ground state of the target Hamiltonian $H$, we first prepare the ground state of a much simpler Hamiltonian $H_0$, then slowly and continuously transform $H_0$ to $H$.  As long as the rate of the transformation is small compared to the many-body energy gap $\Delta_E$ between the ground state and the first excited state, 
\begin{equation}
    \frac{1}{t_\text{ramp}}\ll \Delta_E,
\end{equation}
the adiabatic theorem tells us that we will end up close to the ground state of $H$.  Barring decoherence, in the limit of an infinitely slow ramp and a finite energy gap everywhere along the ramp, we would achieve the exact ground state.  

Importantly, this procedure also works not just for the global ground state, but also for an eigenstate within a specific quantum number sector corresponding to a global symmetry.  In the specific case of the Bose-Hubbard model, Eq.~\eqref{eq:transmon_bhm}, we have particle-number conservation, so we can find an eigenstate of $H$ with filling 1 per site by starting from an eigenstate of $H_0$ with that same filling. 
Furthermore, our target state is the ground state in the filling-1 sector of the repulsive Bose-Hubbard model, $H_{\text{RBM}}$.   Since our physical Hamiltonian is $H=-H_\text{RBM}$, we instead target the highest-excited eigenstate in this symmetry sector.

We therefore begin by preparing the highest-energy state of the simplified $H_0$ that we derive from Eq.~\eqref{eq:transmon_bhm} by tuning all qubit frequencies to be equal, so that $\delta_i=0$ for all sites $i$, and tuning the coupler frequencies to set $g=0$ on all couplers.  Then
\begin{equation}
    H_0 = \frac{\eta}{2}\sum_i n_i (n_i-1)
\end{equation}
with $\eta<0$, so that the highest energy state within the desired particle number sector is simply the product state $|\psi(t=0)\rangle = |1\rangle \otimes|1\rangle\otimes\cdots|1\rangle$ = $|1\rangle^{\otimes N_q}$ where $N_q$ is the number of qubits.

We then simultaneously ramp up $g$ to the desired value for the target Hamiltonian and ramp the frequency of each qubit to its value in a given disorder realization of interest.  We use a ramp of length 100 ns, chosen because it is sufficient to achieve approximate adiabaticity. We hold the Hamiltonian constant for a short period of 10 ns, so that any transient control signals have time to settle to the desired values.  After this, we have prepared a final state that is close to the target state.  

Finally, before performing any measurements we must isolate the individual qudits by ramping the coupling strength $g$ to 0 and the qubit frequencies back to their idle configuration.  Ideally this would be a quench---an instantaneous change to $H$ so that the population on each qubit is frozen at its value from the prepared high-energy state (approximate ground state with $U<0$).  However, a perfect quench is not possible on a physical device, so we simply ramp-down as fast as possible.  This ramp-down substantially alters the state, an effect which we characterize using numerical simulations described in Part~\ref{part:simulation}, starting on page.~\pageref{part:simulation}.

\begin{figure}
    \centering
    \includegraphics[width=0.7\linewidth]{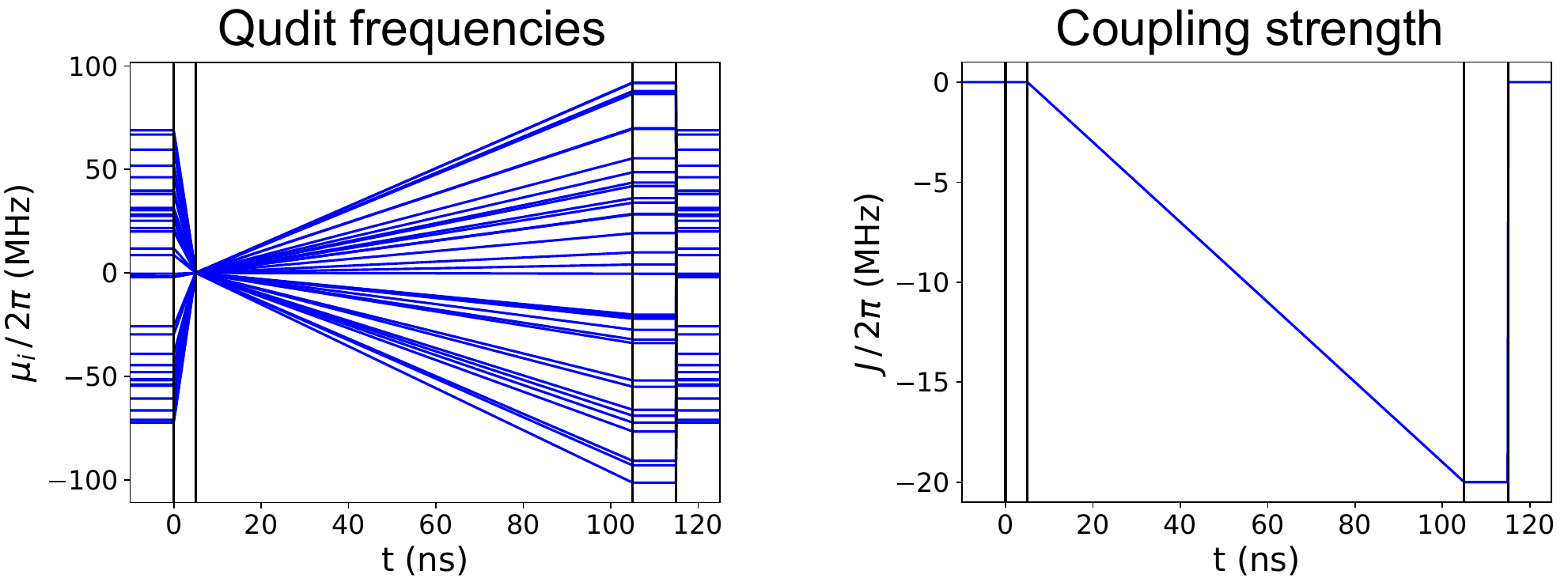}
    \caption{
      Illustration of the adiabatic state preparation procedure for coupling strength $J/(2\pi)=20$ MHz and disorder $W/(2\pi)=200$ MHz, for one randomly chosen disorder realization.  The left panel shows the chemical potentials of all qudits as a function of time; the right panel shows nearest-neighbor coupling strength.  Until time $t=0$ ns the system is at idle, with no coupling.  (The true idle frequencies have been replaced with random values.)  The qubits are brought onto resonance by $t=5$ ns, then over 100 ns the coupling is adiabatically switched on and the qubit frequencies are ramped to their final disorder configuration.  The system is held with this constant Hamiltonian for 10 ns, and then quickly ramped back to idle.  Here we show the ramp-down as a perfect instantaneous quench, though that is not possible in practice; the actual ramp-down takes finite time on the order of ns.  Vertical black lines indicate where ramps begin and end.}
    \label{fig:adiabatic_prep_illust}
\end{figure}

This whole procedure is summarized in Fig.~\ref{fig:adiabatic_prep_illust}.  The qudits are initially at some idle configuration (optimized to achieve zero qudit-qudit coupling), and we use microwave gates to create the initial state $|1\rangle^{\otimes N_q}$.  Starting at time $t=0$ we quickly (5 ns) ramp the qudit frequencies to all be equal; since the coupling is 0, evolution during this time gives only an irrelevant global phase so that the initial state is unchanged\footnote{In practice, there is a tiny amount of evolution during this time, as the coupling includes many-body effects and cannot be perfectly turned off.  But this effect is small enough to ignore.}.  Then we adiabatically ramp up the coupling to the desired value (in the figure, $J/(2\pi) = 200$~MHz) and ramp the qubit frequencies to a specified disorder realization, over 100 ns.  We hold for 10 ns, and then ramp back to the idle configuration as fast as possible (shown in the figure as a quench, but in practice taking a short time that is set by hardware details).  Finally, we perform measurements on the state.

Note that in the figure, the true qubit idle frequencies (start and end) have been replaced with random values.  However, the frequencies between 105 and 115 ns correspond to one of the actual disorder realizations used in our experiments.  Also note that the figure shows a linear ramp for qubit frequency and coupler strength, but in practice a different ramp shape is used.

State preparation in the presence of disorder is hampered by the existence of metastable states. Although it is not entirely possible to prevent the system from settling into a metastable configuration during the adiabatic ramping, we find that the optimal state preparation protocol is one in which (1) we first bring the qudits to a common interaction frequency with no disorder, and then (2) turn on the inter-site coupling and disorder potential at the same time and rate.  In Sec.~\ref{sec:MPS} below, we use tensor network simulations to compare the state prepared by adiabatic state preparation with the true ground state at each coupling strength and disorder strength.  In most of the parameter space considered, the adiabatic procedure successfully prepares a state close to the ground state.  Only when the disorder strength is comparable to or greater than $U$ and the hopping is very small, we find that the procedure fails due to disorder, producing states with local observables very different from those in the ground state. (See Fig.~\ref{fig:MPS_vs_exp_doublon_frac}.) 

\subsection{Energy measurements vs. simulation results}

We estimate the energy of our final state by measuring the expectation value of the Hamiltonian: $E=\langle H\rangle=K+V_{\rm int}+ V_{\rm \delta}$, with
\begin{align}
    K &= -J\sum_{\langle ij\rangle}\langle a^\dagger_i a_j + a^\dagger_j a_i\rangle, \\
    V_{\rm int}&=\frac{U}{2}\sum_i \langle n_i (n_i-1)\rangle, \\
    V_{\rm \delta}&=\sum_i \mu_i (\langle n_i\rangle - 1),
\end{align}
referred to as the kinetic, interaction, and on-site energy, respectively.
Here $J$ and $\mu_i$ are the parameter values from the target Hamiltonian after the ramp, not from the final Hamiltonian at the time of the measurement.  (For example, in the case illustrated in Fig.~\ref{fig:adiabatic_prep_illust}, we use $J/(2\pi) = 20$ MHz to estimate energy, but $J=0$ at the time of measurement.)

\begin{figure}
    \centering
    \subfloat{
        \includegraphics[width=0.46\textwidth]{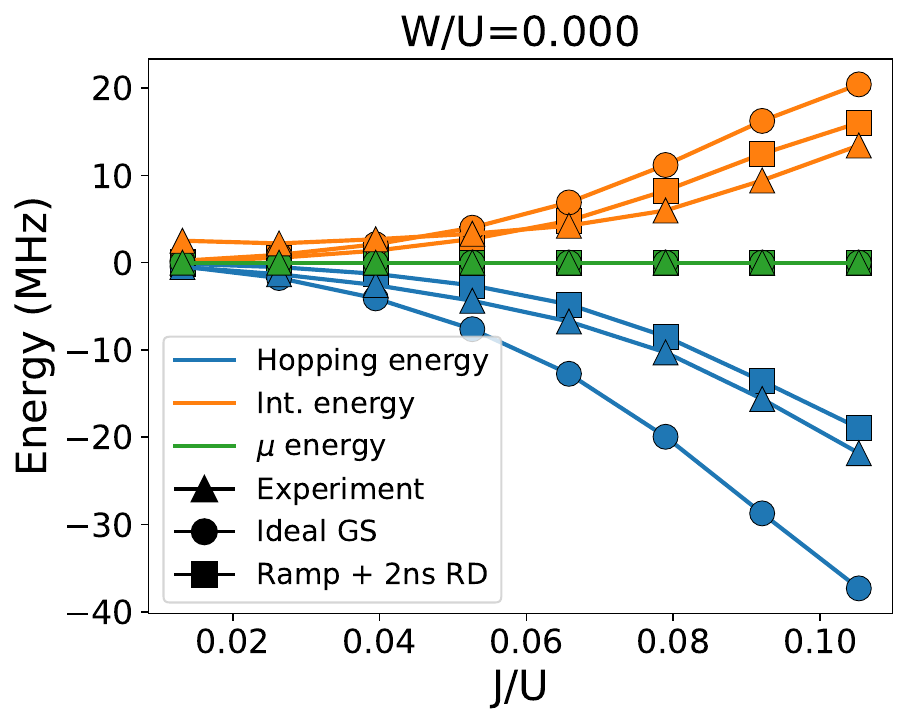}
    }
    \qquad
    \subfloat{
        \includegraphics[width=0.46\textwidth]{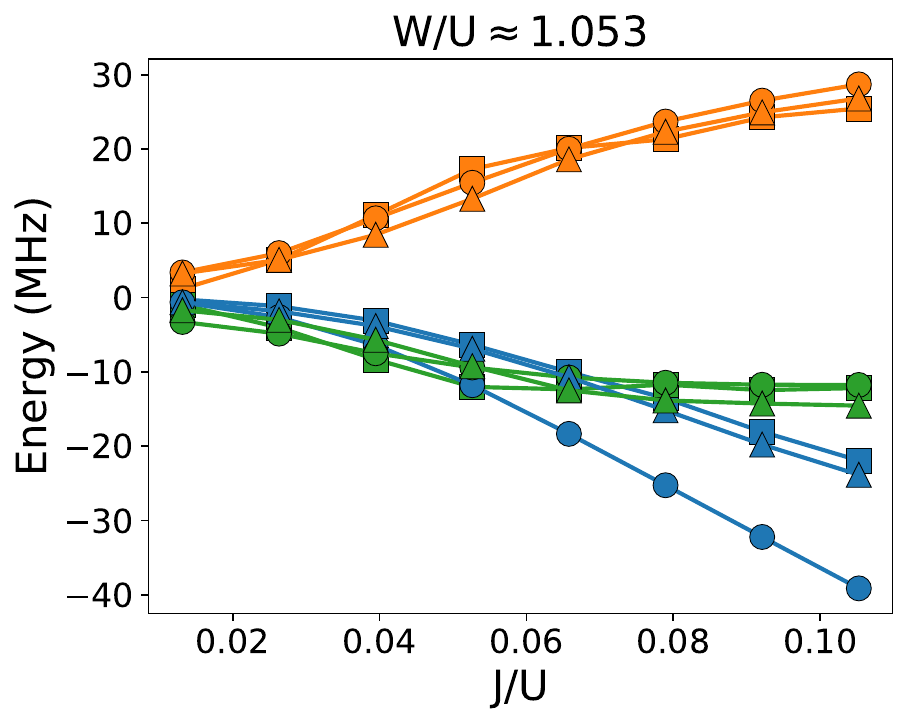}
    }
    \caption{We compare the energy density measured in the experiment with that predicted by tensor network numerical simulations (described in Sec.~\ref{sec:MPS} below), both (a) in the disorder-free case, $W=0$, and (b) for $W/(2\pi)=200$ MHz.  The orange lines correspond to interaction energy ($U$ term in the Hamiltonian), the blue lines to energy from nearest neighbor correlators, $\langle a_i^\dagger a_j\rangle$, and the green lines to energy from fluctuations in particle number $\langle n\rangle$ across the chip.  The latter is exactly 0 in the disorder-free case but is significant when disorder is nonzero. 
    The experimental energy density is compared with energy computed in the ideal target ground state (circles), as found using well-converged density matrix renormalization group (DMRG) simulations~\protect\citeSM{white_1992, white_1993, Schollwock_2011}.  We also show the results of simulating adiabatic state preparation followed by a fast 2 ns ramp-down of the coupling (squares), close to the procedure used in the actual experiment.  This gives a much closer agreement with the experimental energy density, while simulating just adiabatic state preparation (not shown) gives good agreement with the true ground state instead.  Taken together, these data suggest that we indeed prepare a very good approximation to the target state through our adiabatic state preparation procedure. 
    }
    \label{fig:energy_density}
\end{figure}

The measured values of the three energy terms are shown in Fig.~\ref{fig:energy_density} for a range of coupling strengths and for two disorder strengths ($W=0$ and $W\approx U$); in the latter case, the results are averaged over disorder realizations.  The figures also show the same quantities computed by tensor network numerical simulations described in Part~\ref{part:simulation}.  

Notably, the experimental energy measurements (triangles) only roughly agree with the predictions from a direct numerical simulation of the target ground state (circles).  However, the disagreement is primarily due to the finite ramp-down after the state preparation procedure.  To see this, we also simulate the effects of a 2 ns ramp-down, close to the actual fastest ramp-down allowed by our hardware.  The resulting energies (squares) are very close to those found in the experiment.  In comparison, if we simulate the adiabatic state preparation procedure up to the hold time (110 ns in Fig.~\ref{fig:adiabatic_prep_illust}), the results are nearly identical to those of direct ground state search, confirming that the discrepancy between the experiment and the target ground state is due to the ramp-down rather than a failure of the adiabatic state preparation.  See Sec~\ref{sec:sims_GS_prep} for more details on the numerical simulations and further comparisons between experiment and theory.

\section{Experimental protocol: compressibility\label{sec:compressibility}}
The compressibility, defined by 
\begin{equation}\label{kappadef}
    \kappa = \frac{1}{n^2}\frac{\partial n}{\partial \mu},
\end{equation}
where $n$ is the average filling, is an indication of how particle density responds to a change in the chemical potential, $\mu$.    An incompressible system, where $\kappa=0$, is one in which there is no change in the particle density in response to a perturbation in the potential.  This is characteristic of an insulator. Conversely, a compressible system will rearrange its particle density to respond to spatial gradients in the potential. 

\subsection{Compressibility as a response to a potential gradient}

We measure the compressibility as the response of the system to a long wavelength potential, so that Eq.~\eqref{kappadef} applies locally in different parts of the system, while the total particle number remains constant.  Intuitively, $\delta n_i\approx \kappa \delta\phi_i$, where the disturbance has the form
\begin{align}
H^\prime=\sum_i \phi_i n_i.
\end{align}
The potential $\phi_i=\delta \mu_i$ is slowly varying 
and $\sum_i \phi_i=0$.  In particular, we take $\phi_i=\delta \mu \cos(x_i \pi/L_x)$, where $L_x=9$ is the length of the system in the long direction.  We then extract the compressibility as
\begin{equation}
    \kappa \approx \frac{2}{N\delta\mu}\sum_j \langle n_j\rangle \cos \left(\frac{x_j\pi}{L}\right). \label{eq:compressibility}
\end{equation}
We take $\delta\mu=30$ MHz, and we have verified that we are in the linear response regime, where $\kappa$ computed according to Eq.~\eqref{eq:compressibility} is independent of the strength of the disturbance, $\delta \mu$.

One potential complication is that we measure compressibility in the presence of a disordered potential, where the chemical potential $\mu_i$ on each site is independently drawn from a uniform distribution between $-W/2$ and $W/2$.  It could happen that a disorder realization naturally includes a significant ``tilt'' where one side of the system has higher $\mu$ than the other side on average.  Because this would naturally induce a gradient in particle density, and in measuring compressibility we want to see how the controlled tilt $\delta\mu$ perturbs particle density away from a flat configuration, we also subtract off the natural tilt from each disorder configuration.

To be specific, the local potential $V_i$ on each site in the compressibility experiments has two contributions: $h_i$ from the disorder realization and $\phi_i$ from the applied tilt.  We subtract off the natural tilt due to the disorder configuration by modifying $V_i$ to
\begin{equation}
   \begin{aligned}
        V_i &= \mu_i + (\delta\mu-A)\cos(\pi x_i/L);\\
        A&\equiv \frac{2}{N}\sum_i \mu_i \cos(\pi x_i/L).
   \end{aligned}
\end{equation}
Thus we measure response only to the intentionally applied tilt and not to the small random tilt inherent in the disorder configurations.  Note that while averaging over sufficiently many disorder realizations should make this unnecessary, this way we avoid errors from statistical fluctuations that would leave us with a small net tilt among a finite set of disorder realizations.

\subsection{State preparation protocols in the presence of a tilt and disorder}
The state in which we take the average, $\langle n_i\rangle$, in Eq.~\eqref{eq:compressibility} has not been rigorously defined yet.  The key is that we are trying to measure a susceptibility, which inherently involves multiple states---not just the ground state, but also the states that one gets after applying small perturbations.  These states, and hence the measured $\langle n_i\rangle$ and therefore the compressibility, could depend on the way that the perturbations are applied.  In particular, in the setting of adiabatic state preparation, one can tune from the simple Hamiltonian $H_0$ to the target $H$ along any number of paths through parameter space, and the measured compressibility could depend on the choice of path.

In a system that quickly equilibrates, this is not a concern: with reasonably adiabatic state preparation, as long as the final Hamiltonian is the same, any path in Hamiltonian space we take will lead to the same final prepared state and hence a consistent (path-independent) definition of compressibility.  Fast equilibration, i.e. ergodicity, is expected for clean (disorder-free) systems according to the eigenstate thermalization hypothesis (ETH)~\citeSM{Deutsch1991}.

Now, let us consider what happens if disorder is present. If the system is glassy, as is expected in the regime of intermediate to strong disorder, the time scales for equilibration become very long.  When we attempt adiabatic state preparation, instead of reaching the ground state, we end up in a metastable state whose local subsystems are not in equilibrium.  This metastable state depends on the path taken in parameter space, even if the initial state and initial and final Hamiltonians are the same.

With this in mind, we use two distinct state preparation protocols to perform the susceptibility measurement in the presence of disorder, each of which maps onto an experimental paradigm in the study of classical glasses.  They are:
\begin{itemize}
    \item Zero-field cooled (ZFC): Prepare the state with disorder and then apply the cosine tilt to the potential. 
    \item Field cooled (FC): Apply the cosine tilt first and then turn on the couplers.
\end{itemize}
In fact, we use both protocols everywhere in parameter space, including where the disorder is small and the system is ergodic. In that case, as noted above, we expect the state preparation to be path-independent and hence that the compressibility measurement between the two protocols will agree.  On the other hand, in the presence of strong disorder, the two protocols may give different results.  Thus disagreement between the two is a useful diagnostic for non-ergodicity.

Finally, let us address some practical considerations. First, one might be concerned that initializing a state with a tilted potential will create resonances in the system, thus destroying the Mott insulator even for very small values of coupling $J$. However, the time scale associated with this process should be $\tau \sim \frac{1}{J^L}$, which is very slow assuming small $J$. Furthermore, we can always make the tilt small enough to prevent any resonances from appearing. We take $\delta\mu = 30~\text{MHz}$, which we confirm to be in the regime of linear response.

\section{Experimental protocol: Bragg spectroscopy}\label{supp:linear_response}

We probe the excitations of our system by measuring the response to a perturbation of the form
\begin{equation}
    H' = \sum_i \phi_i(t) \hat{n}_i
    \label{Eq:probe}
\end{equation}
We use a sinusoidal drive, $\phi_j(t)= A_d\cos(\omega_d t)\gamma_j$, where $A_d$  and $\omega_d$ are the amplitude and the frequency of the drive, respectively, $\hat{n}_i=a_i^\dagger a_i  $, and the coefficients $\gamma_i$
describe a non-uniform spatial part of the perturbation that will be specified below.

\vspace{0.3cm}

\paragraph{Linear response.} Within the linear response regime one can
use time-dependent perturbation theory at $T=0$ and express density perturbation on site $i$ as:
\begin{equation}
\delta n_i(t) = \sum_{j}\int_{-\infty}^\infty dt' \chi^R_{ij}(t-t')\phi_j(t^\prime),
\label{Eq:delta n_i(t)}
\end{equation}
where $\delta n_i(t)=\langle n_i(t)\rangle -\langle n_i\rangle_0$ is the deviation of the density from its equilibrium value and
\begin{equation}
 \chi^R_{ij}(t) = -i\Theta(t)\langle[ \hat{n}_i(t),\hat{n}_j(0)]\rangle_0, \label{eq:polarizability_density_density}
 \end{equation}
is the retarded density-density correlation function~\citeSM{PinesNozieres1999,NegeleOrland1998}. The angular brackets for any observable stand for its
expectation value  over the ground state $|0\rangle$. Transforming Eq.~\eqref{Eq:delta n_i(t)} into the Fourier domain we get:
\begin{equation}
\label{Eq:deltan_omega}
\langle\delta n_i(\omega_d)\rangle=A_d \sum_j \chi_{ij}(\omega_d)\gamma_j,
\end{equation}
where the dynamical susceptibility 
\begin{equation}
\chi_{ij}(\omega_d)=\sum_{k \ne 0}\langle 0|\hat n_i|k\rangle\frac{2\omega_{k0}}{(\omega_d+i\epsilon)^2-\omega_{k0}^2} \langle k |\hat n_j|0\rangle,
\label{Eq:chi_ij}
\end{equation}
is the Fourier transform of $\chi_{ij}^R(t)$ in Eq.~\eqref{eq:polarizability_density_density}, $\omega_{k0}=\omega_k-\omega_0$ and $\epsilon\rightarrow +0$. Here the summation is taken over all excited many-body states $|k\rangle$.

Our primary goal is to measure the susceptibility $\chi_{ij}(\omega)$ and detect emergence of the collective sound modes with wavelength $\lambda/2\sim L$ where $L$ is the linear dimension of the system. To achieve this goal we perform a Bragg spectroscopy experiment and choose coefficients $\gamma_j$  in the form of standing waves on a rectangular grid:
\begin{equation}
\gamma_{j}^{m}=A_{pq}\cos(k_{x,p}x_r)\cos(k_{y,q}y_s).
\label{Eq:drum_modes}
\end{equation} 
Here $j=\{r,s\}$ and $m=\{p,q\}$ are combined indices, $x_r$ and $y_s$ are $x,y$ coordinates of the qudits on a 2D grid with linear dimensions $L$ and $M$ in $x$ and $y$ directions, respectively, such that 
the total number of sites ${\cal N}=(L+1)(M+1)$. If the space origin is chosen in the left bottom corner of the grid, $x_r$ and $y_s$ are integer numbers running from 0 to $L$ and from 0 to $M$,
respectively. The indices $p$ and $q$ enumerate the wave numbers $k_{x,p}=\pi p/L, p=0\dots L$ and   $k_{y,q}=\pi q/M, q=0\dots M$, but the point $q=p=0$  must be excluded.  

The set of matrix elements $P_m =\left \{\langle m |\hat n_i |0\rangle\right\}$ for a given eigenstate $m$ and with $i$ running over all lattice sites can be considered as a vector with $(L+1)(M+1)$ elements. Similarly, the set $Q_n=\{\gamma_j^n\}$ is also a vector of the same dimension. It may happen that $P_m$ strongly overlaps with only one  $Q=Q_m$ and is orthogonal (or almost orthogonal) to all other vectors $Q$, i.e. the dot product $P_m\cdot Q_n\simeq \text{Const}\times\delta_{mn}$. In this case, if we choose all $\gamma_i$ from the set $Q_m$ in 
Eq.~\eqref{Eq:deltan_omega}, the perturbation will select only one state $|m\rangle$ according to Eq.~\eqref{Eq:chi_ij}:
\begin{equation}
\langle\delta n_i(\omega_d)\rangle=A_d\sum_{k \ne 0}\langle 0|\hat n_i|k\rangle\frac{2\omega_{k0}}{(\omega_d+i\epsilon)^2-\omega_{k0}^2} \sum_j  \langle k |\hat n_j|0\rangle\gamma_j^m
\propto \gamma_i^m \frac{2\omega_{m0}}{(\omega_d+i\epsilon)^2-\omega_{m0}^2}.
\label{Eq:StandingWave}
\end{equation} 
 Here we used the fact that the vectors $Q_m$ and $P_m$ are collinear. Eq.~\eqref{Eq:StandingWave} shows that we were able to produce a collective excitation of the same standing wave shape as the perturbation. This excitation is a sound wave with the same wave number as the perturbation. To find the frequency of this state it is convenient to normalize both $P_m$ and $Q_m$. Then we can project the density wave onto $Q_m$. The resulted {\em mode} susceptibility does not have any spatial dependence and depends only on frequency (or time):
 \begin{equation}
 \sum_i \gamma_i^m\langle\delta n_i(\omega_d)\rangle=A_d\frac{2\omega_{m0}}{(\omega_d+i\epsilon)^2-\omega_{m0}^2}\equiv A_d \chi_m(\omega_d).
 \end{equation}
We can detect the resonance by analyzing both the absorptive and reactive parts of the mode susceptibility, however we found that in practice the latter is more robust and we detect the resonance by finding the frequency at which the reactive part of the response crosses zero, i.e. when the response is exactly in quadrature with the drive.

Our approach is based on the same intuition as the Feynman-Bijl ansatz~\citeSM{Feynman1954} for the excited states of the superfluid $^4$He:
\begin{equation}
|\bm{k}\rangle\propto\sum_ie^{\bm{kr}_i}\hat{n}_i|0\rangle,
\end{equation}
where $|0\rangle$ is the ground state, ${\bm k}$ is the wave vector and ${\bm r}_i$ are the positions of the $^4$He atoms. Then our density probe~\eqref{Eq:probe} and response~\eqref{Eq:StandingWave} have the same shape as the Feynman-Bijl's trial functions in which the plane waves are replaced with the appropriate drum modes~\eqref{Eq:drum_modes} of the finite 2D system.  

The collective sound modes described above emerge only at large $J/U\equiv |g/\eta|$, which is consistent with the crossover to the superfluid phase. Their existence has been confirmed both experimentally and numerically. In the MI phase the coupling with the density is very weak and these modes are not detectable, which is another confirmation that the emergence of the collective sound waves is a signature of superfluidity.  Since we were able to  assign a wave vector to each of these states, we know the dispersion relation  $\omega(k)$  for several discrete k-points. This, in turn, allows us to extrapolate  the dependence to zero and confirm that these are indeed Bogoliubov-Goldstone excitations with the linear dispersion $\omega(k)\propto k$.

\vspace{0.3cm}

\paragraph{Strong driving.} In the regime of a strong driving~Eq.~\eqref{Eq:delta n_i(t)} is no longer valid.  However, in the vicinity of the resonance for a transition between the ground state $|0\rangle$ and an excited state $|k\rangle$ only these two states are important and we have to solve the Scr\"{o}dinger equation, projected onto the two-dimensional subspace spanned by  $|0\rangle$ and $|k\rangle$, non-perturbatively:
\begin{equation}
\label{Eq:SE}
i\left(\begin{array}{c}\dot{c}_0(t)\\\dot{c}_k(t)\end{array}\right)=\left(
\begin{array}{cc}
 -i\Gamma_0  & A_d \cos( \omega_d t) 
   \\
 A_d \cos( \omega_d t) &
   \omega_{k0} -i\Gamma_k  \\
\end{array}
\right)\left(\begin{array}{c}c_0(t)\\c_k(t)\end{array}\right)
\end{equation}
with initial condition $c_0(0)=1$ and $c_k(0)=0$.  Here $\Gamma_0=(\gamma_s-\gamma)/2$,  $\Gamma_k=(\gamma_s+\gamma)/2$ are the decay rates of the state $|0\rangle$ and $|k\rangle$ respectively. The parameter $\gamma$ describes  asymmetry in the $T_1$ decay caused by the spread of $T_1$ times for different qudits. While the average decay $\gamma_s$ can be remedied using post-selection on the total number of particles, the $T_1$
asymmetry depends on the spatial character of the wave function and must be taken into account explicitly.  Solving Eq.~\eqref{Eq:SE} allows us to define the non-linear mode susceptibility in the time domain as:

\begin{equation}
\chi_k(t) = 
\frac{c_0^*(t) c_k(t) + c_k^*(t) c_0(t) + \langle k |\hat \nu_k|k\rangle |c_k(t)|^2}{A_d\left(|c_0(t)|^2+|c_k(t)|^2\right)} 
\end{equation}
where we have defined $\hat \nu_k = \sum_i \gamma_i^{k} \hat n_i$. This susceptibility is non-linear, but it still can be Fourier analyzed, and in most cases this allows us to detect the resonance using the reactive part of the response at the {\em driving frequency},  as was done in the case of the linear response. Technically, we perform the same Bragg spectroscopy experiment as before but the analysis needs to take into account some unwanted shifts of the resonant frequencies. 


\part{\LARGE Simulation and Modeling\label{part:simulation}}

\section{Simulation: preparation and properties of the target ground state\label{sec:sims_GS_prep}}

In the main text, we show experimental results for properties of the Bose-Hubbard ground state as a function of coupling strength and disorder.  The experimental results make sense for the model: we find the expected Mott insulator, superfluid, and Bose Glass phases, with reasonable locations for the transitions between them.  However, there remains the question of how close we get to preparing the true ground state, i.e. whether we experimentally find the ground state phase diagram with quantitative as well as qualitative accuracy.  

By performing extensive tensor network simulations, we verify that our experiment qualitatively reproduces all physical phenomena of the intended Bose-Hubbard model.  On the other hand, there is substantial quantitative disagreement between simulation and experiment for certain observables such as the condensate fraction.  We show that these differences are mostly accounted for by the fact that the experiment includes a finite ramp-down time between adiabatic state preparation and measurement.  We also show that, although the true Hamiltonian realized by the transmon chip is not just the nearest-neighbor Bose-Hubbard model as described in Sec.~\ref{sec:BHM} above but also includes substantial longer-ranged hopping and interaction terms (see Sec.~\ref{sec:eff_model} below for details), this \emph{does not} have a substantial effect on any measured observables.

\subsection{Details of matrix product state simulations\label{sec:MPS}}

For comparison with the experimental results, we perform large-scale matrix product state (MPS)~\citeSM{Ostlund_1995, Schollwock_2011} simulations (using the TeNPy library~\citeSM{tenpy}).  We study the same system size used in the experiments, a $4\times 9$ rectangular array of sites.  To capture the physics of the Bose-Hubbard model at filling 1, we need to include at least three states per site, $|0\rangle$, $|1\rangle$, and $|2\rangle$, giving a total Hilbert space size of $3^{36} \approx 2^{57}$.  Even when using particle number conservation, the Hilbert space size is larger than $2^{53}$, beyond what is accessible via any exact state vector method, hence the use of tensor networks.  We choose not to include the $|3\rangle$ and higher states on each site, as they are not needed to capture the physics of interest.

To gain insight into how well the experiment reproduces the ideal ground state phase diagram, we use a variety of approaches to simulating the system.  We describe each approach to simulation below, making reference to the adiabatic state preparation procedure as illustrated in Fig.~\ref{fig:adiabatic_prep_illust} above.  
With each simulation type, we compute experimental observables such as doublon fraction and condensate fraction, as well as theoretical quantities such as entanglement entropy, throughout the $J$-$W$ phase diagram.

\vspace{0.5cm}

\noindent\textbf{Ideal ground state:} We first compute the ideal ground state phase diagram.  For each coupling strength $J$ and disorder strength $W$, we use the exact same 10 disorder realizations studied in the experiment, using the density matrix renormalization group (DMRG) algorithm~\citeSM{white_1992, white_1993, Schollwock_2011} to find the ground state of the ideal Bose-Hubbard Hamiltonian that is intended to be implemented after the ramp-up and before the ramp-down/quench in the adiabatic state preparation procedure; with times defined as in Fig.~\ref{fig:adiabatic_prep_illust}, the Hamiltonian is taken at $t=110$ ns.  This is an ideal nearest-neighbor Bose-Hubbard model with uniform coupling strength $J$ and interaction strength $U$, but including the disorder in qubit frequencies.

We perform each ground state simulation with a fixed max bond dimension $\chi$, and we observe convergence of all quantities as we increase that bond dimension from 64 to 1024 in powers of 2.  The observables are already nearly converged at $\chi=64$. For example, the relative error in the condensate fraction between $\chi=64$ and 1024 is below 1\% for all but one $(g,W)$ parameter point, and the relative error between $\chi=128$ and 1024 is below $10^{-3}$ for all parameter points.

Also note one technical detail: as discussed in Sec.~\ref{sec:BHM}, on our actual device the interaction $U$ is negative, and the experiment in fact is intended to adiabatically prepare not the lowest but the highest energy state within a fixed particle number sector.  We therefore run DMRG on $-H$ rather than $H$ (defining $H$ with negative $\eta$ as in Eq.~\ref{eq:transmon_bhm}) to find the highest-excited state in the sector with unit filling.

In the figures below, results from the ideal ground state simulations are labeled ``Ideal BH GS.''

\vspace{0.5cm}

\noindent\textbf{Ideal adiabatic state preparation:} In the experiment, we do not directly prepare the ground state but rather try to reach it using adiabatic state preparation, as described in Sec.~\ref{sec:adiabaticity}.  We also use MPS simulations, with the two-site time-dependent variational principle (TDVP) algorithm~\citeSM{Haegeman2016, PAECKEL2019}, to simulate how the ideal Bose-Hubbard model behaves with the adiabatic ramps and 3 of the disorder realizations used in the experiment.  (Because the ramp does not prepare an exact eigenstate, and hence the state continues to evolve during the hold time, we average the properties over the hold window, between $t=105$ and $t=115$ ns for $t$ defined as shown in Fig.~\ref{fig:adiabatic_prep_illust}.)  

All time evolution simulations are performed with a maximum bond dimension of 128; in the ground state search this was sufficient to converge all measured quantities to sufficient accuracy for comparison with the experiment. Here there is slightly higher entanglement due to the wavefunction including components of excited states, but even so the values of observables are almost indistinguishable comparing simulations with $\chi=64$ and with $\chi=128$.  In particular the error is smaller than the typical errors in the experiment, so larger bond dimension are not necessary.

In the figures below, results from the ideal adiabatic state preparation simulations are labeled ``Ramp $\rightarrow$ Hold.''

\vspace{0.5cm}

\noindent\textbf{Effect of ramp-down after ideal adiabatic state preparation:} 
One additional effect in the experiment that has not yet been taken into account is the necessity of including a ramp-down as the coupling strength must be brought back to 0 for readout.  Ideally this would be an instantaneous quench, as shown in Fig.~\ref{fig:adiabatic_prep_illust}, in which case the wavefunction would not change before the measurement.   However, for the physical device, a perfect quench is not possible.  The true minimum ramp-down time is determined by 
details of the experimental hardware.
For comparison with the experimental results, we simulate a linear ramp-down with three different durations: 1 ns, 2 ns, and 3 ns.

As we will see in the figures below, this finite ramp-down time explains most of the discrepancy between the values of observables measured in the experiment and those computed from the target ground states as found using DMRG. 

In the figures below, results from the ideal adiabatic state preparation simulations with different lengths of ramp-down are labeled ``Ramp + [$T$]ns RD'' for $T$ being 1, 2, or 3.

\vspace{0.5cm}

\noindent\textbf{Ground state of effective device Hamiltonian:} 
As discussed in detail in Sec.~\ref{sec:eff_model} below, the physical quantum chip includes longer-ranged couplings, 3-body terms, and more that go beyond the idealized Bose-Hubbard model.  To check what effect these couplings have on the quantities measured in the experiment, we perform direct ground state search with DMRG on the effective model from Schrieffer-Wolff perturbation theory (see Sec.~\ref{sec:eff_model} for details), for comparison with the ground state results in the ideal Bose-Hubbard model.  We simulate 10 disorder configurations from the experiment.

We organize the terms in the effective models by two size metrics.  First, we consider the Manhattan distance (MD) between the farthest-separated sites acted on by operators in the term.  For example, a term like $a_1^\dagger n_2 a_3$ acting on three qubits in a line would have MD=2.  We also sort terms into 1-body (e.g. $a^\dagger a$), 2-body (e.g $n_1 n_2$), 3-body, etc.  Keeping all terms up to MD=3, 3-body, and coefficient larger than 1 kHz, for the largest couplings $J$ we have more than 17000 terms on 36 sites, which we compress (almost losslessly) using deparallelization and delinearlization~\citeSM{Hubig2017} into a matrix product operator (MPO) of bond dimension around 150.  

We investigate the convergence of observables with the number of terms kept, and we find that keeping terms up to MD=2 and up to 2-body gives results indistinguishable at the scale of experimental accuracy from keeping terms up to MD=3 and up to 3-body.  Nevertheless, we run DMRG on the full effective model including terms up to MD=3 and 3-body. 

Some of these thousands of additional terms beyond the ideal Bose-Hubbard model are large, notably the 3-body interactions labeled by $W$ in Eq.~\eqref{eq:H_eff} and Fig.~\ref{Fig:eff_ham} whose coefficients are at the same order of magnitude as the nearest-neighbor hopping.  Despite the fact that the Hamiltonian appears to be very different, as shown in the figures below, the ground state observables of interest are almost indistinguishable from those for the ideal Bose-Hubbard model.  This suggests that the state prepared in the experiment (before the ramp-down) is locally very similar to the target ground state despite the longer-ranged hopping and interactions in the true device Hamiltonian.  

In the figures below, results from the effective device model ground state simulations (with terms up to Manhattan distance 3) are labeled ``$H_\text{eff}$ GS (MD $\leq$ 3).''

\subsection{Results for various observables}

Using each of these approaches to simulation, we compute observables from the wavefunctions in order to determine the ground state phase diagram and compare with the experimental data.  Here we focus on five quantities: doublon fraction, decay of two-point correlators with distance, condensate fraction, entanglement entropy, and inverse participation ratio.

\vspace{0.5cm}

\begin{figure}
    \centering
    \subfloat{
        \includegraphics[width=\textwidth]{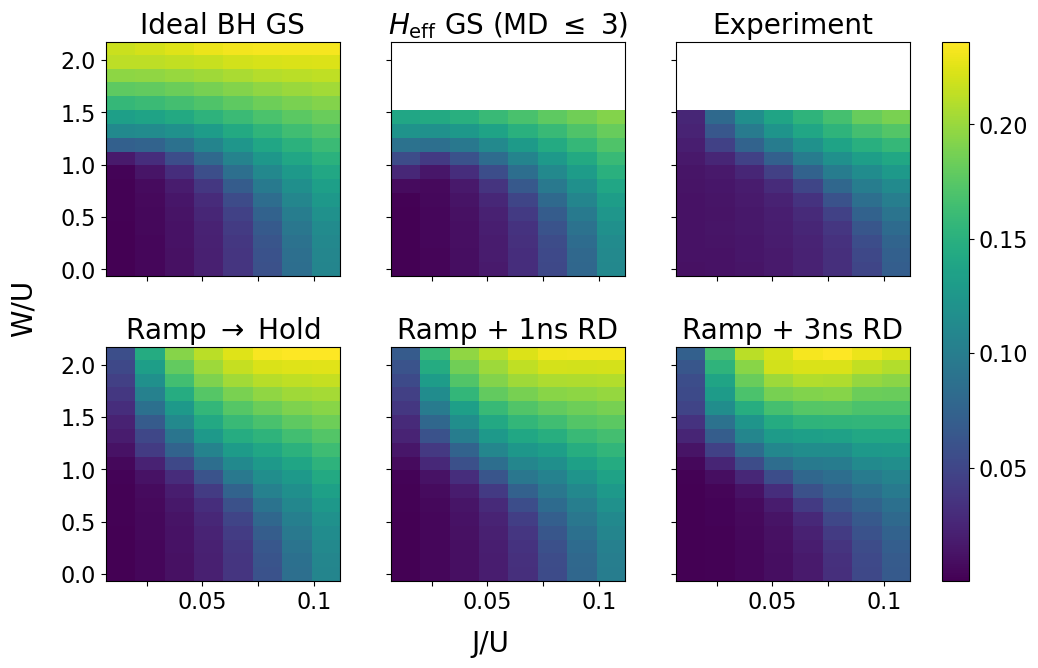}
    }
    \\
    \subfloat{
        \includegraphics[width=0.45\textwidth]{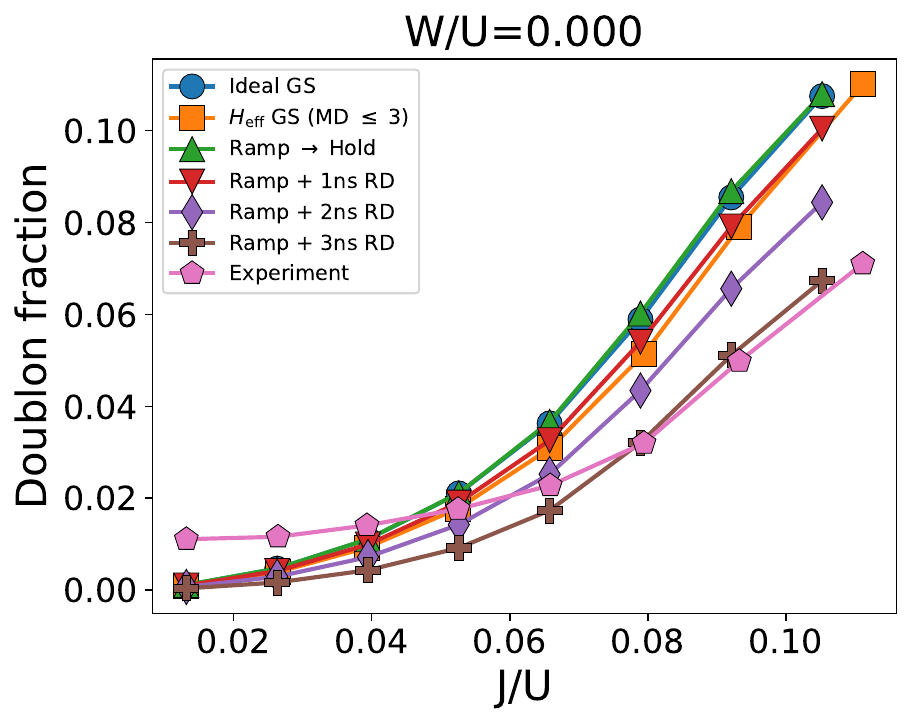}
    }
    \qquad
    \subfloat{
        \includegraphics[width=0.45\textwidth]{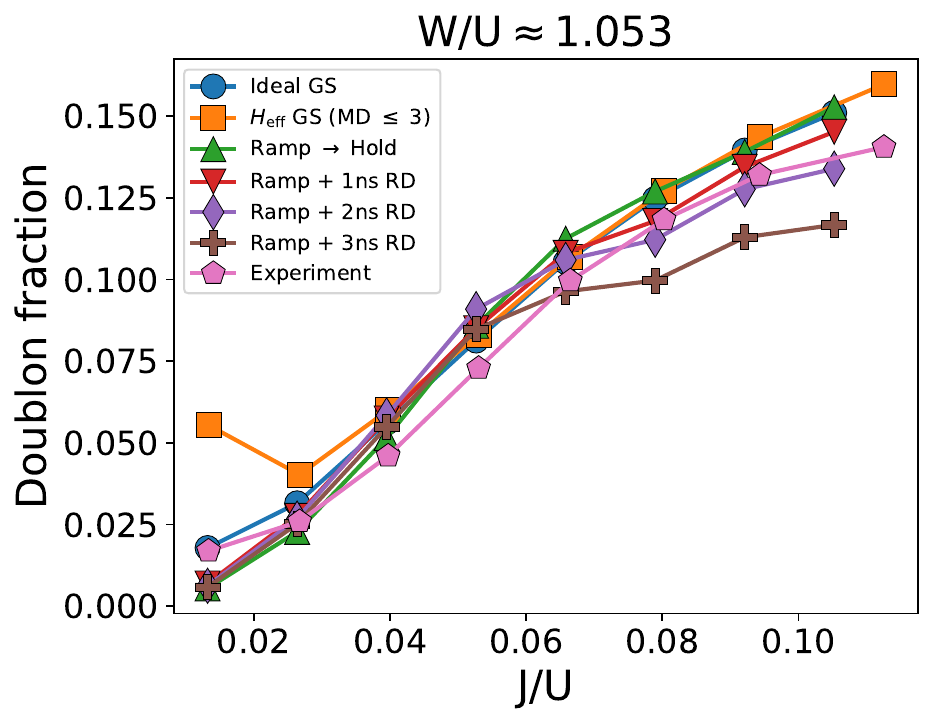}
    }
    \caption{\textbf{Doublon fraction.}
    We compare the doublon fraction found in our experiment (main text, Fig.~2, reproduced here) with that found in each of a variety of MPS simulations described in Sec.~\ref{sec:MPS} in the Supplement, including ground state search for the ideal Bose Hubbard model (``Ideal BH GS'') and for the effective device Hamiltonian with long-ranged terms up to Manhattan distance 3 (``$H_\text{eff}$ GS''), as well as adiabatic state preparation with (``Ramp + [$T$]ns RD'') or without (``Ramp $\rightarrow$ Hold'') a ramp-down before measurement.  Results from simulation are averaged over disorder instances taken directly from the experiment (10 for ground state simulations, 3 for time evolution simulations).  Bond dimensions are up to 1024 for the ideal BH model ground state, and up to 128 for all other simulations; this is sufficient to converge all observables to the level visible in the figures.  In (a) we show the results throughout parameter space, giving a qualitative picture of the extents of different phases and the rough scale of the observables.  The blank regions for experimental data and the effective model are present because the device cannot reach the largest $W$ values for the calibration settings we used.  For a more quantitative picture, in (b) and (c) we show the values as a function of coupling strength $J$ along two horizontal cuts with fixed disorder, $W/(2\pi)=0$ and $200$ MHz, respectively.  In the latter figures, the $J/U$ values for both the effective device model and the experimental data are shifted to the value derived in the effective model, rather than using the bare input $J/U$ value.} 
    \label{fig:MPS_vs_exp_doublon_frac}
\end{figure}

\noindent\textbf{Doublon fraction:} Specifically, we find $\langle a_i^{\dagger 2}a_i^2\rangle$/2 = $\langle n_i^2 - n_i\rangle$/2.  The $|0\rangle$ and $|1\rangle$ states do not contribute, so when truncating the local Hilbert space after the $|2\rangle$ state in the simulations, this quantity is precisely the probability of finding the system in $|2\rangle$ on site $i$.  The result is shown in the full $J$-$W$ parameter space in Fig.~\ref{fig:MPS_vs_exp_doublon_frac}(a), and on cuts at $W=0$ and $W/(2\pi)=200$ MHz in Fig.~\ref{fig:MPS_vs_exp_doublon_frac}(b) and (c).  The figure includes the experimental results (main text, Fig.~2) for easy comparison.

For the fixed-$W$ cuts, we shift the plotted $J/U$ ($x$-axis) values for the lines for the effective device Hamiltonian and for the experiment.  When requesting a certain coupling strength on the device (as shown in the idealized experimental trajectory, Fig.~\ref{fig:adiabatic_prep_illust}, right panel), in the effective model from Schrieffer-Wolff perturbation theory we find that both $J$ and $U$ values are modified.  Calling the new values $J_\text{SW}$ and $U_\text{SW}$, we plot using $J_\text{SW}/U_\text{SW}$ in place of $J_\text{ideal}/U_\text{ideal}$.

Also note that the simulations go to larger disorder strengths than used in the experiment, in order to give a more complete picture of the phase diagram.

\vspace{0.5cm}

\begin{figure}
    \centering
    \subfloat{
        \includegraphics[width=0.45\textwidth]{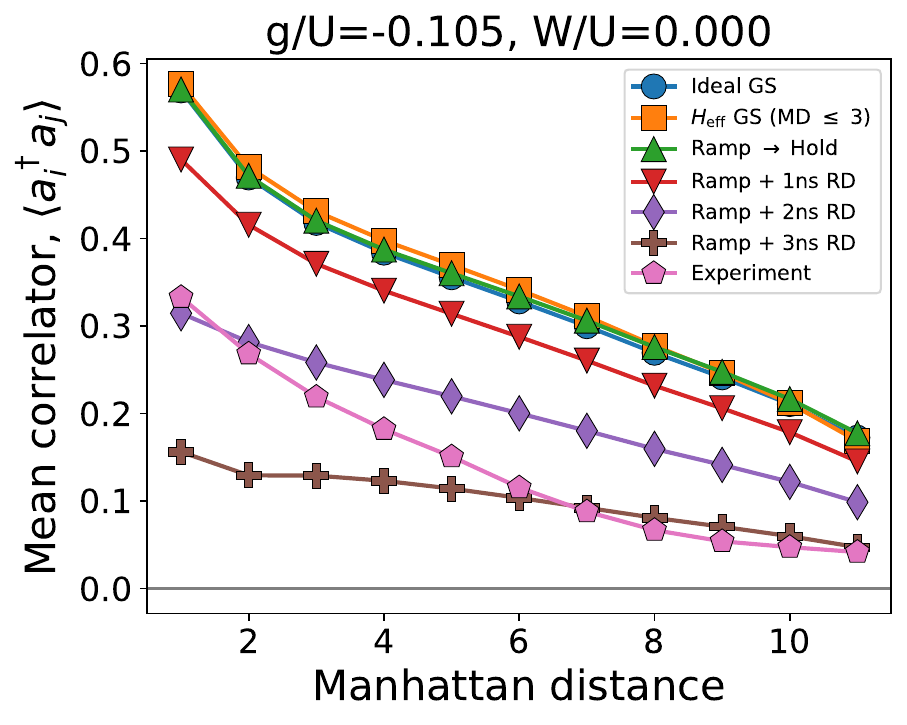}
    }
    \qquad
    \subfloat{
        \includegraphics[width=0.45\textwidth]{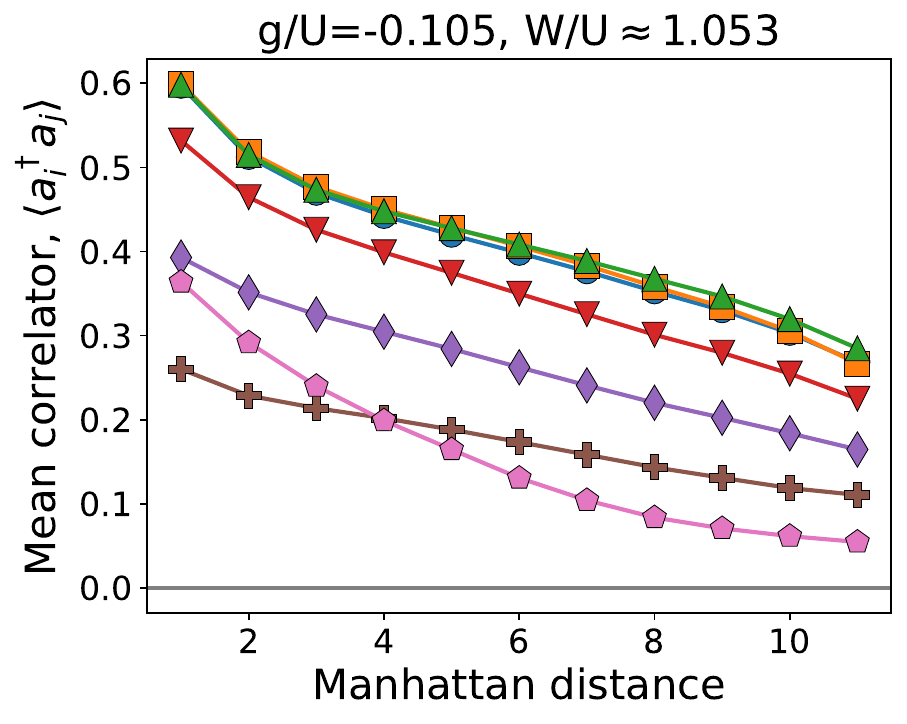}
    }
    \caption{\textbf{Correlator decay.}  We compare the decay of two-point correlators $\langle a_i^\dagger a_j\rangle$ found in our experiment with that found in each of our variety of MPS simulations described in Sec.~\ref{sec:MPS}.  Disorder instances and bond dimensions are the same as in the previous figure. 
    The experimental data are slightly different from the data shown in Fig.~4 of the main text; the correlators here are averaged over the full grid, while the correlators in the main text are taken only from the corners of the grid. 
    Results shown for $J/(2\pi) = 20$ MHz, with (a) $W=0$ and (b) $W/(2\pi) = 200$ MHz.} 
    \label{fig:MPS_vs_exp_corr_decay}
\end{figure}

\noindent\textbf{Decay of two-point correlators:} We compute in the MPS all expectation values $\langle a_i^\dagger a_j\rangle$ between sites $i$ and $j$ on the $4\times 9$ grid.  Then we sort the correlators by Manhattan distance between the sites.  For a given parameter point $(J,W)$, we plot the mean correlator as a function of Manhattan distance (averaged over pairs of sites with that separation, and over disorder realizations).  The results are shown in Fig.~\ref{fig:MPS_vs_exp_corr_decay}.  Note that the experimental data shown here are not exactly the same as those shown in the main text, Fig.~4; the correlators in the main text are taken from the corner sites of the $4\times 9$ grid, while the experimental data shown here are, like the simulation results, averaged over all correlators in the grid at the given Manhattan distance. 

In the ground state simulations (ideal model and effective model), the Hamiltonian and the observable are both real, so each expectation value $\langle a_i^\dagger a_j\rangle$ is real.  On the other hand, for the time evolution simulations the wavefunctions are complex, and $a_i^\dagger a_j$ is non-Hermitian, so the correlator is in general complex.  When we average over all correlators for a given Manhattan distance, every correlator ends up paired with its conjugate, $\langle a_i^\dagger a_j\rangle \rightarrow \langle a_i^\dagger a_j\rangle + \langle a_j^\dagger a_i\rangle$, and hence we get only the real part.  The magnitude of the computed correlator then oscillates in time even when the coupling is 0 and the qubits are in their idle configuration, because in general the sites $i$ and $j$ have different idle frequencies, $\mu_i\neq \mu_j$, so the phase of $\langle a_i^\dagger a_j\rangle$ rotates with frequency $\mu_i - \mu_j$.  Then what is the ``correct'' value of the correlator?  Ideally, between ground state preparation and measurement, no evolution would occur, and the correlator would remain purely real as in the ground state.  To capture this effect, we take the magnitude of the correlator, $|\langle a_i^\dagger a_j\rangle|$, \emph{before} averaging over all contributions at a given Manhattan distance.  Note that the problem and solution discussed here are analogous to the discussion of phase corrections in Sec.~\ref{sec:correlators} above, but the solution is much simpler for the simulations since we have direct access to both the real and imaginary parts of the correlator.

\vspace{0.5cm}

\begin{figure}
    \centering
    \subfloat{
        \includegraphics[width=\textwidth]{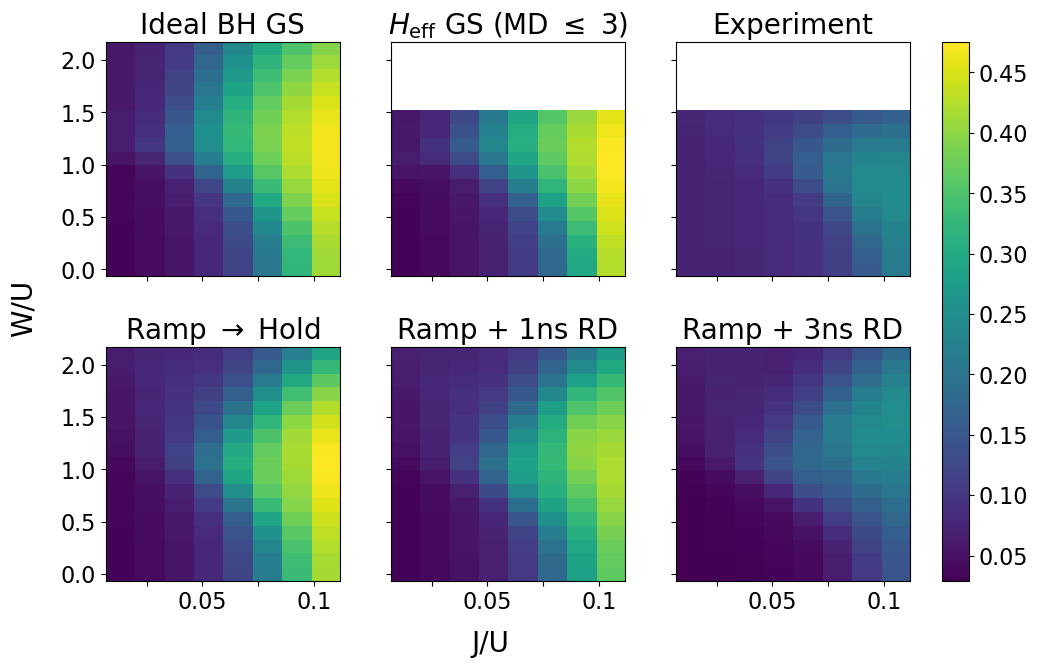}
    }
    \\
    \subfloat{
        \includegraphics[width=0.45\textwidth]{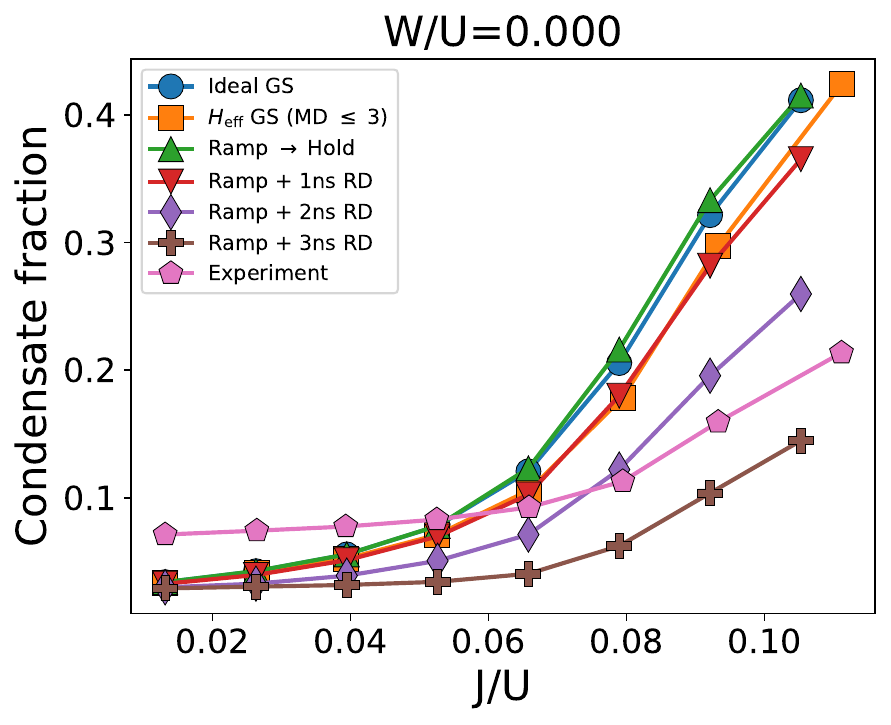}
    }
    \qquad
    \subfloat{
        \includegraphics[width=0.45\textwidth]{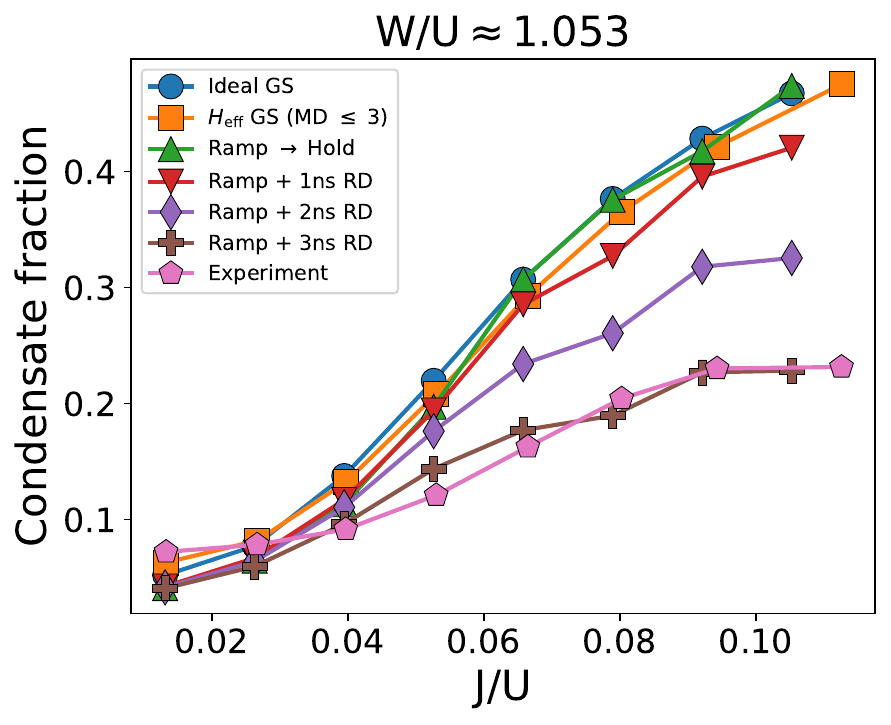}
    }
    \caption{\textbf{Condensate fraction.}
    We compare the condensate fraction found in our experiment (main text, Fig.~4, reproduced here) with that found in each of our variety of MPS simulations described in Sec.~\ref{sec:MPS}.  Disorder instances and bond dimensions are the same as in the previous figures.  (a) Results throughout $J$-$W$ parameter space.  The blank regions are present because the device cannot reach the largest $W$ values for the calibration settings we used.  (b) and (c): Results for $W/(2\pi)=0$ and $200$ MHz, respectively.  } 
    \label{fig:MPS_vs_exp_cond_frac}
\end{figure}

\noindent\textbf{Condensate fraction:} We group all computed correlators into the SPDM, with matrix elements $C_{ij} = \langle a_i^\dagger a_j\rangle$.\footnote{Note that here too we replace $\langle a_i^\dagger a_j\rangle$ by its magnitude in the case of the time-evolution simulations.  However, the effect is negligible because separate time evolution of each qubit when the hopping is turned off just multiplies each raising and lowering operator by a time-dependent phase, $a_i\mapsto a_i e^{i\theta(t)}$.  This results in a unitary conjugation of the SPDM, which leaves the eigenvalues invariant.}  The condensate fraction is then the largest eigenvalue of the SPDM, divided by the number of qubits $N_q$.  The result is shown alongside the experimental data in the full $J$-$W$ parameter space in Fig.~\ref{fig:MPS_vs_exp_cond_frac}(a), and on cuts at $W=0$ and $W/(2\pi)=200$ MHz in Fig.~\ref{fig:MPS_vs_exp_cond_frac}(b) and (c).  We also include the experimental data (shown in the main text in Fig.~4).

\vspace{0.5cm}

\begin{figure}
    \centering
    \subfloat{
        \includegraphics[width=\textwidth]{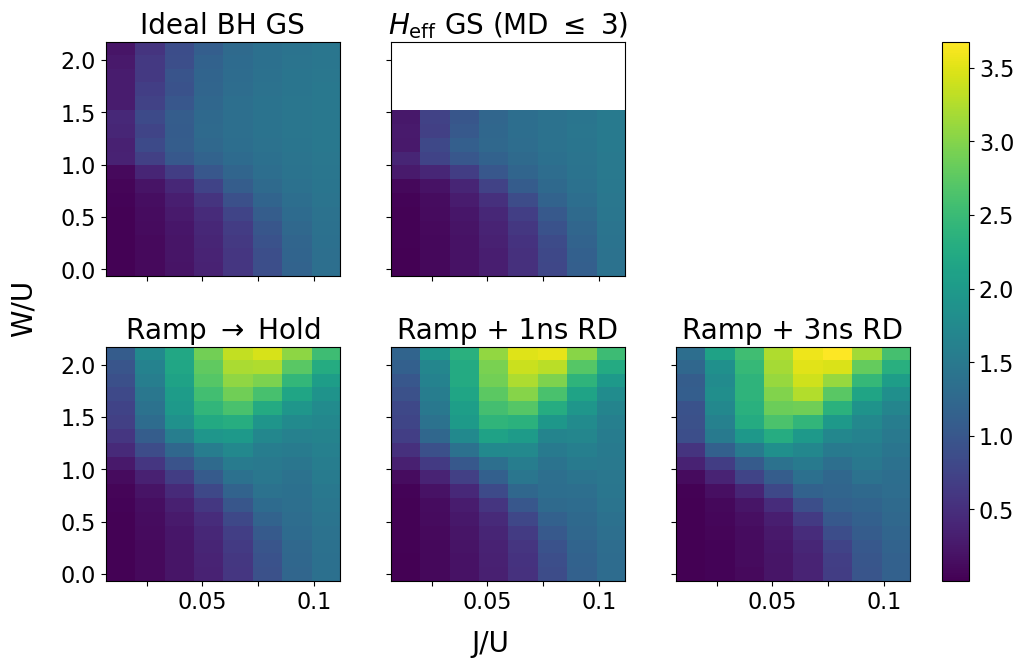}
    }
    \\
    \subfloat{
        \includegraphics[width=0.45\textwidth]{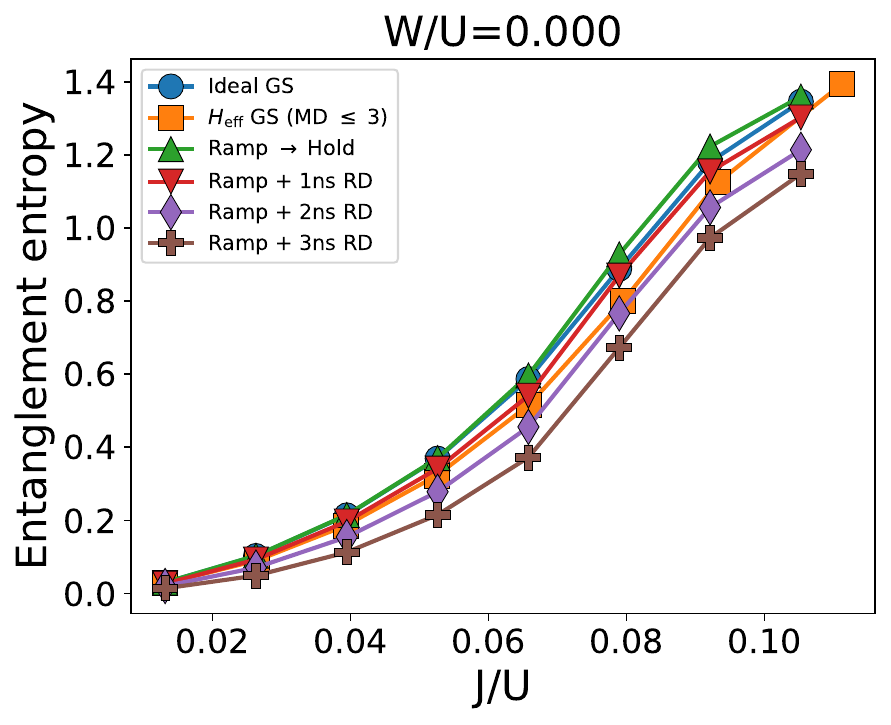}
    }
    \qquad
    \subfloat{
        \includegraphics[width=0.45\textwidth]{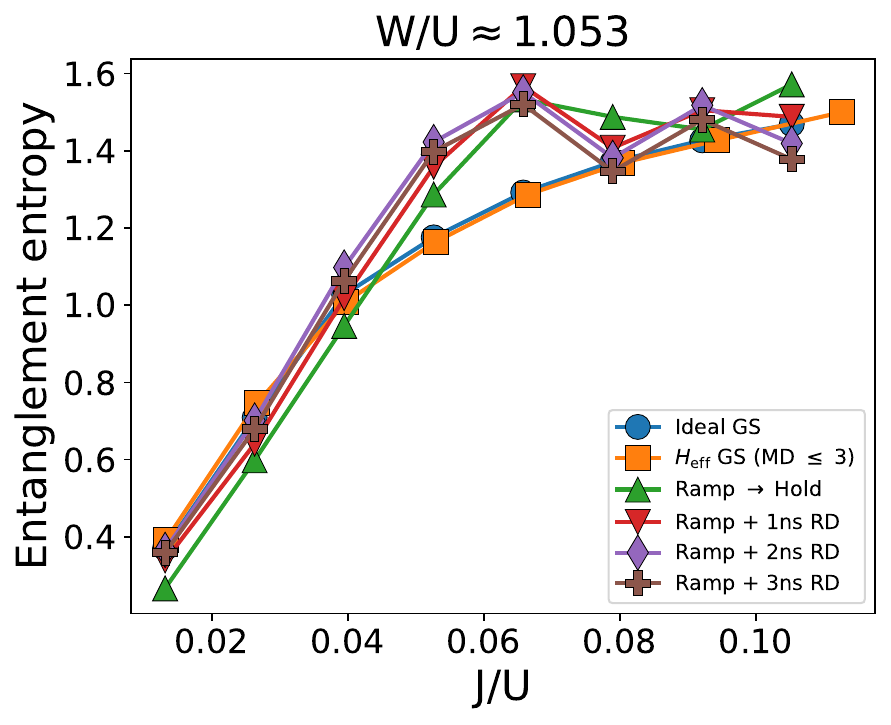}
    }
    \caption{\textbf{Entanglement entropy.}
    We compute the entanglement entropy in each of our variety of MPS simulations described in Sec.~\ref{sec:MPS}.  Disorder instances and bond dimensions are the same as in the previous figures.  (a) Results throughout $J$-$W$ parameter space.  The blank region is present because the device cannot reach the largest $W$ values for the calibration settings we used. (b) and (c): Results for $W/(2\pi)=0$ and $200$ MHz, respectively.  The entanglement clearly shows that, especially for intermediate coupling and large disorder, the ramp procedure produces a different state from the ground state of the Bose Hubbard model.} 
    \label{fig:MPS_vs_exp_entanglement}
\end{figure}

\noindent\textbf{Entanglement entropy:} We compute the entanglement entropy between left and right halves of the system.  Highly localized phases like the Mott insulator have low entanglement, while the superfluid phase is highly entangled.  The result is shown in the full $J$-$W$ parameter space in Fig.~\ref{fig:MPS_vs_exp_entanglement}(a), and on cuts at $W=0$ and $W/(2\pi)=200$ MHz in Fig.~\ref{fig:MPS_vs_exp_entanglement}(b) and (c).  Entanglement is not measured in the experiment.

\vspace{0.5cm}

\begin{figure}
    \centering
    \subfloat{
        \includegraphics[width=\textwidth]{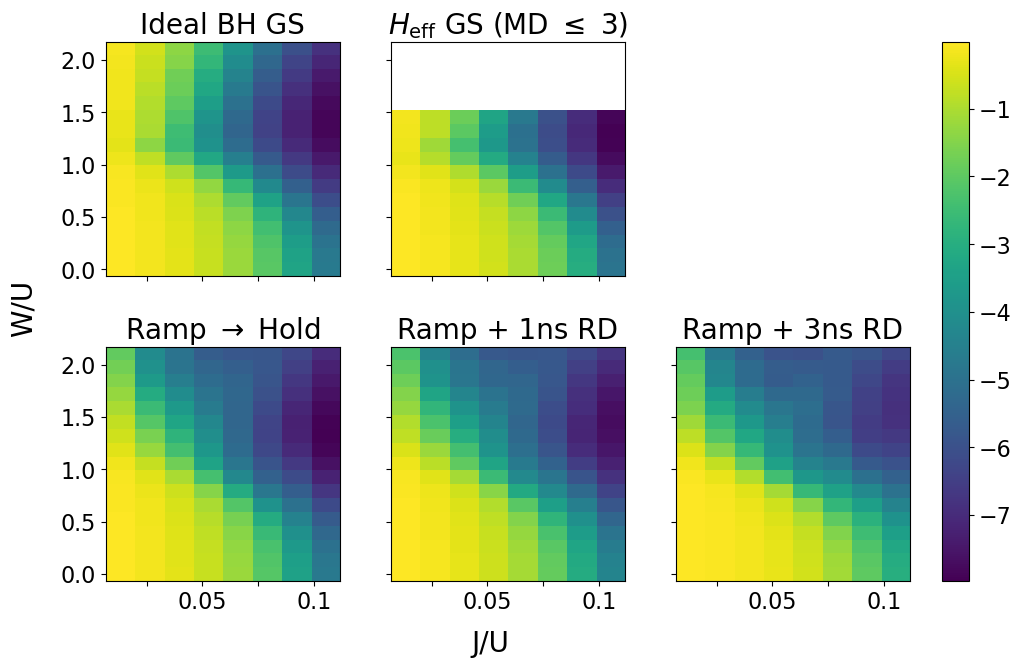}
    }
    \\
    \subfloat{
        \includegraphics[width=0.45\textwidth]{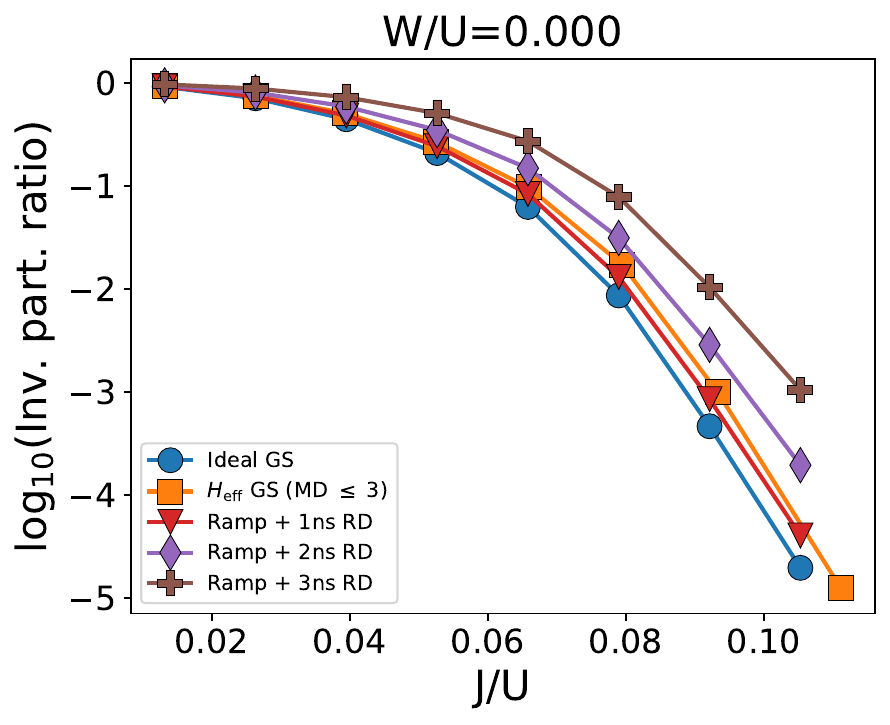}
    }
    \qquad
    \subfloat{
        \includegraphics[width=0.45\textwidth]{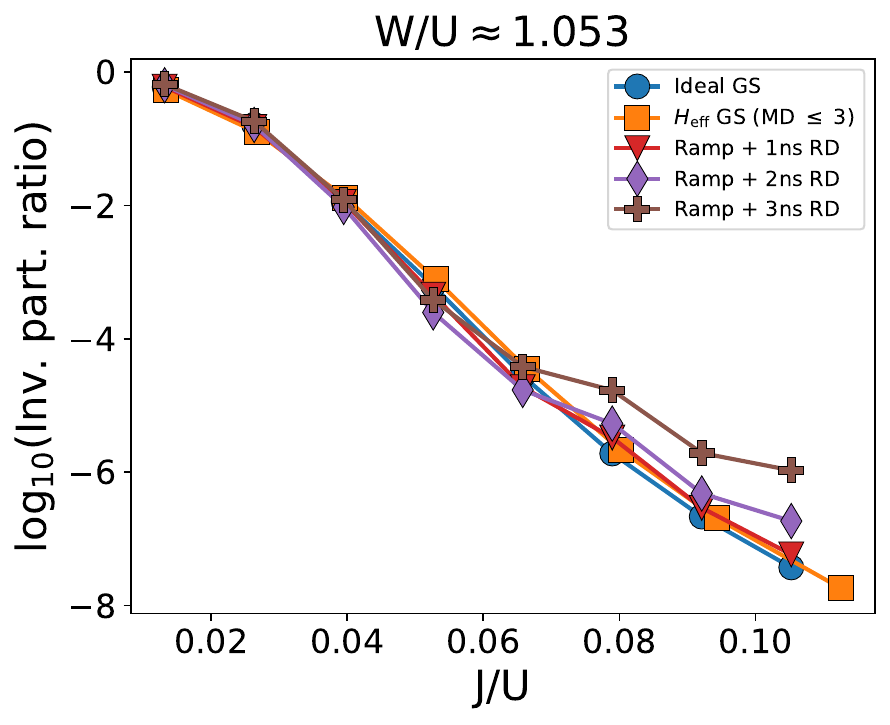}
    }
    \caption{\textbf{Inverse participation ratio.}
    We compute the inverse participation ratio (IPR) in each of our variety of MPS simulations described in Sec.~\ref{sec:MPS}; the IPR is computed directly from the wavefunction, not via sampling.  Because the value varies over more than seven orders of magnitude across the parameter regime we consider, we plot $\log_{10}(\text{IPR})$.  Disorder instances and bond dimensions are the same as in the previous figures.  (a) Results throughout $J$-$W$ parameter space.  The blank region is present because the device cannot reach the largest $W$ values for the calibration settings we used. (b) and (c): Results for $W/(2\pi)=0$ and $200$ MHz, respectively.  Like the entanglement entropy, the IPR shows that the ramp procedure prepares a state quite different from the ground state at intermediate $J$ and large $W$.} 
    \label{fig:MPS_vs_exp_IPR}
\end{figure}

\noindent\textbf{Inverse participation ratio:} The inverse participation ratio (IPR), $\sum_i |\psi_i|^4$, where $|\psi\rangle = \sum_i \psi_i |b_i\rangle$ summed over all computational basis bitstrings $b_i$.  This is computed directly from the MPS with a cost $\mathcal{O}(\chi^5)$ where $\chi$ is the MPS bond dimension.  A high value indicates a high degree of localization, corresponding to insulating and glassy phases.  The results are shown in the full $J$-$W$ parameter space in Fig.~\ref{fig:MPS_vs_exp_IPR}(a), and on cuts at $W=0$ and $W/(2\pi)=200$ MHz in Fig.~\ref{fig:MPS_vs_exp_IPR}(b) and (c).  Note that we use a log scale because the IPR varies over about 7 orders of magnitude in the parameter space we consider.

\vspace{0.5cm}

\subsection{Insights from comparison between experiment and simulation}

Comparing the ground state of the ideal model with the experimental data, the phase diagram implied by the observables is qualitatively similar: the Mott insulator at small $J$ and $W$ is characterized by small doublon and condensate fractions, as well as low entanglement and a large inverse participation ratio, indicating that the state is nearly a product.  At large $J$ there is the superfluid phase characterized by large condensate fraction and high compressibility, and at intermediate $J$ and large $W$ there is the Bose Glass phase, which has doublons and no condensate.  

However, there are some significant discrepancies.  Of particular note, at small $J$ and large $W$, we find a large density of doublons in the true ground state, which is not seen in the adiabatically prepared state in the experiment.  Furthermore, the peak condensate fraction in the experiment is reduced by approximately a factor of 2 relative to the ground state, and the peak in the condensate fraction occurs at lower values of $W$ in the experiment.

One possible source of these effects is that the ramp-up in the experiment is not sufficiently adiabatic to prepare the ground state.  Comparing the ``Ideal BH GS'' and ``Ramp $\rightarrow$ Hold'' panels in Figs.~\ref{fig:MPS_vs_exp_doublon_frac}, \ref{fig:MPS_vs_exp_cond_frac}, \ref{fig:MPS_vs_exp_entanglement}, and \ref{fig:MPS_vs_exp_IPR}, we see that the ramp procedure \emph{does} in fact explain the lower doublon count seen in experiment at low $J$ and large $W$.  When directly finding the ground state, when $W > 2U$, there will be some sites for which the energy gap (setting inter-site coupling to 0) between $|1\rangle$ and $|2\rangle$ is smaller than the average gap between $|0\rangle$ and $|1\rangle$ across all sites, so that a substantial number of doublons are expected due purely to on-site terms.  On the other hand, these states are dynamically inaccessible with small coupling strength and the ramp times used in our experiment (and hence also are not seen in the simulations of the adiabatic procedure).

The simulated adiabatic procedure still disagrees with the experiment on the magnitude of the condensate fraction in the superfluid phase, as well as on the extent of that phase in parameter space.  Adding a 1 ns ramp-down in the simulation has no significant effect on any measured observable.  On the other hand, adding a 2 ns or 3 ns ramp-down, which are probably more accurate for the real device, does bring the magnitude of the condensate fraction much closer to what is seen experimentally.  This suggests that the discrepancy from the true ground state may be due primarily to the need to non-instantaneously turn off coupling before measurement.

Finally, we can ask whether long-ranged and many-body terms in the true device Hamiltonian also provide a possible explanation for the deviation of the experiment from the ideal ground state.  However, as seen in the data, we find that none of the observables we consider are strongly impacted by these extra terms; the difference between observables in the ground state of the ideal model and the ground state in the effective device model is much smaller than the difference with the experimentally measured values.


\section{Simulation: compressibility experiments}

As discussed in Sec.~\ref{sec:compressibility} above, to measure compressibility we apply a ``tilt'' to the system via a gradient in chemical potential along the long (9-site) direction and then measure the response in the local occupations $\langle n_i\rangle$.  Because glassy phases are non-ergodic, the resulting measured susceptibility can depend on the path in Hamiltonian parameter space taken during the state preparation.  We therefore consider two different procedures for the adiabatic preparation.  In the ``field-cooled'' (FC) case, we first ramp the qubit frequencies to their tilted configuration, then ramp on the coupling, so that the system is ``cooled'' to the ground state through the adiabatic ramp with the tilt in chemical potential (the ``field'') already applied.  In the ``zero-field-cooled'' (ZFC) case, we turn on the couplings first, and only then add the tilt.  

\begin{figure}[h]
    \centering
    \subfloat{
        \includegraphics[width=0.66\textwidth]{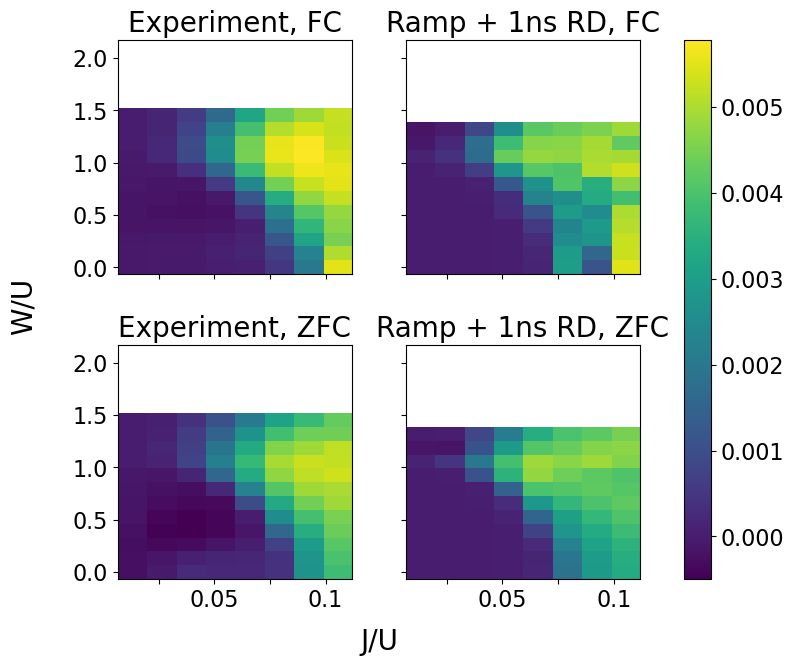}
    }
    \\
    \subfloat{
        \includegraphics[width=0.45\textwidth]{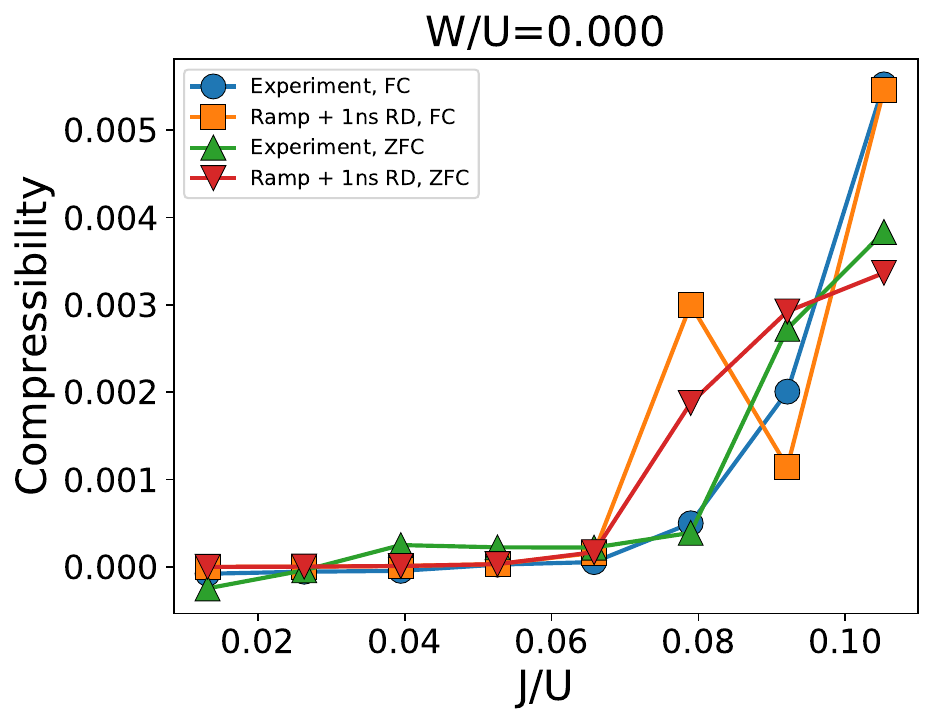}
    }
    \qquad
    \subfloat{
        \includegraphics[width=0.45\textwidth]{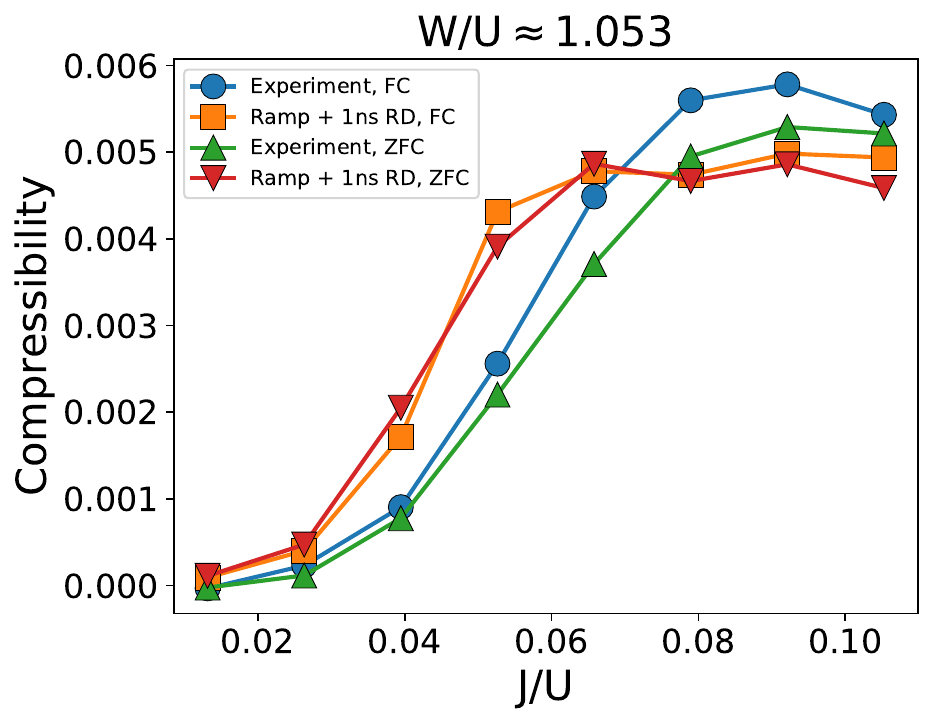}
    } 
    \caption{\textbf{Compressibility.}  
    We run MPS time evolution simulations to emulate the compressibility experiments (Fig.~3 in the main text).  The setup and computation are exactly as for the experiment, except that the experiment uses 150 disorder realizations and the simulation uses only 5.  Dynamics simulations are performed at bond dimension 128.  (a) Results throughout $J$-$W$ parameter space.  We keep the same extent in $W$ as in the previous figures for easier comparison, leaving blank the large-$W$ region that we could not access with the calibration settings we used. (b) and (c): Results for $W/(2\pi)=0$ and $200$ MHz, respectively. } 
    \label{fig:MPS_vs_exp_compressibility}
\end{figure}

Using an ideal Bose-Hubbard model (including qudit frequency disorder), we run MPS time evolution simulations with TDVP, precisely following the experimental procedure.  (All ramp times are the same, etc.)  We include a 1 ns ramp-down time, so the compressibility simulations are analogous to the simulations labeled ``Ramp + 1ns RD'' for the case of ground state preparation.  Results are shown in Fig.~\ref{fig:MPS_vs_exp_compressibility}. 

(Note that we must run time evolution simulations here, unlike for state preparation where we started with a simulation of the ideal target ground state.  Compressibility cannot meaningfully be computed from the ground state throughout much of the phase diagram, since in the ground state with large tilt and even moderate disorder, it is often favorable to put doublons on the sites with lower chemical potential and holons on the sites with higher chemical potential, even as the coupling $J$ goes to 0; the system then artificially appears compressible when in fact that state could not have been reasonable reached by an adiabatic procedure starting from uniform filling.)

Especially in the FC case, we see that the simulation results actually look noisier than the experimental data.  This is primarily due to the fact that we use just 5 disorder realizations in the simulations, compared with 150 in the experiment.  All the other observables discussed above have only small variations in value between disorder realizations, whereas the compressibility varies widely.


\section{Simulation: Bragg spectroscopy experiments}

Here we present results from simulation and theory for comparison with the response-to-a-drive experiment (see Sec.~\ref{supp:linear_response} above).  We use two different approaches: 
\begin{enumerate}
    \item In Sec.~\ref{subsec:Gutzwiller}, mean-field theory based on the Gutzwiller ansatz to obtain analytical expressions for the dispersion curves $\omega(k)$ and use them as a fitting model for our experimental data, and 
    \item In Sec.~\ref{subspace:response_TN}, time evolution within a subspace spanned by eigenstates of $H$ when no drive is applied.  The eigenstates are found as matrix product states using a multi-target DMRG algorithm.
\end{enumerate}

\subsection{Mean field theory} 
\label{subsec:Gutzwiller}
To illustrate qualitative behavior of the collective excitations in a lattice superfluid we will use mean-field theory based on the Gutzwiller ansatz for the Bose-Hubbard model: 
\begin{equation}
\label{Eq:Gutzwiller}
\left|\Psi\right\rangle=\prod_l\otimes \sum_n c_{l,n}\left |n\right\rangle_l\,
\end{equation}
where $\left|n\right\rangle_l\,$ is the local Fock state with $n$ excitations on site $l$. Since we are interested in the long-wave excitations in the vicinity of the MI-SF transition we will consider a uniform system and a truncated version of the ansatz restricting the sum in Eq.~\eqref{Eq:Gutzwiller} to $|0\rangle$, $|1\rangle$, and 
$|2\rangle$ Fock  states. As a first step, we start with the static mean-field Hamiltonian \citeSM{Sheshadri1993,vanOosten2005}: 
\begin{equation}
\label{Eq:Hmf}
H_\text{MF}=\frac{U}{2}a^\dagger a\left(a^\dagger a -1\right)-\mu a^\dagger a -J \psi (a^\dagger +a),
\end{equation}
where $a$ and $a^\dagger$ are truncated annihilation and creation operators, $\mu$ is the chemical potential, and $\psi=\frac{1}{2}\langle \chi_0|a^\dagger +a |\chi_0\rangle$ is the equilibrium superfluid order parameter. Without loss of generality we choose the equilibrium order parameter and the mean-field eigenstates to be real. 

Since the original Hamiltonian is number conserving, the mean-field ground state in the unit-filling sector, $|\chi_0\rangle$, can be expressed as follows:
\begin{equation}
|\chi_0\rangle=\cos(\varphi)|1\rangle+\frac{1}{\sqrt{2}}\sin(\varphi)\left(|0\rangle+|2\rangle\right).
\end{equation}
Here $\varphi$ is the variational parameter that minimizes the expectation value of the Hamiltonian. The minimization procedure is straightforward and allows us to find $\varphi$ as:
\begin{equation}
\label{Eq:phi}
\varphi=\frac{1}{2}\arccos(\alpha_cU/4J)\equiv \frac{1}{2}\arccos(1/\gamma),
\end{equation}
where $\alpha_c=3-2\sqrt{2}$ is the mean-field value of the critical point $4J_c/U$ for the MI-SF transition. 
Therefore the superfluid state emerges when $\gamma=4J/(\alpha_cU)\equiv 4|g|/(\alpha_c|\eta|)>1$, i.e. the parameter $\gamma -1$ is a  measure of the departure from the critical point to the superfluid phase.

Using Eq.~\eqref{Eq:phi} we can find the equilibrium values of the chemical potential 
\begin{equation}
\mu=\frac{U}{2}\left(1-\alpha_c-\frac{\alpha_c}{2}\left(\gamma -1\right)\right)
\end{equation}
and the order parameter
\begin{equation}
\psi=\frac{1}{\sqrt{8\alpha_c}}\sqrt{1-\left({1}/{\gamma}\right)^2}.
\end{equation}

To find the excitation spectrum we use the method of Refs~\citeSM{Trefzger2008,Krutitsky2011} (see also~\citeSM{Altman2002,Huber2007}). As shown in~\citeSM{Krutitsky2011} we take into account  quantum fluctuations around the mean-field solution by applying a small plane-wave perturbation to the time-dependent Gutzwiller equations  and representing coefficients $c_{l,n}$  in Eq.~\eqref{Eq:Gutzwiller} in the form of plane waves:
\begin{equation}
c_{l,n}(t)=u_{\bm k n}e^{i(\bm k\bm l-\omega(\bm k)t)}-v_{\bm k n}^*e^{-i(\bm k\bm l-\omega(\bm k)t)}
\end{equation}
This results in the eigenvalue problem for the effective non-Hermitian Hamiltonian $H_\text{eff}$:
\begin{equation}
\label{Eq:eigen}
\omega(\bm k)\left(\begin{array}{c} u_{\bm k}\\
v_{\bm k}\end{array}\right)=H_\text{eff}(\bm{k})\left(\begin{array}{c} u_{\bm k}\\
v_{\bm k}\end{array}\right),
\end{equation}
where
\begin{equation}
\label{Eq:Heff}
H_\text{eff}(\bm{k})=\sigma_z\otimes\left(H_\text{mf}-\omega_0\mathds{1}\right)-J\left(1-\xi(\bm k)\right)
\left(\sigma_z \otimes X+i\sigma_y\otimes Y\right).
\end{equation}
Here $\omega_0=(U/2)\left(1-\sqrt{2}-(1+\sqrt{2})\alpha_c\gamma\right)$ is the ground state energy, $\mathds{1}$ is the $3\times 3$ unit matrix, $\sigma_x$ and $\sigma_y$ are the Pauli matrices, $\xi({\bm k})=\sin^2(k_x/2)+\sin^2(k_y/2)$, and
\begin{subequations}
\begin{align}
X&=\chi_a\otimes\chi_a+\chi_c\otimes\chi_c\\
Y&=\chi_a\otimes\chi_c+\chi_c\otimes\chi_a
\end{align}
\end{subequations}
where $\chi_c=\left(\cos (\varphi ),\sin (\varphi ),0\right)$, and $\chi_a=\left(0,\sin (\varphi )/\sqrt{2},\sqrt{2} \cos (\varphi )\right)$ are the vectors comprised of the matrix elements $\langle n|a^\dagger|\chi_0\rangle$ and  $\langle n|a|\chi_0\rangle$, respectively. 

The eigenvalue problem \eqref{Eq:eigen} can be solved using a Bogoliubov transformation. The matrix $H_\text{eff}(\bm{k})$ in Eq.~\eqref{Eq:Heff} is non-diagonalizable and must be reduced to the Jordan form. After separating out two linearly dependent solutions corresponding to the ``condensate mode''  with $\omega=0$, the remaining $4\times 4$ matrix can be diagonalized using the Bogoliubov transformation. There are two pairs of eigenvalues 
$\pm \omega(\bm k)$ describing the excited states in question.  The positive and negative solutions in each pair are equivalent and it is sufficient to consider only the positive eigenvalues. Thus the frequencies of the two excited states at a given ${\bm k}$ take the form:
\begin{equation}
\label{omega_k}
\omega_{\mp} ({\bm k})=\frac{U}{4} \sqrt{A(\bm k)\mp \sqrt{A^2(\bm k)-B(\bm k)}},
\end{equation}   
where ``-'' stands for the lower (Goldstone) and ``+'' for the upper (Higgs) mode respectively. The functions $A(\bm k)$ and
$B(\bm k)$ can be found analytically: 
\begin{subequations}
\label{AB}
\begin{align}
A(\bm k)&=8 \alpha _c \xi (\bm k) \left(2 \sqrt{2}+\alpha _c \xi
   (\bm k)\right)+\frac{\alpha_c}{2} (\gamma -1) \left(16
   \sqrt{2} (1+\gamma )+2 \left(9+\alpha _c+3
   \left(1+\alpha _c\right) \gamma \right) \xi (\bm k)+\alpha
   _c (7+\gamma ) \xi^2 (\bm k)\right),\\
B(\bm k)&=16 \alpha_c ^2 \xi (\bm k) \left(\sqrt{2} \alpha _c
   \left(\gamma ^2-1\right)^2 (\xi (\bm k)-3)+2 (1+\gamma
   )^2 \left(\sqrt{2} \alpha _c (\gamma -1)^2+4
   \left(\xi (\bm k)+\gamma ^2-1\right)\right)\right).
\end{align} 
\end{subequations}  
In the long-wavelength limit, $\xi(\bm k)\simeq (k_x^2+k_y^2)/4$ and the spectrum is isotropic. Using a small
$k$ expansion of $\eqref{omega_k}$ we can find the speed of sound
\begin{equation}
c_s=\frac{U}{4}   \left(\sqrt{2}
   \alpha _c\right)^{1/2} (\gamma +1 )
   \left(1-\frac{\alpha _c (\gamma
   -1)}{4 \sqrt{2} (\gamma
   +1)}\right)^{1/2}\simeq\frac{U}{4}   \left(\sqrt{2}
   \alpha _c\right)^{1/2} (\gamma +1 ).
\end{equation} 

The dispersion curves overlaid with the data are shown in Fig.~5 of the main text. Quite surprisingly, after rescaling $\alpha_c$ and $\gamma$ to more realistic values $\alpha_c=0.3$ and $\gamma=1.1$, the dispersion curve for the Goldstone mode is remarkably close to the experiment.

\subsection{Tensor network subspace modeling}\label{subspace:response_TN}

In our second modeling approach, we make use of the fact that the wavefunction evolution in response to the applied drive remains within a relatively small subspace of the full Hilbert space.  In particular, we expect that the evolution mostly remains within the space of low-excited states of the Hamiltonian from before the drive is applied.

We therefore construct a subspace by finding a number of low-energy eigenstates, in the form of matrix product states.  We then project the Schrodinger equation into this subspace and solve the projected equations exactly to numerical precision.  Once the Hamiltonian and the drive are specified, the only approximation in this method is the number of eigenstates used to define the subspace.  (In practice, the Hamiltonian we use for these simulations is also approximate, since we neglect the long-ranged terms discussed in Sec.~\ref{sec:eff_model} below.)

\vspace{0.5cm}

\noindent\textbf{Excited state calculations with MPS:} The first step is to find many low-lying eigenstates of the Bose-Hubbard Hamiltonian in order to build a subspace.  Note that this way we also get interesting information about energies and other properties of the various phonon modes.  We find the excited states using the multi-targeted density matrix renormalization group (MTDMRG) method~\citeSM{Baker:2021fvd,Gonzalez2025MTDMRGX}, in which we optimize all the eigenstates at once as a multi-target MPS (MTMPS).  The MTMPS is just like a normal MPS, except that one tensor has an extra leg, of size equal to the number of states represented; when we specify the index value on that leg, the result is an MPS corresponding to one of the states.  

The multi-targeted DMRG algorithm presents a strategy to optimize the MTMPS, to simultaneously find an approximation to the ground state and to some set number of excited states.  It becomes exact as the MTMPS bond dimension $\chi\to\infty$.  We use the ITensor library~\citeSM{itensor} to implement this algorithm.  We have also verified that we find the same states when we use a standard single-targeted DMRG algorithm with an energy penalty to target the excited states (see, for example, \citeSM{Stoudenmire2012}).
\begin{figure}[h]
    \centering
    \includegraphics[width=0.7\linewidth]{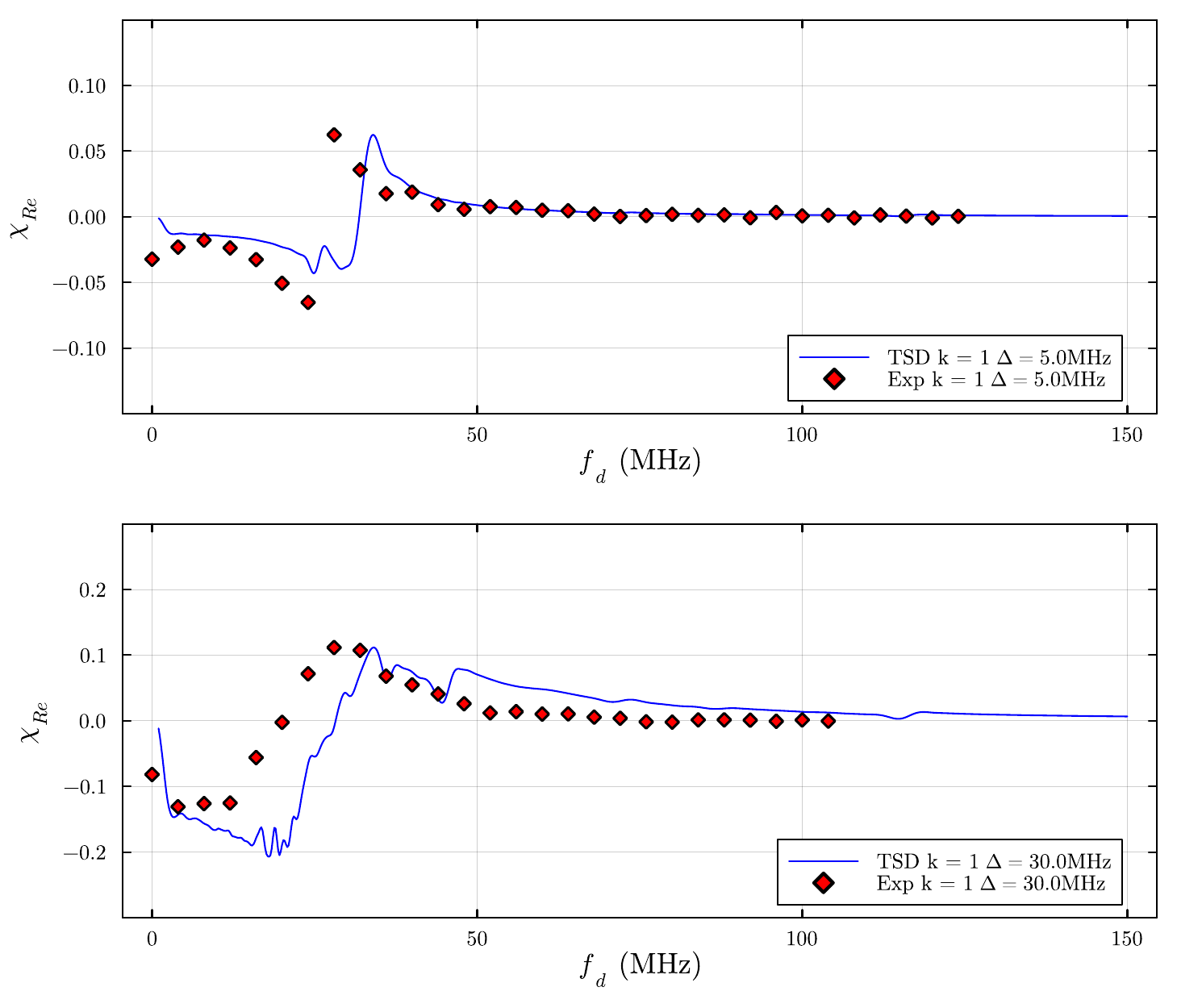}
    \caption{Real part of the density response $\chi(\omega_d)=\int_0^T e^{-i\omega_d t} \langle \psi(t)|\hat \rho_k |\psi_t\rangle$, where $\hat \rho_k = \sum_i \hat n_i \cos(k x_i)$, in the presence of a an oscillating potential,  $\hat H=\hat H_0+\hat V_k(t)$ with $\hat V_k(t)=(\Delta/2)\sin(\omega_d t) \hat \rho_k$. Here $k=\pi/L$ is the lowest spatial mode.  The zero crossing can be interpreted as a resonant frequency.  Solid blue lines are calculated by solving the exact equations of motion projected into a subspace spanned by the 36 lowest energy many-body eigenstates of $\hat H_0$.   Red diamonds are experimental data. The numerical values have been rescaled to match the peak height of the experiment.}
    \label{fig:tsd_v_exp}
\end{figure}

We obtain the $m=36$ lowest energy states of the Bose-Hubbard model for the case of zero disorder, $W=0$. The states computed using this method yield $\text{var}(\hat{H}/U) \sim10^{-4}$ and the collective bundled MTMPS has $\chi = 3912$. 

\vspace{0.5cm}

\noindent\textbf{Spectroscopy using multi-targeted states:} 
We truncate the many-body Hilbert space to the $m$ states that we have constructed, writing
\begin{equation}
    |\psi(t)\rangle =\sum_{\alpha=1}^m c_\alpha(t) |\psi^{(\alpha)}\rangle.
\end{equation}
The Schrodinger equation $i\partial_t |\psi\rangle = (H_0+V(t))|\psi\rangle$, when projected into this space, becomes a linear first order time dependent differential equation for the $c_\alpha(t)$, which can be efficiently solved numerically.  The expectation value of an operator is $\langle\psi(t)| \hat O |\psi(t)\rangle =\sum_{\alpha\beta} O_{\alpha \beta} c_\alpha^*(t) c_\beta(t)$ where $O_{\alpha\beta}=\langle \psi^{(\alpha)}|\hat O | \psi^{(\beta)}\rangle$.

We numerically follow the same procedure as in the experiment.  For each wave-vector $k$ we take the drive to be  $\hat{V}_k(t) = \frac{\Delta}{2}\sin\left(\omega_dt\right)
\hat{O}_k$
and measure the response of $\hat{O}_k = \sum_i\hat{n}_i\cos\left(kx_i\right)$ at the drive frequency, calculating
\begin{align}
    \chi(\omega_d)=\frac{1}{T}\int_0^T e^{-i \omega_d t} \langle\psi(t)| \hat O_k|\psi(t)\rangle.
\end{align}
Figure~\ref{fig:tsd_v_exp} shows the real part of the resulting spectrum, using the experimental time window, $T=250$ns, for two different drive strengths.  We rescale the vertical axis so that the peak height matches the experiment.

In both cases the line-width is largely set by nonlinearities -- and the stronger drive yields a broader response.  
The nominal resonance location, corresponding to where the real part of $\chi$ vanishes, shifts to lower frequencies with increasing drive.
The model appears to capture the shape of the response curve, but there are some quantitative discrepancies.  

We have repeated these calculations using half as many excited states, and find almost no change to the numerical spectra.  This suggests that we have captured the most important many-body states for the dynamics.  When we increase the drive strength further, we find we need to include more states.


\section{Details of device modeling: deriving the effective Hamiltonian \label{sec:eff_model}}

In this section, we describe our approach for accurate device calibration and modeling, building upon the techniques introduced in Ref.~\citeSM{Andersen2025}.
Our analysis consists of two steps: first, we characterize the \emph{bare} microscopic degrees of freedom on our device---i.e.~the active qudits and the couplers; second, we ``integrate out'' the couplers---by combining the exact Schrieffer-Wolff transformation and a linked-cluster expansion---to obtain an \emph{effective} Hamiltonian acting on the qudits alone. 
Notably, in the latter step, we take into account high-order, coupler-mediated processes which generate a variety of corrections beyond the standard Bose-Hubbard model (see Eq.~\ref{eq:H_eff}). 
Nevertheless, as shown in Section \ref{sec:MPS}, these high-order corrections have a relatively small impact on the physical observables considered in this study; thus, the standard Bose-Hubbard model remains a good approximation for capturing the physics of interest.

\vspace{0.5cm}

\noindent \textbf{Bare parameter calibration:} As in Ref.~\citeSM{Andersen2025}, we first model the \emph{bare} Hamiltonian of our device as a set of coupled transmons:
\begin{equation} \label{eq:H_bare}
H_\textrm{bare} = \sum_{i \in Q,C} \left (\omega_i a_i^\dagger a_i +  \frac{\eta_i}{2} a_i^\dagger  a_i^\dagger  a_i  a_i \right) + \sum_{i, j \in Q,C} \frac{k_{ij}}2 \sqrt{\omega_i \omega_j} ( a_i +  a_i^\dagger) ( a_j +  a_j^\dagger)
\end{equation}
where $ a_i$ are bosonic operators acting on the combined set of qudits  and couplers, $\omega_i$ are the frequencies, $\eta_i$ are the anharmonicities, and $k_{ij}$ are the capacitive coupling efficiencies. 
For the coupling efficiencies, we include nearest-neighbor qudit-qudit  and qudit-coupler 
couplings (as depicted in Fig.~\ref{Fig:circuit_coupler}), as well as next-nearest-neighbor qudit-qudit and coupler-coupler couplings which arise in larger-scale lattices [Fig.~\ref{Fig:cluster_SW}(a)].

The parameters in this model ($k_{ij}$, $\eta_i$, and $\omega_i$) are determined through a combination of spectroscopy experiments and device modeling, following the procedures established in Ref.~\citeSM{Andersen2025}.
While $k_{ij}$ and $\eta_i$ are static parameters intrinsic to the device fabrication, the frequencies $\omega_i$ are calibrated for a specified experimental sequence to set the nearest-neighbor interaction strength, $|J|$, and onsite disorder.

\vspace{0.5cm}
\noindent \textbf{Effective modeling:} We next turn to deriving an \emph{effective} Hamiltonian acting only on the qudit degrees of freedom.
This simplification is motivated by the large detuning between the qudits and couplers, $\Delta \equiv |\omega_c - \omega_q| \sim (2\pi) 2$ GHz, relative to the interaction strengths, $g_{qc} \equiv k_{qc} \sqrt{\omega_q \omega_c} \sim (2\pi) 100$ MHz. 
The separation of energy scales ensures that the coupler excited states remain weakly hybridized with the qudit sector and can be decoupled (or ``adiabatically eliminated'') through a basis transformation.

To accurately derive the effective Hamiltonian, we introduce a scalable technique which combines the exact Schrieffer-Wolff transformation and a linked-cluster expansion.
These two components are described in more detail below: 

\begin{figure*}[t]
\centering
\includegraphics[width=0.6\textwidth]{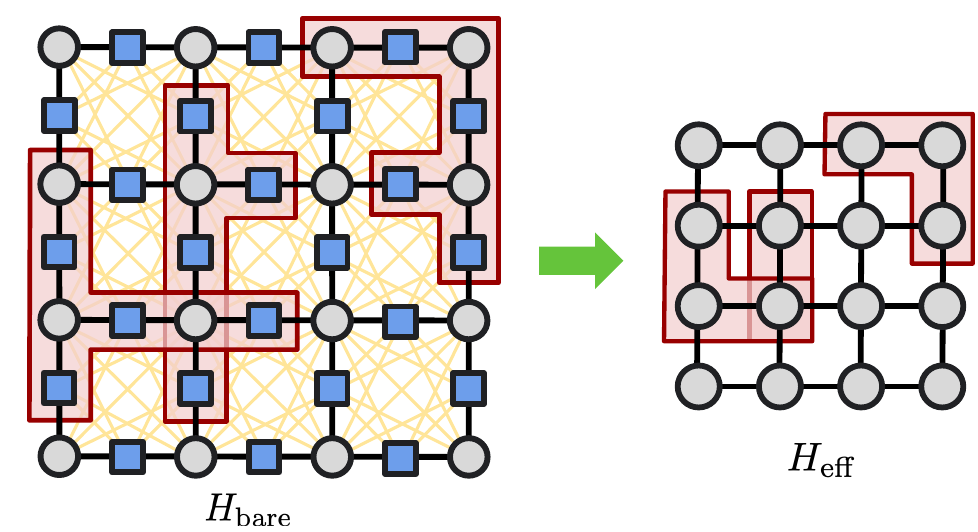}
\caption{\textbf{Device modeling procedure.} The bare Hamiltonian, $H_{\textrm{bare}}$, is modeled as a lattice of interacting qudits (grey circles) and couplers (blue squares). Capacitive couplings, $k_{ij}$ are included for nearest-neighbor qudit-coupler and qudit-qudit pairs (black lines), along with next-nearest-neighbor pairs (yellow lines). The effective Hamiltonian, $H_{\textrm{eff}}$---acting on the qudits alone---is obtained by combining the exact Schrieffer-Wolff transformation and a linked-cluster expansion. Three representative clusters of size $k=7$ are shown in the bare Hamiltonian (red regions); their corresponding support is highlighted in the effective Hamiltonian. } \label{Fig:cluster_SW}
\end{figure*}

\vspace{0.5cm}

\noindent \emph{Schrieffer-Wolff transformation}---Formally, the Schrieffer-Wolff transformation identifies the minimal unitary rotation, $U$, that maps an unperturbed subspace $\mathcal{P}_0$ to a ``dressed'' subspace $\mathcal{P}$ \citeSM{bravyi2011schrieffer}.
In our context, the unperturbed subspace $\mathcal{P}_0$ is the set of states with excitations only within the qudit degrees of freedom, i.e.~for which the couplers remain in the ground state.
These are approximate eigenstates of the system  when the couplers are set to their ``idle'' frequencies (see Eq.~\ref{eq:omega_idle}). 
The dressed subspace $\mathcal{P}$ is the corresponding set of eigenstates of the bare Hamiltonian when the couplings are activated---i.e.~set to their ``interaction'' frequencies. 
Owing to adiabaticity, the system is expected to smoothly evolve from the bare subspace to the dressed subspace as the coupler frequencies are ramped to their interaction frequencies.  
Thus in practice one can identify $\mathcal{P}$ by exactly diagonalizing $H_\text{bare}$ and determining the eigenstates with maximal overlap with $\mathcal{P}_0$.

For a finite-sized system, the exact form of $U$ can be computed through a straightforward numerical procedure. 
Specifically, the exact transformation is obtained by $U = \sqrt{(2 P - I)(2P_0 - I)}$, where $I$ is the identity operator, $P_0$ is the projector onto $\mathcal{P}_0$, and $P$ is the projector onto $\mathcal{P}$~\citeSM{bravyi2011schrieffer}.
The effective Hamiltonian within the dressed subspace is then given by:  $H_\textrm{eff} = P_0 U^{\dagger} H_\textrm{bare} U P_0$. 
In practice, however, this numerical procedure is restricted by the high computational cost of exact diagonalization to systems with up to $\sim$ 10-20 degrees of freedom.

\vspace{0.5cm}

\noindent \emph{Linked-cluster expansion}---To scale our analysis to large systems, we integrate the Schrieffer-Wolff transformation into a linked-cluster expansion framework~\citeSM{bravyi2011schrieffer,gelfand2000high,hormann2023projective}. 
In particular, we first utilize the exact SW transformation to derive effective Hamiltonians for a series of overlapping ``linked clusters''. 
We then aggregate these local contributions---according to the inclusion-exclusion principle---to obtain a global effective Hamiltonian. 
This combined approach ensures a computational cost that scales with the system size while systematically capturing high-order processes. 

More explicitly, the total effective Hamiltonian is given by the sum over linked clusters~\citeSM{gelfand2000high}:
\begin{equation} \label{eq:cluster expansion}
H_{\textrm{eff}} = \sum_{c} W(c),
\end{equation}
where $W(c)$ is the so-called cluster weight defined recursively by
\begin{equation}
W(c) =  H_{\textrm{eff}}(c) - \sum_{c^\prime \subset c} W(c^\prime).
\end{equation}
Here, $H_{\textrm{eff}}(c)$ is the effective Hamiltonian on cluster $c$, and the final sum is over subclusters of $c$.
Physically, the cluster weight describes the fully ``connected'' piece of the effective Hamiltonian which cannot be reduced to smaller subclusters.

In our implementation, we define linked clusters to be sets of sites---qudits or couplers---in which each site is adjacent to at least one other site in the cluster (according to the Manhattan distance). 
We perform the cluster expansion with clusters up to size $k=7$, where $k$ is the total number of sites within the cluster. 
A few examples of $k=7$ clusters are shown in Fig.~\ref{Fig:cluster_SW}. 
In practice, we enumerate \emph{all} linked clusters of size $k=7$ within the system; then, by taking the intersections between the clusters, we determine all subclusters recursively.  

For each cluster $c$, we compute the effective Hamiltonian $H_\textrm{eff}(c)$ as follows:
\begin{itemize}
\item We isolate the bare cluster Hamiltonian $H_\textrm{bare}(c)$ with support on $c$ and truncate onto states with up to 4 excitations per qudit, 3 excitations per coupler, and 7 excitations in total. (These dimensions were determined to lead to negligible truncation errors.)
\item We perform the exact SW transformation on $H_\textrm{bare}(c)$ with respect to particle-conserving qudit sectors with up to 3 excitations. In particular, the $n$-particle effective Hamiltonian is given by $H_{\textrm{eff}}^{(n)} = P_0^{(n)} U^{(n) \dagger} H_\textrm{bare} U^{(n)} P_0^{(n)}$, where $P_{0}^{(n)}$ is the projection onto states with $n$ excitations in the qudits (and no excitations in the couplers), and $U^{(n)}$ is defined analogously. The effective Hamiltonian for the cluster is the direct sum of the individual sectors, i.e.~$H_\textrm{eff}(c)$ = $H_\textrm{eff}^{(0)}(c) \oplus H_\textrm{eff}^{(1)}(c) \oplus H_\textrm{eff}^{(2)}(c) \oplus H_\textrm{eff}^{(3)}(c)$. 
\end{itemize}

By summing the effective cluster Hamiltonians (according to Eq.~\ref{eq:cluster expansion}), we derive a global effective Hamiltonian with up to 3-body bosonic terms on 4 adjacent qudits.
To check convergence, we also perform the cluster expansion for varying maximum cluster sizes. 
As expected for a weakly hybridized system, we observe that the terms in the effective Hamiltonian converge as the cluster size increases; based on this convergence, we estimate the accuracy of each term to be $\lesssim 10$ kHz.

\begin{figure*}[th!]
\centering
\includegraphics[width=\textwidth]{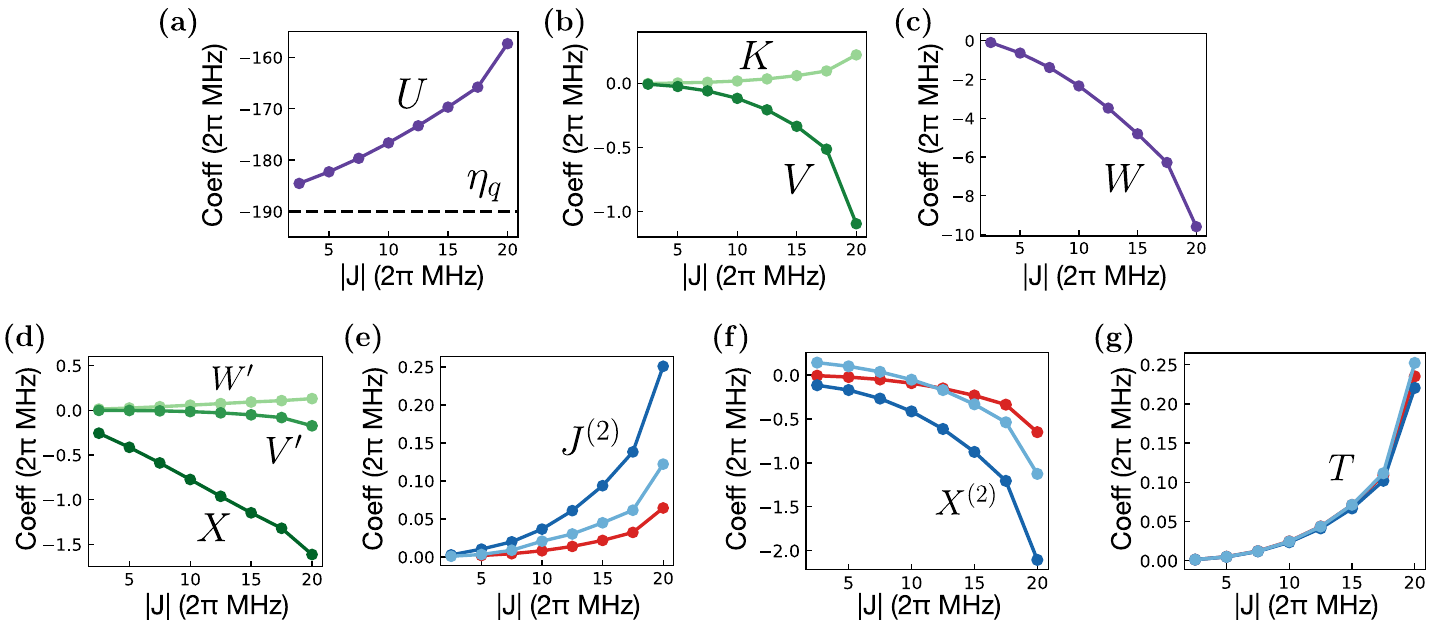}
\caption{Average coefficients in the extended Bose-Hubbard Hamiltonian, Eq.~\eqref{eq:H_eff}, as a function of $|J|$ for \textbf{(a)} 2-body onsite interactions, $U$; \textbf{(b)} density-density nearest-neighbor interactions, $V$ and $K$; \textbf{(c)} 3-body onsite interactions, $W$; \textbf{(d)} density-dependent nearest-neighbor hoppings, $X$, $V^\prime$, and $W^\prime$; \textbf{(e)} next-nearest neighbor hoppings, $J^{(2)}$; \textbf{(f)} next-nearest neighbor density-dependent hoppings, $X^{(2)}$; and \textbf{(g)} density-mediated 3-site hoppings. For (a), we compare the effective onsite interaction to the intrinsic qudit anharmonicity, $\eta_q$ (dashed). For (e), (f) and (g), we plot the three relative next-nearest-neighbor orientations: along the two diagonal directions (light, dark blue) and along a line (red). For (g), the hopping occurs between next-nearest sites along the same three orientations and the density operator acts on the site in the middle.} \label{Fig:eff_ham}
\end{figure*}

\vspace{0.5cm}

\noindent \textbf{Extended Bose-Hubbard model:} The resulting effective models are described by the following extended Bose-Hubbard Hamiltonian:
\begin{equation}\label{eq:H_eff}
\begin{split}
H_{\textrm{eff}} &= -\sum_{\langle i, j \rangle} J_{ij} (a^\dagger_i a_j + a^\dagger_j a_i ) -\sum_{i} \mu_i n_i + \sum_{i} \frac {U_i} 2 n_i(n_i-1) 
\\ &+ \sum_{\langle i, j \rangle } V_{ij} n_i n_j + \sum_{\langle i, j \rangle} K_{ij} \left[n_i (n_i-1) n_j + h.c.\right] + \sum_i \frac{W_i}{6} n_i (n_i-1) (n_i -2) 
\\ & - \sum_{ \langle i, j \rangle} X_{ij}  \left[a^\dagger_i (n_i + n_j) a_j + h.c. \right]  - \sum_{ \langle i, j \rangle} V^\prime_{ij} \left[  a^\dagger_i  n_i n_j a_j + h.c. \right ] - \sum_{ \langle i, j \rangle} W^\prime_{ij} \left(  a^\dagger_i  \left[n_i(n_i-1) + n_j(n_j-1) \right ]a_j + h.c. \right )
\\ &- \sum_{\langle \langle i, j \rangle \rangle} J^{(2)}_{ij} (a^\dagger_i a_j + a^\dagger_j a_i ) - \sum_{\langle \langle i, j \rangle \rangle} X^{(2)}_{ij}  \left[a^\dagger_i (n_i + n_j) a_j + h.c. \right] - \sum_{\langle i, j, k \rangle} T_{ijk} n_i (a_j^\dagger a_k + a_k^\dagger a_j) 
\end{split}
\end{equation}
The first line contains terms that appear in the leading-order analysis described in Section \ref{sec:BHM} (i.e.~up to second order in $g_{qc}/\Delta$). 
The remaining lines contain terms beyond the standard Bose-Hubbard Hamiltonian. 
We note that the effective Hamiltonian contains additional terms beyond those explicitly listed, but their magnitude is generally below $\sim 50$ kHz and their contribution to physical observables of interest is negligible.
    
In Fig.~\ref{Fig:eff_ham}, we plot the average coefficient for each term in the above Hamiltonian for the 4x9 array of qudits as a function of the nearest-neighbor hopping strength, $|J|$. 
The largest correction in the effective parameters is the deviation of the Hubbard interaction ($U$) from the intrinsic qudit anharmonicity ($\eta/(2\pi) \approx -190$ MHz).
This correction arises from a dispersive shift between an adjacent qubit and coupler that occurs at the same order as the interaction $J$ itself (i.e.~$\mathcal{O}(g_{qc}^2/\Delta$)).
Other terms that vary linearly in $|J|$ are the density-dependent two-site hopping $X$ and the three-body onsite term $W$.
The remaining terms originate from higher-order perturbative processes; consequently, their magnitudes increase non-linearly in $|J|$ and diminish rapidly as a function of both interaction order (e.g.~3-body vs.~2-body) and spatial distance (e.g.~next-nearest-neighbor vs. nearest-neighbor). 

As shown in Section \ref{sec:MPS}, the higher-order terms in the extended Bose-Hubbard model shift the values of the observables studied in this work by approximately 1–10\%. 
This variation roughly matches the relative magnitude of the higher-order terms compared to the primary coupling strength, $|J|$. 

\bibliographystyleSM{apsrev4-2}
%


\end{document}